\documentclass[final]{siamltex}
\usepackage{amsmath,amssymb,graphicx}
\pdfoutput=1
\usepackage{hyperref}

 \title{Numerical study of the transverse stability of NLS soliton solutions in several classes of NLS type 
 equations\thanks{
 We thank J.C. Saut and C. Klein for helpful discussions and hints. 
 This work has been supported by the Austrian Science Foundation FWF, project SFB F41 (VICOM) and project I830-N13 (LODIQUAS). 
 We are grateful for access to the HPC resources from GENCI-CINES/IDRIS (Grant 2013-106628) on which part of the 
 computations in this paper has been done, the CRI (Centre de Ressources Informatiques) of the university of Bourgogne and to the Vienna Scientific Cluster (VSC).
 }}

\author{K.~Roidot\thanks{Fak. Mathematik, Univ. Wien,
Oskar-Morgenstern-Platz 1, 1090 Wien, Austria
    ({\tt kristelle.roidot@univie.ac.at})}
    \and
N. Mauser\thanks{WPI c/o Fak. Math., Univ. Wien,
Oskar-Morgenstern-Platz 1, 1090 Wien, Austria, 
    ({\tt mauser@courant.nyu.edu})}
}
\begin{document}
\maketitle
\begin{abstract}
Dispersive PDEs are important both in applications (wave phenomena e.g. in hydrodynamics, 
nonlinear optics, plasma physics, Bose-Einstein condensates,...) and a mathematically very challenging class of partial differential equations, especially in the time dependent case. \\
An important point with respect to applications is the stability of exact solutions like solitons. 
Whereas the linear or spectral stability can be  addressed analytically in some situations,
the proof of full nonlinear (in-)stability remains mostly an open question.
In this paper, we numerically investigate the transverse (in-)stability 
of the solitonic solution to the one-dimensional cubic NLS equation, the well known \emph{isolated soliton}, 
under the time evolution of several higher dimensional models, 
being admissible as a tranverse perturbation of the 1d cubic NLS.
One of the recent work in this context \cite{RT09} allowed to prove the instability of the soliton, under 
the flow of the classical (elliptic) 2d cubic NLS equation, for both localized or periodic perturbations.
The characteristics 
of this instability stay however unknown. Is there a blow-up, dispersion..?
We first illustrate how this instability occurs for the elliptic 2d cubic NLS equation
and then show that the elliptic-elliptic 
Davey Stewartson system (a (2+1)-dimensional generalization of the cubic NLS equation) behaves 
as the former 
in this context.
Then we investigate hyperbolic variants of the above models, for which no theory in this context is available. 
Namely we consider the hyperbolic 2d cubic NLS equation and the Davey-Stewartson 
II equations.
For localized perturbations, 
 the isolated soliton appears to be unstable for the former case, but seems to be 
orbitally stable for the latter.
For periodic perturbations the soliton is found to be unstable for all transversally perturbed models
considered.

\end{abstract}

\section{Introduction}

Nonlinear Schr\"odinger (NLS) equations have many applications, e.g.~in quantum physics, 
hydrodynamics, plasma physics and nonlinear optics where they can be 
used to model the amplitude modulation of weakly nonlinear, strongly 
dispersive waves. \\
The Cauchy problem for the standard ("elliptic") NLS is given by
\begin{equation}
\left\{
\begin{array}{ccc}
i u_t + \frac{1}{2} \Delta u -\frac{\rho}{\sigma}\left|u \right|^{2\sigma}u & = & 0 \\
u(\mathbf{x}, t=0) & = & u_0(\mathbf{x}),
\end{array}
\right.
\label{NLSellgen}
\end{equation}
where $u: \mathbb{R}^d \times \mathbb{R} \to \mathbb{C}$ is a complex-valued function 
of time and space, $\Delta$ is the $d$-dimensional Laplacian in the space variables, 
$\rho=\pm1$ and $\sigma \geq 0$.
The cases $\rho=-1$ or $\rho=1$ are known as the \emph{focusing} and \emph{defocusing} equations, respectively. \\
For $u_0 \in H^1(\mathbb{R}^d)$, the
local (in time) existence  
of a unique solution of (\ref{NLSellgen}) in $H^1(\mathbb{R}^d)$ holds 
for $0 \leq \sigma < \frac{2}{d-2}$ if $d \geq 3$ (there is no condition when $d=1$ or $2$), 
and this local solution extends to all times if $\sigma< \frac{2}{d}$ 
in the focusing case (no further condition in the defocusing case) \cite{Bourg99, Cazenave12003, SS}. 
At or above this critical value, finite time blow up can occur.  
This defines the notion of criticality
for existence of blowup solutions for (\ref{NLSellgen}).
The situations when $\sigma d < 2$ are then referred to as \emph{subcritical dimensions}, 
$\sigma d = 2$, the \emph{critical dimension}, and $\sigma d \geq 2$ the \emph{supercritical dimensions}.
When a solution exists, conserved quantities for (\ref{NLSellgen}) are the mass (\ref{massgenNLS}) 
and the energy (or Hamiltonian) (\ref{nrjgenNLS}),
\begin{eqnarray}
\label{massgenNLS}
& N:=  \underset{\mathbb{R}^d}{\int} |u|^2 d\mathbf{x} & \\
 & E:=  \underset{\mathbb{R}^d}{\int} \left( |\nabla u|^2 - \frac{\rho}{\sigma + 1} |u|^{2\sigma + 2} \right) d\mathbf{x} &
 \label{nrjgenNLS}
\end{eqnarray}
\\
In the focusing case ($\rho=-1$) and with the condition $u \to 0$ as $|\mathbf{x}| \to \infty$, the initial value problem 
(\ref{NLSellgen}) admits the well known \emph{solitary waves} solutions of the form   
$u(\mathbf{x},t) = e^{i\lambda^2t} \Phi(\mathbf{x})$, where $\Phi$ satisfies
\begin{equation}
 \Delta \Phi - \lambda^2 \Phi + |\Phi|^{2\sigma} \Phi = 0, \,\, \lambda^2>0, \,\,  \underset{|\mathbf{x}| \to \infty}{\lim} \Phi(\mathbf{x}) = 0.
\end{equation}
An important question with respect to applications concerns the stability or instability of such solitary waves.
Their \emph{orbital stability}\footnote{
Orbital stability here refers to stability up to the transformations keeping the equation invariant.
}
 has been intensively studied. It turns out that the solitary waves of the elliptic NLS equation (\ref{NLSellgen})
are orbitally stable only in the subcritical dimensions, see \cite{CazenaveLions, W86}. 
In the critical and supercritical cases, 
instability occurs trough the apparition of blow 
up, see \cite{Weins83, CazenaveBere}.  
\\
\\
The one-dimensional cubic NLS, ($d=\sigma=1$ in (\ref{NLSellgen})) has the property to be completely integrable
by inverse scattering techniques (IST), as it was shown first by 
Zakharov and Shabat \cite{ZS72}. This pioneer result yielded to the existence of a variety of exact solutions 
for it in the literature, mainly solitons and breathers solutions.
In particular,  
in the focusing case ($\rho=-1$),  
exponentially localized in space solitary waves that travel without change of 
shape and velocity, the well known \emph{solitons solutions} 
(balance between the cubic nonlinearities and wave dispersion) can be derived. They are also called 
\emph{isolated solitons} and can be written in 
the form
\begin{equation}
u_I = \lambda e^{i \left( \phi_0 + vx + \frac{\lambda^2 - v^2}{2}t \right) } \,\frac{1}{\cosh (\lambda(x-x_0-vt))},
\label{trav}
\end{equation}
where the parameters $(\lambda, v, \phi_0, x_0)$ represent respectively the amplitude,
velocity, phase and spatial location of the solitary wave. The constants 
$x_0, \Phi_0$ reflect the invariance of NLS by space translation and phase shift, and the velocity $v$
is associated to the invariance by Galilean transformation.
The stability of the isolated soliton (\ref{trav}) have been intensively studied by various techniques, 
such as 
numerical experiments, but also formal asymptotics \cite{KM77}, 
PDE analysis \cite{W86, GSS87a, GSS87b}, and it turns out that it is stable under 
unidimensional perturbations of both initial data and the equation (orbital stability). 
In this paper, we are rather/however interested in the \emph{transverse stability} of (\ref{trav}),
i.e., the stability or instability of (\ref{trav}) under general (typically localized or periodic) 
two dimensional perturbations. 
\\

The question of transverse stability of (\ref{trav}) was first addressed by 
Zakharov and Rubenchick \cite{ZakRu74} and Yajima \cite{Yajima01091974} and reviewed in \cite{ASDS}.
Typically, the problem is first linearized around the soliton solution,
and then one deals with the detection of unstable modes of the resulting problem, leading to conditions 
for the \emph{spectral stability} of the solution.
Many results in this context are available in the literature. However, as 
pointed out for instance in \cite{RT09}, the relevance of this linear analysis results with respect 
to the fully nonlinear problem 
stay in some situations unclear, 
due to the lack of understanding of the whole spectrum of
the linearized problem.

The question of nonlinear transverse (in-)stability stays consequently in some situations an open question.
A recent advance in this context was given in \cite{RT08}, in which 
the authors present a theory which allows to reduce the problem of the transverse 
nonlinear instability of 1d solitary waves for Hamiltonian PDEs for both periodic
or localized transverse perturbations to the linear instability of the latter. 
In particular, 
the nonlinear
 transverse instability of (\ref{trav}) under the time evolution of the 2-d elliptic NLS equation was 
 proved in \cite{RT09}.
This analytical method however requires some conditions to be fulfilled by 
the transversally perturbed system considered, and 
does not allow to consider a large class of perturbed transverse systems as we intend to do in this paper.
In addition, such kind of analytical results do not provide any insight on the 
characteristics of the instability.

In this paper, we address numerically the question of
the transverse (in-)stability 
of the isolated soliton to the one-dimensional cubic NLS equation (\ref{trav})
under the time evolution of several higher dimensional models, 
being admissible/ as a transverse perturbation for/of the 1d cubic NLS, namely 
the 2d NLS equations, both elliptic and hyperbolic variants, and 
their equivalent when coupled with a mean field satisfying an elliptic equation, known as the 
Davey Stewartson systems.

We first illustrate the features of the instability of (\ref{trav}) in 2-d elliptic NLS equation,
which occurs via a $L_{\infty}$-blow up of the solution and then show that the elliptic elliptic 
Davey Stewartson system behaves as the former in this context.
Then we investigate hyperbolic variants of the above models, namely the hyperbolic 
2d cubic NLS equation and the Davey-Stewartson 
II equation.
Whereas the isolated soliton appears to be unstable for the former case, it appears to be 
orbitally stable for the latter.
For periodic perturbations the soliton is found to be unstable for all tranversally perturbed models
considered, for elliptic equations this instability occurs via multiple-point blow up.

The paper is organized as follow, in section 2, we present the different models (NLS type equations) 
we consider
and discuss
briefly the different issues we face to study these systems numerically.
In section 3 we present the numerical methods used to deal with the above identified issues. 
In section 4 numerical simulations concerning localized perturbations of (\ref{trav}) are reported 
and in section 5 
periodic perturbations of (\ref{trav}) are considered. Some concluding remarks are given in section 6.

\section{Several NLS type equations}

We consider the general form of the focusing 2-dimensional cubic NLS equation,
\begin{equation}
 iu_t + u_{xx} + \mu u_{yy} + 2|u|^2 u = 0, \,\, \mu=\pm1
 \label{cubNLSHE}
\end{equation}
which allows also the study of the hyperbolic (also known as the non-elliptic) NLS equation when $\mu= - 1$. 
In the following 
we will denote by NLS$^+$ the 2d elliptic cubic NLS ((\ref{cubNLSHE}) with $\mu = 1$) and by 
NLS$^-$ the hyperbolic variant when $\mu=-1$.
The Hamiltonian (or energy) for this equation is given by
\begin{equation}
E_{NLS2d}[u]:=\frac{1}{2}\underset{\mathbb{R}^2}{\int} \left(
| u_{x}|^2 + \mu |u_{y}|^2 - |u|^4
\right) dx dy.
 \label{NLS2dEHnrj}
\end{equation}

The 2d elliptic NLS equation NLS$^+$ is known to not be completely integrable, and 
to allow blow up phenomena (we are namely in the critical case defined above).
One can prove rigorously \cite{SS} 
that, for initial conditions for which the Hamiltonian (\ref{NLS2dEHnrj})
is negative,
there exists a time $t^*$ such that 
\begin{equation}
 \underset{t\to t^*}{\lim} \underset{\mathbb{R}^2}{\int} |\nabla u|^2 dxdy = \infty
 \label{blow}
\end{equation}
yielding to a $L_{\infty}$-blow up of $u$ when $t \to t^*$.  This means that the solutions lose after 
finite time the regularity of the initial data, a norm of the 
solution or of one of its derivatives becomes infinite. 
This phenomenon is referred to as \emph{self-focusing} in the context of nonlinear
optics and as \emph{collapse} when applied to problems on turbulence.
It is also known that 
blowup is possible if the energy of the initial data is greater than 
the energy of the ground state solution, see e.g.~\cite{SS} and 
references therein,  and 
\cite{MR} for an asymptotic description of the blowup profile.
Moreover, the existence of blow-up solutions that have more than one spatial blow up points
have been proved in \cite{Merlekblow}.
\\

The NLS$^-$ equation describes the evolution of gravity-capillary waves in deep water that may be 
two-dimensional, nearly monochromatic, and are slowly modulated. A derivation of this equation   
can be found in \cite{ASDS}. It has been used also to investigate the evolution 
of optical pulses in normally dispersive (quasi discrete) optical waveguide array structures
\cite{DHMC,LFSDHMC}, as well as more generally in normally dispersive optical media \cite{CTDTVPJT,DTVPJTCT}.
The NLS$^-$ equation is in fact related to both the Ishimori and Davey-Stewartson systems, to be discussed in 
the following.
More precisely, the Davey-Stewartson systems (DS)
can be written as

 \begin{equation}
\label{DSgen}
\begin{array}{ccc}
i u_{t}+ \alpha u_{xx}+ u_{yy}  & = & b \Phi_x u +\chi \left|  u \right|^{2}u,
\\
\Phi_{xx}+\beta \Phi_{yy} & = & \left(\left| u\right|^{2}\right)_{x},
\end{array}
\end{equation}
where $\beta$ and $b$ can have both signs, $\alpha$ and $\chi$ take the values $\pm1$, 
and $\Phi$ is a mean field. 
These systems describe the amplitude modulation of weakly 
nonlinear, strongly dispersive $2 + 1$-dimensional waves in 
hydrodynamics and nonlinear optics, and appear also in plasma physics to describe the evolution of a plasma under the action of a magnetic field.
They have been classified in \cite{GS}, , 
according to the signs of $\alpha$ and $\beta$, as elliptic-elliptic (E-E) for $\alpha > 0 $ and $\beta > 0$, 
hyperbolic-elliptic (H-E) for $\alpha < 0 $ and $\beta > 0$, elliptic-hyperbolic (E-H) for $\alpha > 0 $ and $\beta < 0$ and hyperbolic-hyperbolic (H-H) for $\alpha < 0 $ and $\beta < 0$.

In \cite{GS} Ghidaglia and Saut showed that NLS$^-$ satisfies the same Strichartz estimates
as its elliptic variant, and based on them they proved some well-posedness results 
in $L^2(\mathbb{R}^2)$ for the (H-E) 
DS system. As well as for the latter, the same argument shows that NLS$^-$ 
is locally well posed in $L^2(\mathbb{R}^2)$ with time of existence depending on the
profile of the initial data, and globally well posed for sufficiently small initial data.
Moreover, in \cite{GS96}, they 
showed that there are no nontrivial localized traveling wave solutions to NLS$^-$.
It is in fact the consequence of the defocusing effect due to the energy  which can 
in this case spread along
the transversal direction to the main propagation. 

However, obviously any $y$-independent solution of the focusing 1d cubic NLS equation
is a solution of NLS$^-$, allowing the question of the transverse (in-)stability
of (\ref{trav}) in such model. An overview of previous work related to this question was presented in \cite{Pelinovsky2001585}.
For the best of our knowledge, all results here deal only with the linear spectral stability problem.
\\

Similarly, 
DS reduces to the cubic NLS in one 
dimension if the potential is independent of $y$, and if $\Phi$ satisfies 
certain boundary conditions (for instance rapidly decreasing at 
infinity or periodic), providing DS as an admissible model as transverse perturbation of the one dimensional cubic NLS equation. 
When
the mean field $\Phi$ in (\ref{DSgen}) is governed by an elliptic equation ($\beta>0$), the latter can 
be solved uniquely with some fall off condition at infinity, 
$\Phi = \mathcal{M}\left( |u|^2  \right)$, where the operator $\mathcal{M}$ is defined in Fourier space by  
$$ \widehat{\mathcal{M}(f)} = \frac{i k_x}{k_x^2 + \beta k_y^2} 
\widehat{f} (k_x, k_y),$$ where $k_x$ and $k_y$ represent the wave 
numbers, in the $x$ and $y$ directions, respectively, and where 
$\hat{f}$ denotes the Fourier transform of a function $f$.
Therefore in the following we will consider only elliptic equations for $\Phi$. 
In this case, the ($\mathbf{\cdot}$-elliptic) versions of DS enjoy the conservation of several
functionals, in particular the Hamiltonian (energy)
\begin{equation}
\label{DSenergy}
    E_{DS}[u]  :=  \int_{\mathbb{R}^2} \alpha | u_x|^2 + | u_y|^2 +\frac{\chi|u|^{4}+ b\left(\Phi_x^{2}+\beta \Phi_y\right)^{2}}{2}
   d x d y.
   \end{equation}
For both cases (H-E) and (E-E),
Ghidaglia and Saut proved local existence and uniqueness of a solution in $L^2(\mathbb{R}^2)$,  
and global existence under a small norm assumption on the initial data.
For the (E-E) case, they also proved the existence of solutions which blow up in finite time.
Notice that numerical evidence for the occurrence of muli-blow up phenomena in (E-E) DS, 
as in the case of elliptic NLS equation, has been 
given in \cite{BMS}.
\\
In the following, we will consider an elliptic-elliptic case of DS, 
which can be re-written as 
 \begin{equation}
\label{DSeeform}
\begin{array}{ccc}
i u_{t}+ u_{xx}+ u_{yy}  & = &  \Phi_x u +\chi \left|  u \right|^{2}u,
\\
\Phi_{xx}+\beta \Phi_{yy} & = & -\gamma \left(\left| u\right|^{2}\right)_{x},
\end{array}
\end{equation}
which is non integrable and
very similar to the NLS$^+$, see for instance 
\cite{GS} for a study of the blow up phenomena and \cite{BMS} for numerical 
simulations. We denote this system by DS$^{++}$ in the following.
We expect the same behavior as in NLS$^+$, i.e. the same kind of instability 
for (\ref{trav}) in the focusing case, i.e., $\chi=-1$.
We also consider the so called DS II equation, 
which can be re-written as 
\begin{equation}
\label{DSII}
\begin{array}{ccc}
i u_{t}+\left(u_{xx}- u_{yy}\right)-2\left(\Phi+\left|  u \right|^{2}\right)u & = & 0,
\\
\Phi_{xx}+ \Phi_{yy}+2\left| u\right|_{xx}^{2} & = & 0,
\end{array}
\end{equation}
and which has in addition the property to be completely integrable by IST. 
\\

Note that the hyperbolic Laplacian in DS II leads to a 
different dynamics compared to the standard elliptic NLS equations 
(\ref{NLSellgen}).  
Therefore many PDE techniques successful for NLS could not be applied 
to the DS system. Using integrability,
Fokas and Sung \cite{FokS,Sun} studied the existence and long-time 
behavior of the solutions of the initial value problem for DS II (with $u(x,y,0)=u_0$). 
They proved the following
\begin{theorem} \label{theosung}
If $u_0$ belongs to the Schwartz space $\mathcal{S}(\mathbb{R}^2)$, 
then there exists  in the defocusing case ($\rho=1$) a unique global 
solution $u$  to DS II such that $t\mapsto u(\cdot,t)$ is a 
$C^\infty$ map from $ \mathbb{R}\mapsto \mathcal{S} (\mathbb{R}^2) $.
The same holds for the focusing case ($\rho=-1$) if
the initial data $u_0 \in L_{q}$ for some $q$ with $1\leq q < 2$ 
has a Fourier transform $\widehat{u_0} \in L_{1} \cap L_{\infty}$ 
such that 
  $  \|  \widehat{u_0} \|_{L_{1}} \|\widehat{u_0}\|_{L_{\infty}}
    <\frac{\pi^{3}}{2}\left(\frac{\sqrt{5}-1}{2}\right)^{2}$. The 
    unique global solution $u$ to DS II  satisfies the decay 
    estimate $\| u(t) \|_{L_{\infty}} < \frac{const}{t}$.\\
Furthermore, if $u_0$ belongs to the Schwartz space, then there is an infinite number of conserved quantities.  
\end{theorem}

The smallness condition in Theorem \ref{theosung} indicates that in general
there might be a blow-up in solutions to the focusing DS II equations. 
In fact, as recalled above, $2+1$ dimensions constitute the critical dimension for 
focusing cubic NLS equations where 
blow-up can occur. But due to the hyperbolic Laplacian in DS II,
this cannot be directly generalized to the latter. Therefore it is important 
in this context
that  Ozawa gave an exact blow-up solution in \cite{Oza}. The 
solution is similar to  the well known lump solutions \cite{APPDS}, 
traveling solitonic wave solutions with an algebraic fall off at 
infinity.  Note that Theorem 1.1 does not hold for lumps due to the 
small-norm assumption imposed on the initial data. It is thus not 
known whether there is generic blow-up for initial data not satisfying 
this condition, nor whether the condition is optimal. Numerical 
studies in \cite{MFP,KRM} indicate, however, that blow-up can 
occur in solutions to the focusing DS II. In fact it was conjectured 
in \cite{MFP} that generic localized initial data are just radiated 
away to infinity or blow up for 
large $t$. 
The first numerical 
studies of DS were done by White and Weideman \cite{WW} using 
Fourier spectral methods for the spatial coordinates and a second 
order time splitting scheme. Besse, Mauser and Stimming \cite{BMS} 
used an advanced parallelized version of this method to study the Ozawa 
solution and blowup in the elliptic-elliptic DS equation. 
\\
\\

As mentioned in the introduction, Rousset and Tzvetkov presented an analytical
method allowing to reduce the problem of the transverse 
nonlinear instability of 1d solitary waves for Hamiltonian PDEs for both periodic
or localized transverse perturbations to the linear instability of the latter.
This theory requires a lot of assumptions on the problem, and it is not the scope
of this paper to review the whole method. However, one can easily check that this method is 
not applicable to the hyperbolic Laplacian, so the question of the nonlinear transverse instability 
of (\ref{trav}) under the flow of NLS$^-$ and DS II remains an open problem. Notice that 
the linear question for NLS$^-$ has been intensively studied, and it turns out that 
(\ref{trav}) is spectrally unstable. 

On the contrary, the method \cite{RT09} was successfully applied to NLS$^+$, providing
the transverse nonlinear instability of (\ref{trav}) under localized and periodic perturbations.
The features of these instabilities however remain unknown. Moreover, due to the 
contribution of the mean field $\Phi$ in the nonlinear part of the elliptic-elliptic DS system, 
it is unclear that the method of Rousset and Tzvetkov can be applied there. We anyway expect 
the same behavior for these both elliptic cases.

We thus investigate these questions in the present paper numerically.
This is a highly non-trivial problem for several reasons: first these
NLS equations are purely dispersive equations, which means that the introduction 
of numerical dissipation has to be avoided as much as possible to 
preserve dispersive effects such as rapid oscillations. This
makes the use of spectral methods attractive since they are known for 
minimal numerical dissipation and for their excellent approximation properties for smooth 
functions. In addition they allow for efficient time integration algorithms 
which should be ideally of high order to avoid a pollution of the 
Fourier coefficients due to numerical errors in the time integration. 

An additional problem is the modulational instability of the focusing 
NLS equations, i.e., a self-induced amplitude modulation of a continuous 
wave propagating in a nonlinear medium, with subsequent generation of 
localized structures, see for instance \cite{Agr,CH,FL} for the NLS 
equation.

Thus to address numerically questions of stability and blowup of 
their solutions, a quite high resolution is needed which can only be achieved by
 computations on high performance parallel computers. 
The use of Fourier spectral methods is also very convenient in this context,
since for a parallel spectral code only existing
optimized serial FFT algorithms are necessary. In addition, such codes 
are not memory intensive, in contrast to other approaches 
such as finite difference or finite element methods.

Furthermore, as already mentioned, solutions to elliptic NLS equations, as well as the DS systems 
considered here can have blow up. 
Obviously it is non-trivial to 
decide numerically whether a solution blows up or not. Note that the criteria to 
determine the appearance of such phenomena in practice are somewhat arbitrary. 
We will use asymptotic Fourier analysis, 
as proposed in \cite{SSF}, and applied in \cite{dkpsulart, DSdDS} to numerically 
prove the appearance of a blow up or the all-time regularity 
of the solution. 
In \cite{DSdDS} the 
efficiency of this method to detect blow up has been illustrated on the well understood example of the 
1d quintic NLS equation. Moreover $(2+1)$-dimensional situations are also studied here for the DS II equation. 
We describe the numerical methods used in the next section.

\section{Numerical Methods}

We consider a periodic setting for the spatial coordinates, which allows
the use of a Fourier spectral method for the space discretization. We 
treat the rapidly decreasing functions we are studying as essentially 
periodic analytic functions within the finite numerical precision. 
For such functions, spectral methods are known for their excellent 
(in practice exponential) approximation properties, see for instance 
\cite{can,tref}. In addition they 
introduce only very little numerical dissipation which is important 
in the study of dispersive effects. Last but not least we use the Fourier 
coefficients to identify the appearance of singularities (blow up) in  
the solution as in \cite{SSF,dkpsulart, DSdDS}.

In all cases,
the numerical precision is controlled via both the good decay of the Fourier
coefficients and the numerically computed energy for each system considered.
More precisely, given $E$, 
 a conserved quantity of the system, defined in the previous section
 for the models we consider here, the numerically computed conservation 
 of 
 $E$ (which will always depend on time due to unavoidable numerical errors)
can be used as a reliable indicator of numerical accuracy \cite{ckkdvnls,KR}
, provided that 
there is sufficient spatial resolution (generally the accuracy of the 
numerical solution is overestimated by two orders of magnitude), by
considering the conservation of the quantity 
\begin{equation}
\Delta_E = \left|\frac{E(t)}{E(0)} - 1\right|.
\label{delE}
\end{equation}
We always aim at a $\Delta_E$ smaller than $10^{-6}$ to ensure an accuracy well beyond the plotting accuracy $\sim 10^{-3}$.

\subsection{Numerical Integration}
So we proceed approximating the spatial dependence via 
truncated Fourier series for the studied equations. 
This leads to large \textit{stiff}\footnote{We use the word stiffness  
to indicate that there are largely different scales to be resolved in 
this system of ODEs which make the use of explicit methods inefficient 
for stability reasons.} 
systems of  ODEs in Fourier space of the form   
\begin{equation}
    v_{t}=\mathbf{L}v+\mathbf{N}(v,t)
    \label{utrans},
\end{equation}
where $v$ denotes the (discrete) Fourier transform of $u$, 
and where $\mathbf{L}$ and $\mathbf{N}$ denote linear and nonlinear 
operators, respectively. These systems of ODEs are classical examples 
of stiff equations where the 
stiffness is related to the linear part $\mathbf{L}$ (it is 
a consequence of the distribution of the eigenvalues of 
$\mathbf{L}$), whereas the 
nonlinear part contains only low order derivatives. 

There are several approaches to 
deal efficiently with equations of the form (\ref{utrans}) with a 
linear stiff part such as implicit-explicit (IMEX), time splitting, 
integrating factor (IF) as well as 
sliders and exponential time differencing. 
By performing a comparison  of stiff integrators for the 
1+1-dimensional cubic NLS equation in semiclassical limit 
(\ref{NLSellgen}) in  \cite{ckkdvnls}, and for the semiclassical limit 
of the DS II equation in  \cite{KR}, 
it was shown that Driscoll's composite Runge-Kutta (DCRK) method 
\cite{Dris} is very efficient in this context. 
We thus use this scheme for the time integration here. 

The basic idea of the DCRK method is inspired by IMEX methods, i.e.,  
the use of a stable implicit method for the linear part of the 
equation (\ref{utrans}), which introduces  the stiffness into the 
system, and an explicit scheme for the nonlinear part which is assumed to be non-stiff. 
Classic IMEX schemes do not perform in general satisfactorily for 
dispersive PDEs \cite{KassT}.  Driscoll's \cite{Dris}  more 
sophisticated variant consists in splitting the linear part of the 
equation in Fourier space into regimes of high and low frequencies, 
and to use the fourth order RK integrator for the low frequencies and 
the nonlinear part, and the linearly implicit RK method of order three for the high frequencies. 
He showed that this method is in practice of fourth order over a wide range of step sizes. 

An additional problem here is the modulational instability of the 
focusing NLS equations, i.e., a self-induced amplitude modulation of 
a continuous wave propagating in a nonlinear medium, with subsequent 
generation of localized structures, see for instance \cite{Agr,CH,FL} 
for the NLS equation. This instability leads to an artificial 
increase of the high wave numbers which eventually crashes the simulation code, 
if not enough spatial resolution is provided (see for instance 
\cite{ckkdvnls} for the focusing NLS equation). 
To allow high resolution simulations, the codes are parallelized as explained below.

Moreover, as recalled in the previous section, some of the models studied here can have blow up solutions, even  
 for smooth initial data. From the point of view of applications, a blow up of a solution does not necessarily mean that
the studied equation is not relevant in this context. It just indicates the limit of its use as an
approximation.  This breakdown of the model can indicate how to amend the
approximation, e.g. by additional terms in the equations, which is a challenging task both from the mathematical and the application point of view.\\
We will use here an asymptotics Fourier analysis 
proposed in \cite{SSF} to numerically detect the appearance of a blow up or the all-time regularity 
of the solution. 
We discuss this method with more detail in the next section.

\subsection{Tracking of the singularities}
To identify numerically the blow up time of the solution with 
sufficient accuracy, we will use asymptotic Fourier analysis as  first 
applied numerically by Sulem, Sulem and Frisch in \cite{SSF}. The 
basic idea here is that functions  analytic in a 
strip around the real axis in the complex plane have a characteristic 
Fourier spectrum for large wave numbers. Thus it is in principle 
possible to obtain the  
 width of the analyticity strip from the asymptotic behavior of the 
 Fourier transform of the solution (in one spatial dimension), or 
 from the angle averaged energy spectrum in higher dimensions. 
 It is thus important here that we treated the coordinates by 
Fourier series. 
 This 
 allows in particular to identify the time when a singularity in the 
 complex plane hits the real axis and thus leads to a singularity of 
 the function on the real line. 
Singular solutions of the two-dimensional cubic NLS equation have been studied with this approach in
\cite{SSP}, and an application of the method to the two-dimensional Euler equations can be found 
in \cite{FMB, MBF}. The method has also been applied to 
the study of complex singularities of the three-dimensional Euler equations
in \cite{CR}, in thin jets with surface tension \cite{PS98}, the 
complex Burgers' equation \cite{SCE96} and the Camassa-Holm equation \cite{RLSS}. 
More recently, its efficiency has been investigated quantitatively for the 
Hopf equation and it was shown that the method can be efficiently used in 
practice to describe the critical behavior of solutions to dispersionless equations. 

More precisely, one makes the use of the following analytical result \cite{asymbook},
\begin{theorem}
 Let $u(z)$ an analytic function of one variable $z\in\mathbb{C}$ such that $|u(z)| \to 0$ uniformly as $|z| \to \infty$.
 Assuming that the singularities of $u(z)$  are isolated one from
another and are of one of the following type: pole, algebraic- or logarithmic-
branch point, then the behavior of the Fourier transform of $u$, denoted by $\hat{u}$, is asymptotically
(when $k \to \infty$) governed by the singularity of the lower half-space closest to
the real domain that is not a multiple pole.
If this singularity is located at $z_{j}=\alpha_{j}-i\delta_{j}$ , with $\delta_{j} \geq 0$, 
and has an exponent
$\mu_{j}\notin \mathbb{Z}$, then in a neighborhood of $z_{j}$, $u(z)\sim 
(z-z_{j})^{\mu_{j}}$ and
\begin{equation}
    \hat{u}\sim 
    \sqrt{2\pi}\mu_{j}^{\mu_{j}+\frac{1}{2}}e^{-\mu_{j}}\frac{(-i)^{\mu_{j}+1}}{k^{\mu_{j}+1}} e^{-ik\alpha_{j}-k\delta_{j}}.
    \label{fourierasym}
\end{equation}
\end{theorem}
It implies that for a single such singularity with positive $\delta_{j}$, the modulus of the Fourier 
coefficients decreases exponentially for large $k$. For 
$\delta_{j}=0$, i.e., a singularity on the real axis,  
the modulus of the Fourier coefficients has an algebraic dependence 
on $k$, and thus the location of 
singularities in the complex plane can be obtained from a given Fourier 
series computed on the real axis. 

More precisely, from (\ref{fourierasym}) several situations are possibles:
If a singularity reaches the real domain after a finite time $t_c$ , i.e. 
$\delta_{j}(t_c)=0$, the
solution looses analyticity and becomes singular.
If the width of the analyticity strip is bounded away from zero, i.e. if there exists 
$\gamma \in \mathbb{R}$ such that $\delta_{j}(t) > \gamma, \,\, \forall t$, then
the solution is
uniformly analytic. 
If the width of the analyticity strip goes to zero without vanishing, (e.g., 
exponential decay), the solution remains smooth for all times.
Finally, if several singularities are relevant asymptotically, then  $\hat{u}$ displays an
oscillatory behavior.
This last case however implies also that only the first singularity occurring can
be recovered in this way.

In practice, 
to numerically compute a Fourier transform, it has to be approximated 
by a discrete Fourier series which can be done efficiently via a 
 fast Fourier transform (FFT), see e.g.~\cite{tref}. The discrete Fourier transform  of the 
 vector $\mathbf{u}$ with components $u_{j}=u(x_{j})$, where 
 $x_{j}=2\pi L j/N$, $j=1,\ldots,N$ (i.e., the Fourier transform on 
 the interval $[0,2\pi L]$ where $L$ is a positive real number) 
 will be always denoted by $v$ in 
 the following. There is no obvious analogue of 
 relation (\ref{fourierasym}) for a discrete Fourier series, but it 
 can be seen as an approximation of the latter, which is also the 
 basis of the numerical approach in the solution of the PDE. It is possible to 
 establish bounds for the discrete series, see for 
 instance \cite{arnold}.
 \\
 \\
 According to (\ref{fourierasym}), $v$ is assumed to be of the form
$v(k,t) \underset{k \to \infty}{\sim} e^{A(t)} k^{-B(t)}e^{-\delta(t)k},
$
and one can trace the temporal behavior of $\delta(t)$ 
via some fitting procedure in order to obtain evidence for 
or against
blow-up (the problem of blow-up reduces to check if delta vanishes in a finite time
$t_c$, which indicates a loss of regularity).
In order to determine $\delta(t)$ from direct numerical simulations, a least-square fit is performed on the logarithm of the
Fourier transform using the functional form
\begin{equation}
    \ln |v|\sim A- B\ln k-k\delta.
    \label{abd}
\end{equation}
The fitting is done for a given range of wave numbers $k_{min}<k<k_{max}$ (we only consider positive $k$), that have to be controlled, 
as explained in detail in \cite{dkpsulart}. The critical time 
$t_c$ is determined by the vanishing of $\delta$, and the type of the 
singularity is  given by the parameter $B(t_c)$ which is equal to $\mu_j+1$.
As explained in \cite{dkpsulart} the reliability of the fitting can be also 
inferred from the value of the fitting error, defined as 
$p=\|   \ln |v| - (A - B 
\ln k - k \delta)   \|_{\infty}$. 
In the case of blow up phenomena, it turns out (see \cite{DSdDS}) that the study of the 
Fourier coefficients in only one direction is sufficient to determine the blow up appearance, and that 
a fitting error of the order of $\sim 0.5$ can be reached.
Typically in this case the bounds $k_{min}, k_{max}$ for the fit interval have no real impact on the determination of the 
blow up time, and one usually considers the classical interval $10<k<2 \max(k)/3$, following the dealiaising rule.
\\
In addition,  one can also determine the real part of the location of the singularity by doing a least square fitting on the 
 imaginary part of the logarithm of $v$ for which one has asymptotically
 \begin{equation}
    \phi:=\Im \ln v\sim C-\alpha k
    \label{phi}.
\end{equation}
Since the logarithm is 
branched in Matlab/Fortran at the negative real axis with jumps of $2\pi$, the computed $\phi$ 
will in general have many jumps.
Thus one has first to construct a 
continuous function from the computed $\phi$, see also \cite{dkpsulart} . The analytic 
continuation is done in 
the following way: starting from the first value (largest $k$), we 
check for all other values of $\phi(k_{j})$ whether 
$|\phi(k_{j+1})-\phi(k_{j})|>|\phi(k_{j+1})-\phi(k_{j})\pm\pi|$. If 
this is the case, we put $\phi(k_{j+1})\to \phi(k_{j+1})\pm \pi$.  
The result of this procedure will be a continuous function which will 
be fitted with a least square approach to a linear function.
Then the location of the singularity on the real 
axis is given by $\alpha(t_c)$.

\subsection{Parallelization of the codes}
As explained before, to be able to provide the high space resolution needed, 
the numerical codes have been parallelized.  This can be conveniently done for 
two-dimensional Fourier 
transforms where the task of the one-dimensional FFTs is performed 
simultaneously by several processors. This reduces also the memory 
requirements per processor with respect to alternative approaches such as finite 
difference or finite element methods. We consider periodic (up to 
numerical precision) solutions in $x$ and $y$, i.e., solutions on 
$\mathbb{T}^2 \times \mathbb{R}$. The computations are carried out with $N_x \times N_y$
points for $(x, y) \in [-L_x\pi, L_x\pi] \times  [-L_y\pi, L_y\pi] $.
In the computations, $L_x = L_y$ is chosen large enough such that the 
numerical solution is of the order of machine precision ($\sim 
10^{-16}$ here) at the boundaries.

A prerequisite for parallel numerical algorithms is that sufficient 
independent computations can be identified for each processor,
that require only small amounts of data to be communicated between 
independent computations.
To this end,
we perform a data decomposition, which makes it possible to 
do basic operations on each object in the data domain (vector, matrix...) 
 to be executed safely in parallel by the available processors.
Our domain decomposition is implemented by developing a
code describing the local computations and local data structures for a single process. 
Global arrays are divided in the following way:
denoting by 
$x_n = 2 \pi n L_x/N_x,\,\, y_m = 2 \pi m L_y/N_y$, $n=-N_x/2,...,N_x/2, \,\, m=-N_y/2,...,N_y/2,$
the respective discretizations of $x$ and $y$ in the corresponding 
computational domain, 
$u$ (respectively $\Psi$) is then represented by a $N_x \times N_y$ matrix.
For programming ease and for the efficiency of the Fourier transform, 
$N_x$ and $N_y$ are chosen to be powers of two. The number  $n_p$ of processes is chosen to divide $N_x$ and $N_y$ perfectly, 
so that 
each processor $P_i, i=1...n_p$, will receive $N_x \times \frac{N_y}{n_p}$ elements of 
$u$ corresponding to the elements 
\begin{equation}
u\left(1:N_x, (i-1).\frac{N_y}{n_p}+1 : i.\frac{N_y}{n_p} \right) 
\end{equation}
in the global array, and then 
each parallel task works on a portion of the data.

While processors execute
an operation, they may need values from other processors. The above domain decomposition 
has been chosen such that the distribution of operations is balanced 
and that the communication
is minimized. 
The access to remote elements has been implemented
via explicit communications, using  sub-routines of the MPI (Message Passing Interface) library \cite{GTL}.

Actually, the only part of our codes that requires communications is 
the computation of the two-dimensional FFT and 
the fitting procedure for the Fourier coefficients.
For the former we use the transposition approach. The latter allows to 
use highly optimized single processor one-dimensional FFT routines, 
that are normally found in most architectures, and a transposition 
algorithm can be easily ported to different distributed memory 
architectures. We use the well known FFTW library because its 
implementation is close to optimal for serial FFT computation, see 
\cite{FJ}. Roughly speaking, a two-dimensional FFT does one-dimensional FFTs on all rows and 
then on all columns of the initial global array.        
We thus first transform in $x$ direction, each processor transforms all the dimensions of the data that are completely local to it, and
the array is transposed once this has been done by all processors. 
Since the data are evenly distributed among the MPI processes, this transpose is efficiently implemented using MPI ALLTOALL communications of the MPI library.

The
asymptotic fitting of the Fourier coefficients in one spatial 
direction requires in addition two local communications.

\subsection{Tests of the codes on the exact solution}
\begin{figure}[htb!]
\centering
\includegraphics[width=0.4\textwidth]{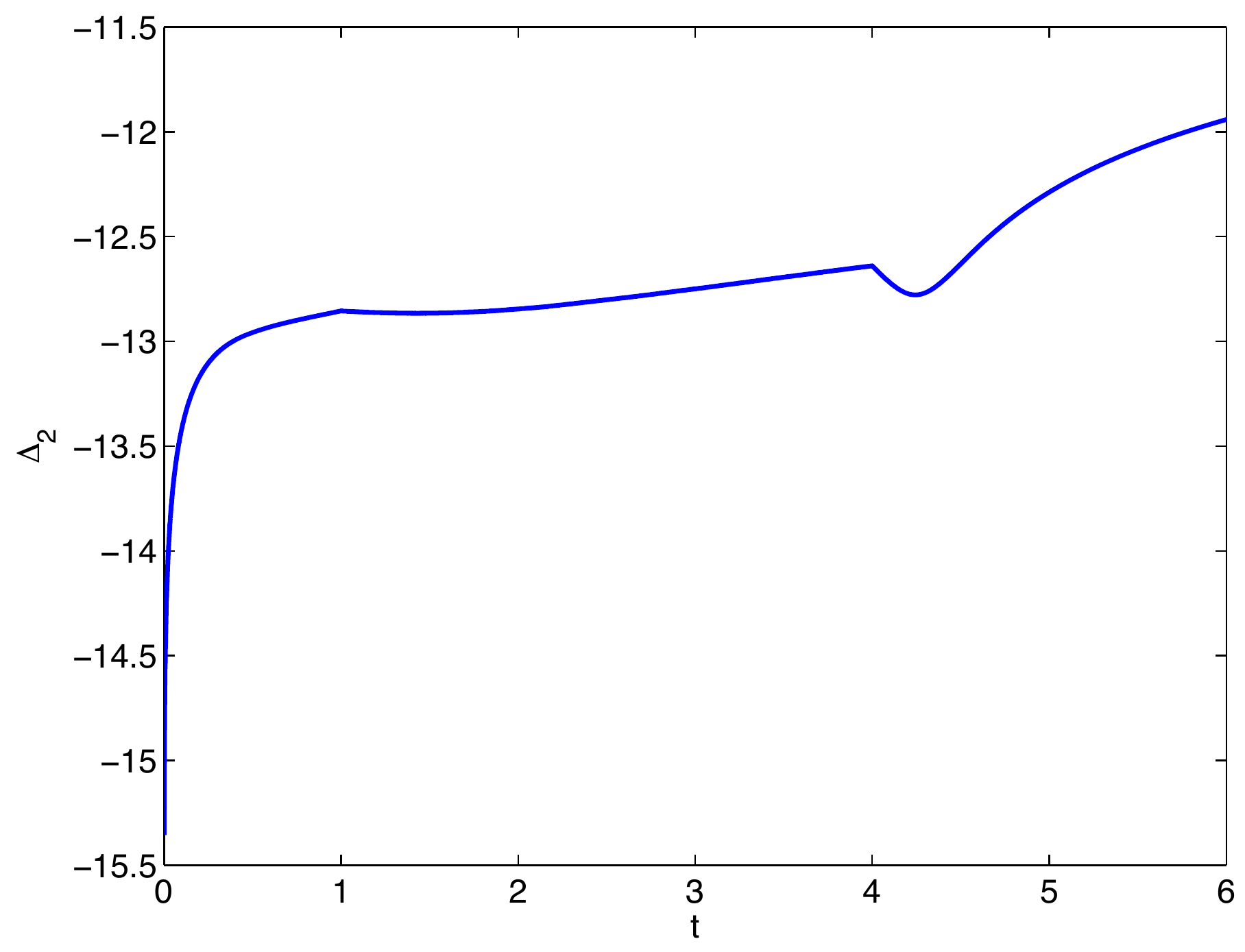} 
\includegraphics[width=0.4\textwidth]{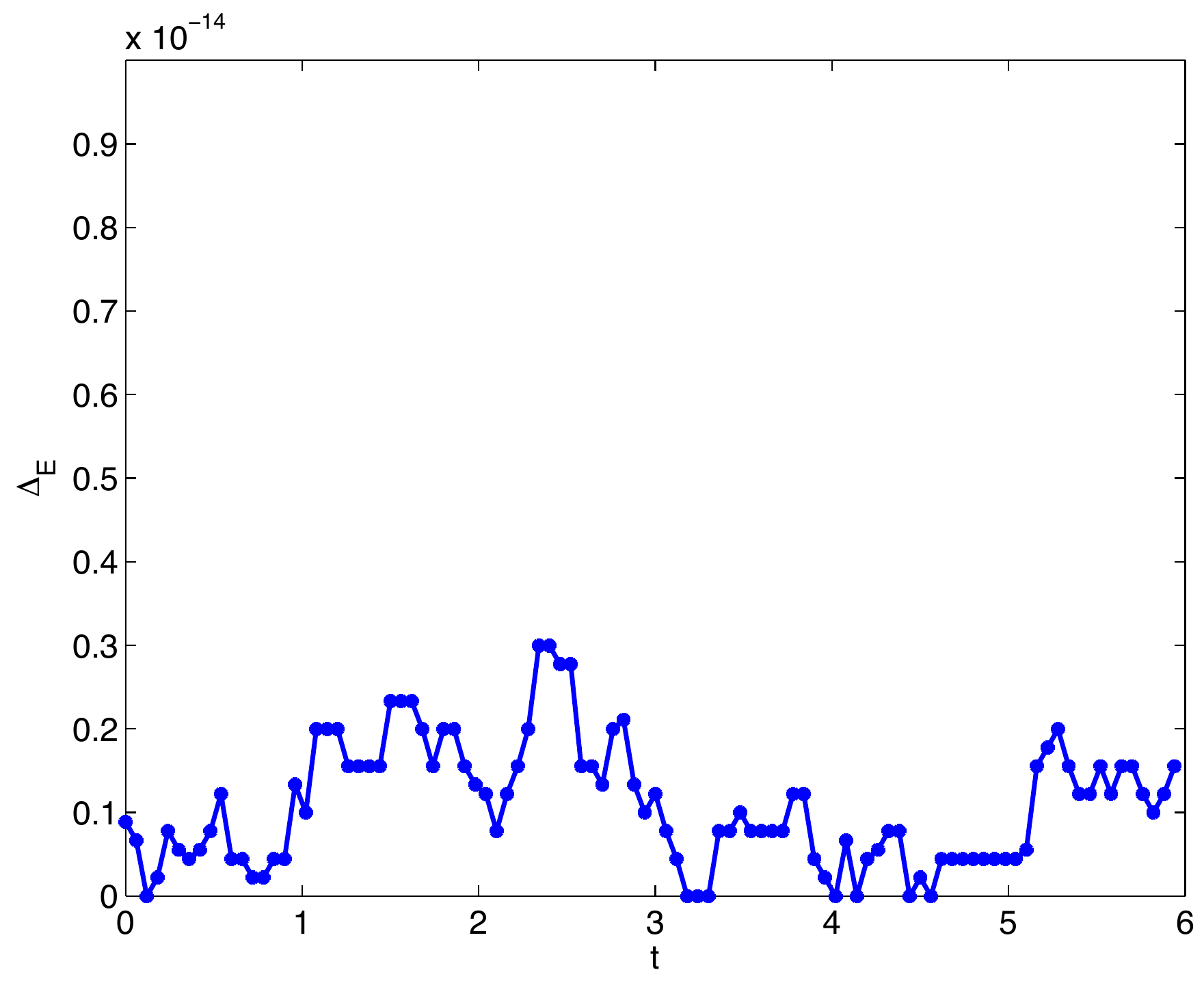} 
 \caption{Time evolution of the $L_2$ norm of the difference between the numerical and 
 the exact solutions $\Delta_2$ for the isolated soliton on the left, and of
 the numerically computed energy $\Delta_E$ on the right}
 \label{numaccuracyexsol}
\end{figure}

\begin{figure}[htb!]
\centering
\includegraphics[scale=0.5]{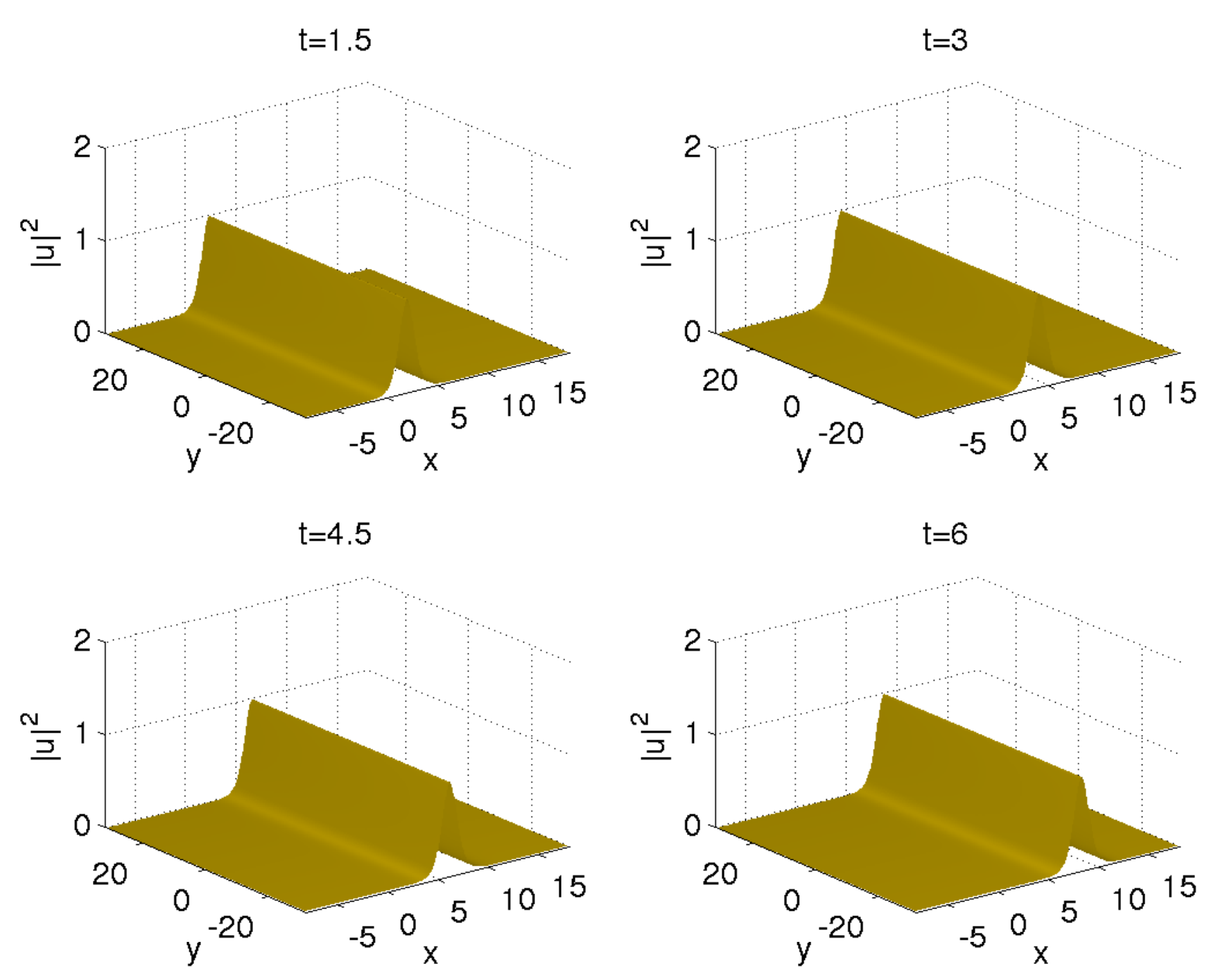} 
 \caption{Isolated soliton at several times in $(2+1)$-dimensional models}
 \label{cont1}
\end{figure}

\begin{figure}[htb!]
\centering
\includegraphics[width=0.45\textwidth]{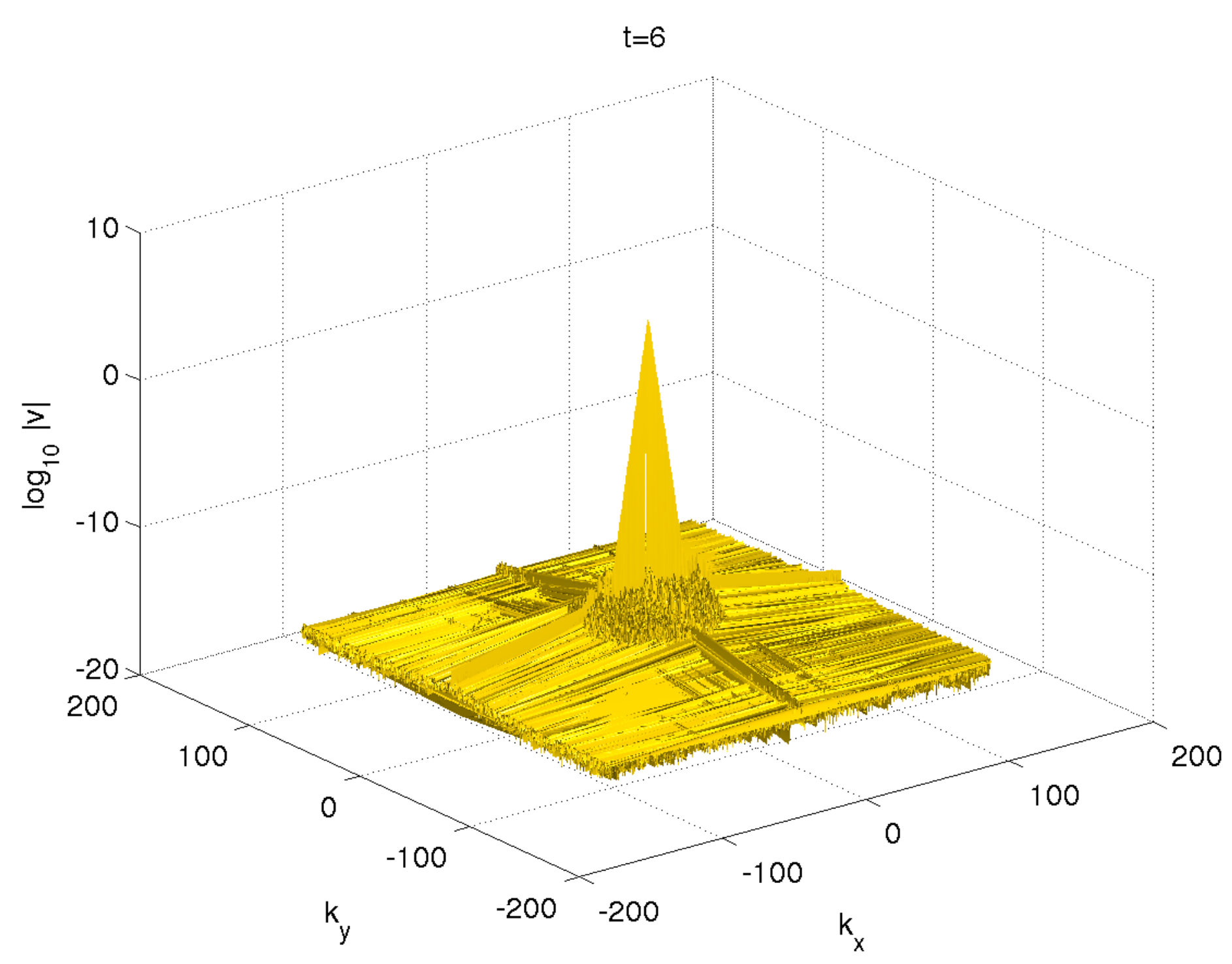} 
 \caption{Fourier coefficients of the numerically computed isolated soliton (\ref{trav})
 with parameters $(\lambda, v, \phi_0, x_0)=(1,\sqrt{2}, 0, 0)$ at $t=t_{max}=6$.}
 \label{four1}
\end{figure}

The isolated soliton (\ref{trav}) is automatically a solution to the four models described in section 2, without explicit $y$-dependence and can be seen as a line soliton for these models. As recalled before, it is known to be unstable for the 2d cubic elliptic NLS equation.  
To test our numerical codes, we propagate the isolated soliton (\ref{trav}) in the four differents models we consider. 
The numerical accuracy is controlled by both the conservation of the numerically computed energy, 
and also the $L_{2}$-norm of the difference between the numerical and the exact solutions, 
denoted in the following by $\Delta_2 := \| u_{num} - u_{ex} \|_2$. 
The computations are carried out with $N_x =N_y= 2^{12}$ points for
$ x \times y \in [-15 \pi, 15 \pi] \times [-15 \pi, 15 \pi]$ for $t\leq 6$ and $\delta_t=6*10^{-4}$.\\
\\
We chose the following parameters $(\lambda, v, \phi_0, x_0)=(1,\sqrt{2}, 0, 0)$ in (\ref{trav}).
The situation is similar for the different models studied: 
We found that the $L_2$ norm of the difference 
between the numerical and the exact solutions reach a value of $\Delta_2 \sim 10^{-12}$ at $t_{max}=6$ in all 
cases, see Fig. \ref{numaccuracyexsol}, and $\Delta_E \sim 10^{-14}$. 
The numerical solution is shown at several times in Fig. \ref{cont1}, it travels with constant speed $\sqrt{2}$,
and the Fourier coefficients are shown in Fig. \ref{four1} at $t=t_{max}=6$. For all models 
the Fourier coefficients decay to machine precision ($\sim 10^{-14}$), 
and no modulational instability occurs up to the maximal time of computation.
This shows that our codes are able to propagate the exact solution even for the unstable case (2d elliptic NLS). It also 
allows to use the quantity $\Delta_E$ as an indicator for the numerical accuracy as in \cite{ckkdvnls, KRM, KR}.

\begin{figure}[htb!]
\centering
\includegraphics[width=0.43\textwidth]{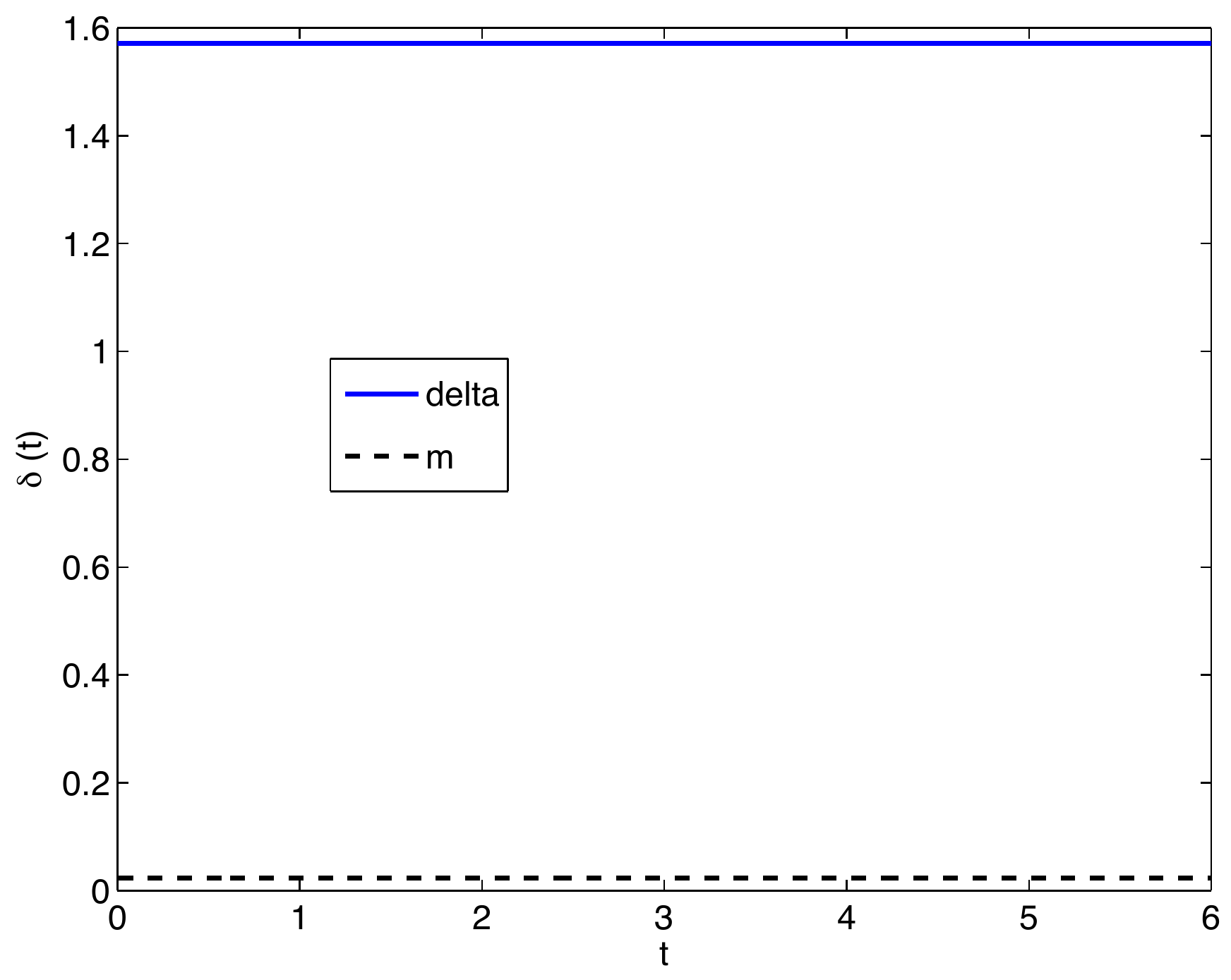} 
\includegraphics[width=0.45\textwidth]{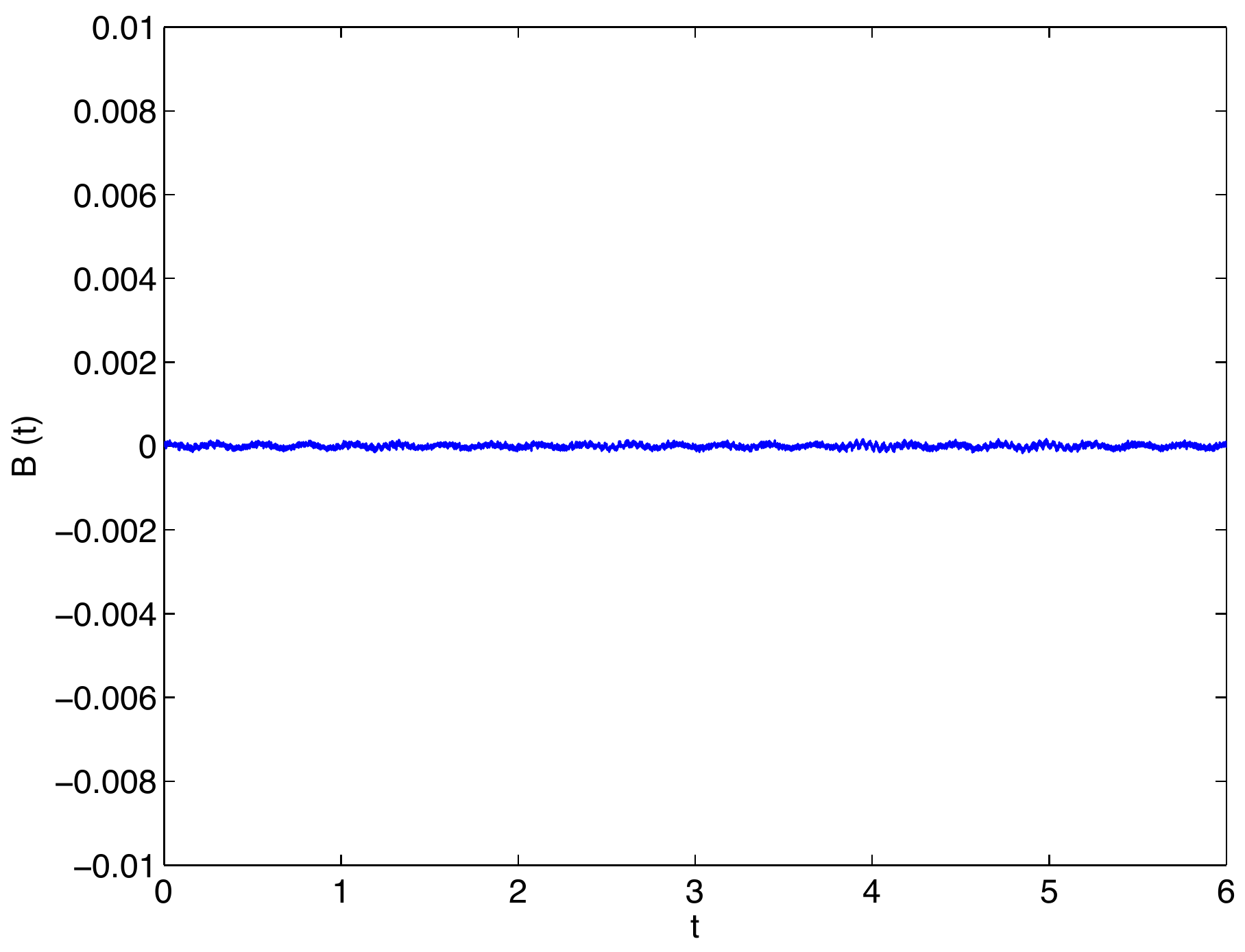} 
 \caption{Time evolution of the fitting parameters in (\ref{abd}), $\delta$ on the left and 
 B on the right, for the numerically computed isolated soliton (\ref{trav})
 with parameters $(\lambda, v, \phi_0, x_0)=(1,\sqrt{2}, 0, 0)$. The fitting is done for $10<k_x<2 max(k_x)/3$}
 \label{solexdel}
\end{figure}

On such travelling waves, it is expected and actually found that the ASM (\ref{abd}) produces 
a constant value for $\delta(t)$, see Fig. \ref{solexdel}, where we find that $\delta(t) \sim 1.57$
for all times studied, and the Fourier coefficients show an exponential decay so that one gets $B=0$ 
for all times studied.

In this section, we explained the numerical tools used for the numerical study of the transverse (in)-stability 
of the isolated soliton of the 1d cubic NLS equation in higher dimensional models. 
We showed that we can numerically efficiently reproduce the exact solution
under the flow of all models considered, typically with spectral accuracy.
In the next section, we investigate the transverse stability of the isolated soliton under 
localized perturbations.

\section{Localized Perturbations (by a Gaussian function)}

In this section we consider perturbations of the isolated soliton with a Gaussian function, and investigate 
its transverse (in-)stability under the flow of different higher models.
More precisely, it is known that the isolated soliton (\ref{trav}) is nonlinearly unstable 
under the 2d cubic NLS flow \cite{RT09}. 
Such kind of analytical results however do not provide any idea about the qualitative
behavior of the perturbed solutions. The cubic 2d NLS equation (NLS$^+$) being in addition known to 
allow blow up phenomena, it is also interesting to simulate such situations in this context.
The similitudes between NLS$^+$ and DS$^{++}$ allow one to expect the same kind of results for 
both models. We will see that in both cases, the instability of the isolated soliton occurs via a $L_{\infty}$-blow up 
of the solution in one spatial point.
On the other hand, we perform a similar study for the hyperbolic variants, 
NLS$^-$ and DS II.
Here theoretical analysis as in \cite{RT09} no longer hold, and we will see that 
(\ref{trav}) appears to be unstable in NLS$^-$ and 'orbitally' stable for DS II.

We thus propagate initial data of the form
\begin{equation}
u(x,y,0) = u_I(x,y,0) + A \exp(-(x-x_1)^2 -(y-y_1)^2),
\label{Agauss}
\end{equation}
where $u_I$ denotes the isolated solution (\ref{trav}), $A \in \mathbb{R}$ and $(x_1,y_1) \in [-L_x \pi, L_x \pi] \times [-L_y \pi, L_y \pi]$. 

\subsection{Elliptic NLS equations}

In this section, the computations are carried out with $2^{13}\times 2^{13}$ points for
$ x \times y \in [-15 \pi, 15 \pi] \times [-15 \pi, 15 \pi]$. 

\begin{figure}[htb!]
\centering
\includegraphics[width=0.7\textwidth]{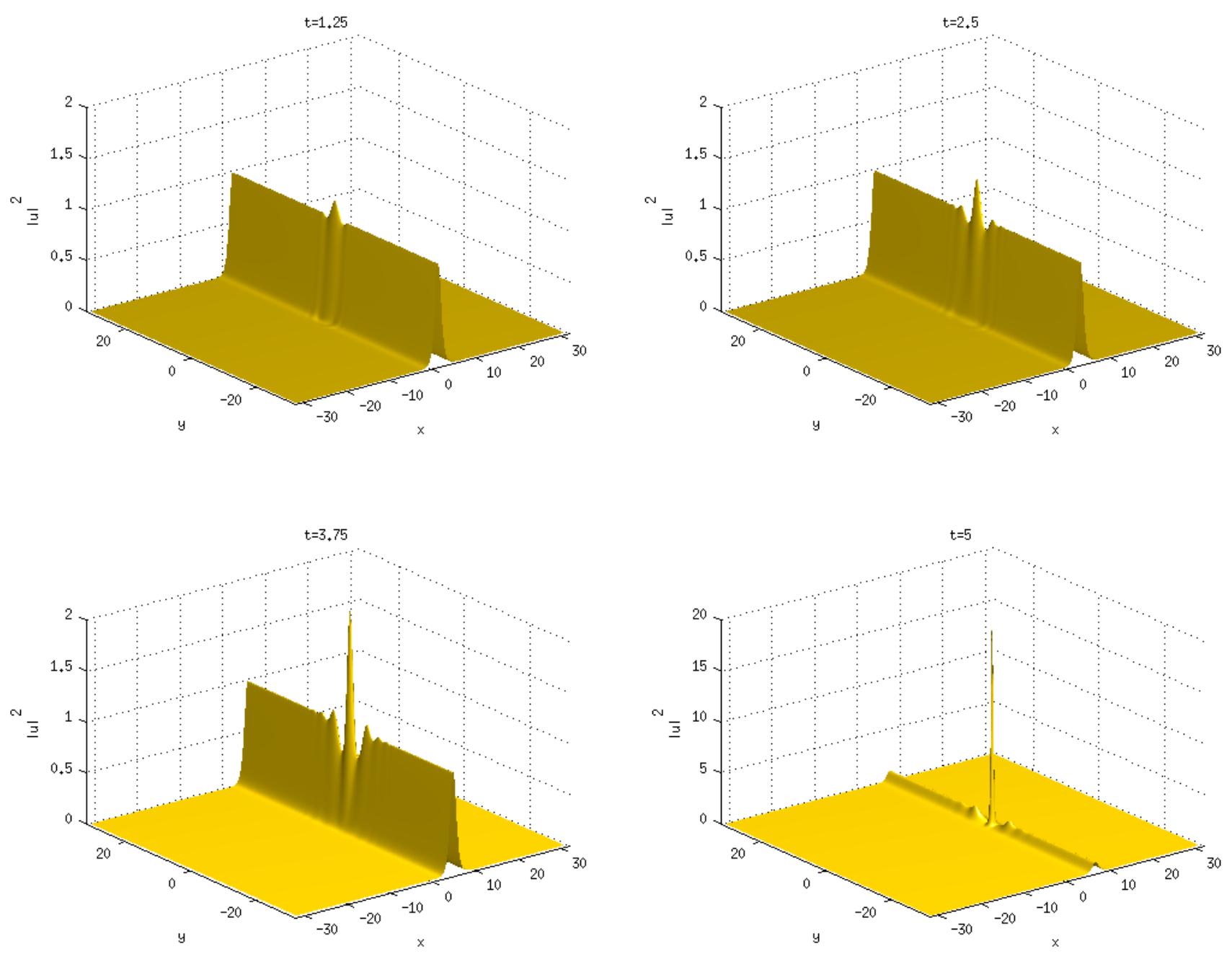} 
 \caption{Solution of the 2d cubic NLS$^+$ equation
 at several times for an initial data of the form (\ref{Agauss}) with $A=0.1$ and $(x_1,y_1)=(0,0)$.}
 \label{nlseA01uts}
\end{figure}

\begin{figure}[htb!]
\centering
\includegraphics[width=0.5\textwidth]{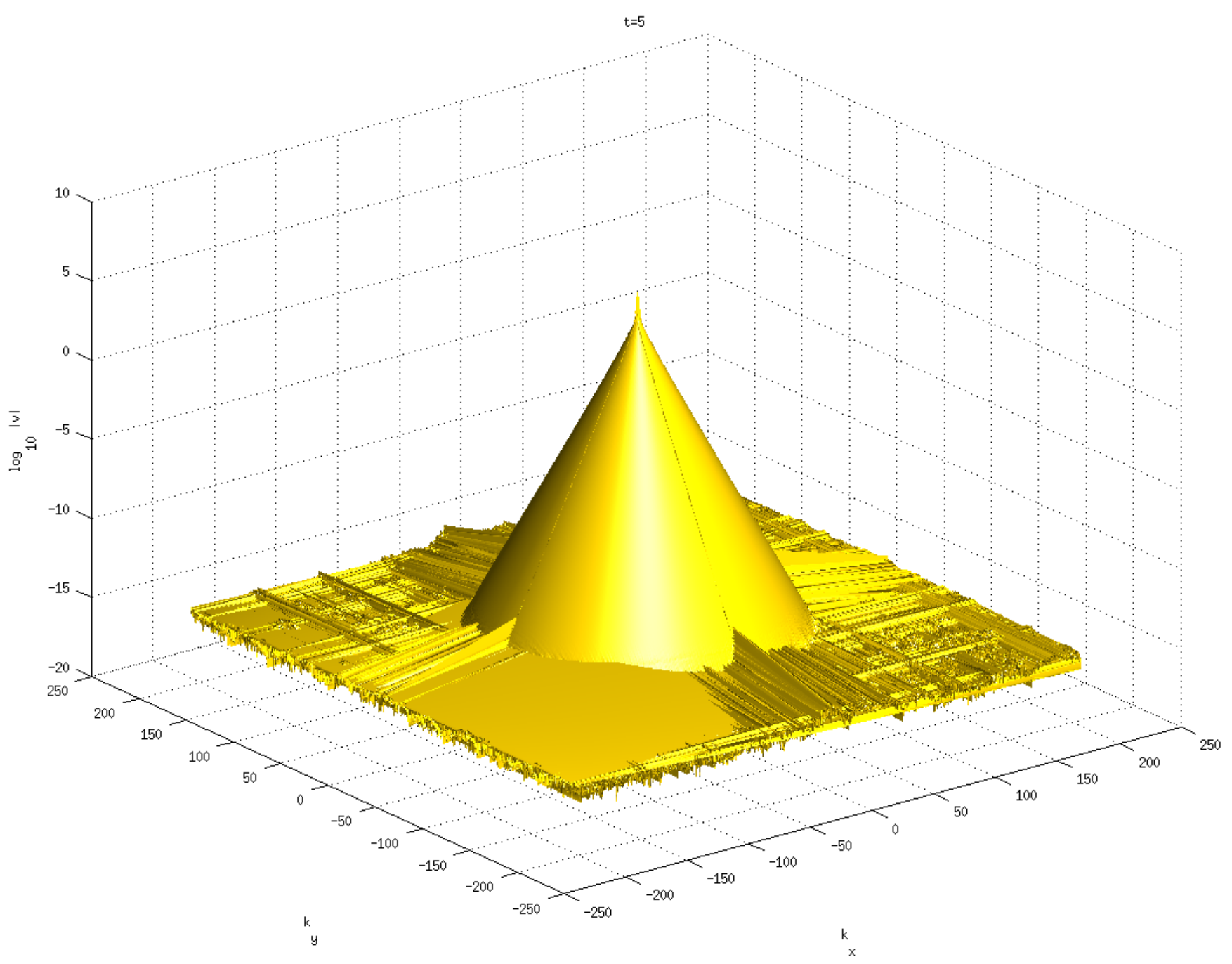} 
 \caption{Fourier coefficients of the solution shown in Fig. \ref{nlseA01uts} at $t=5$.}
 \label{nlseA01amplcoefs}
\end{figure}

\begin{figure}[htb!]
\centering
\includegraphics[width=0.45\textwidth]{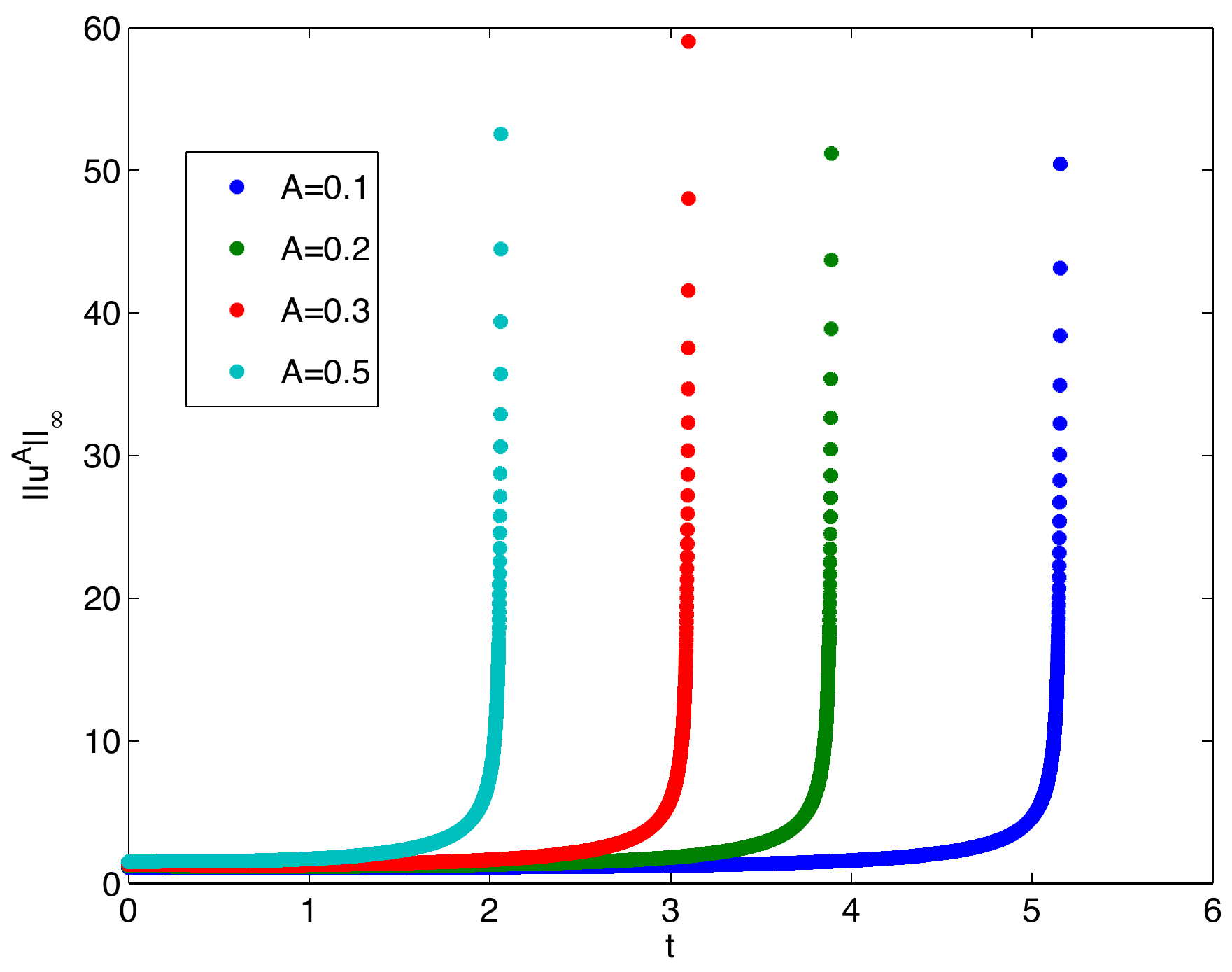} 
\includegraphics[width=0.45\textwidth]{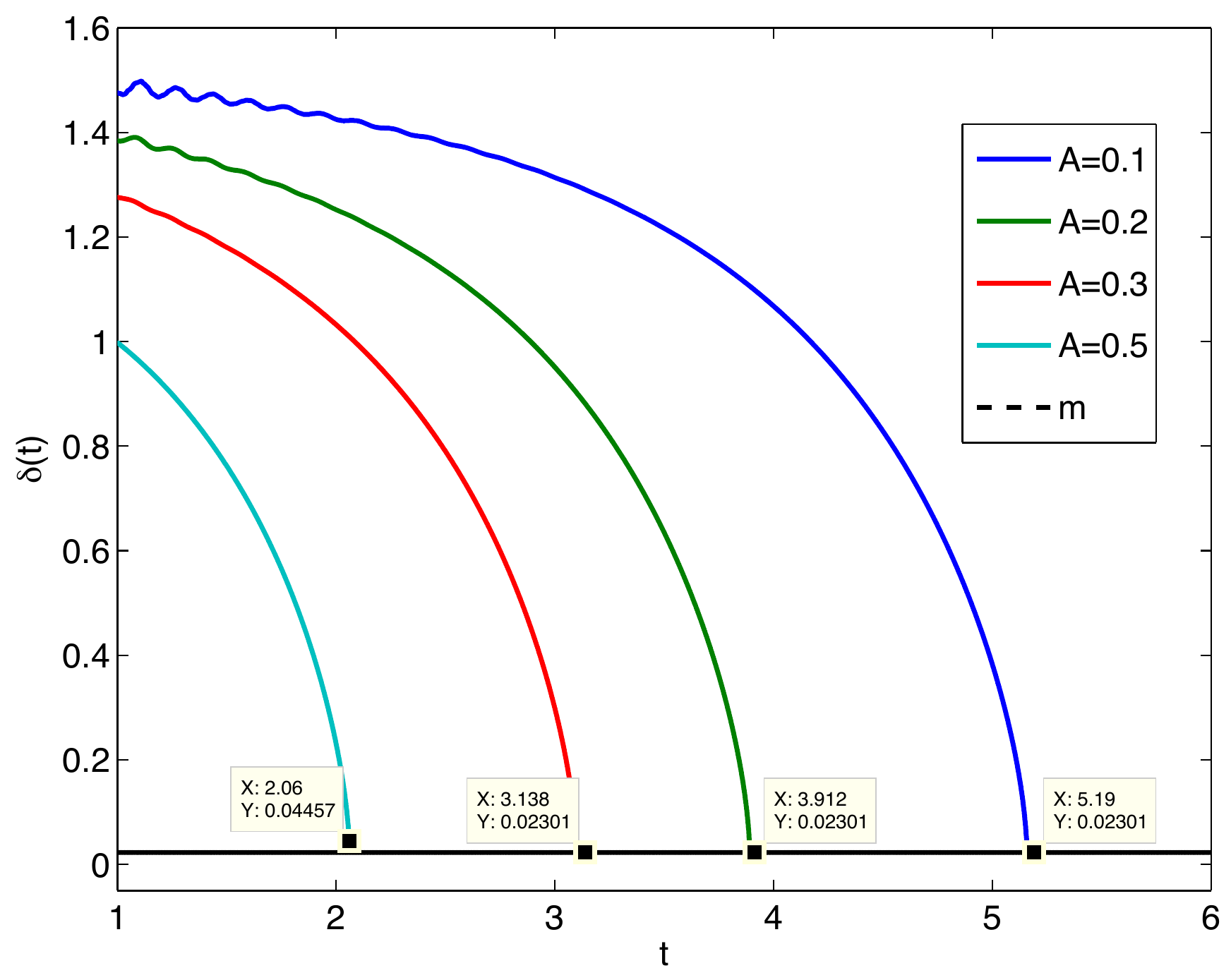} 
 \caption{Time evolution of the $L_{\infty}$-norm of the solutions $u^A$ to the
NLS$^+$ equation for an initial data of the form (\ref{Agauss}) with $(x_1,y_1)=(0,0)$ for several values of $A$ 
on the left,
and the corresponding fitting parameter $\delta(t)$ in (\ref{abd}) on the right.}
 \label{nlseamplAs}
\end{figure}

We first consider an initial data of the form (\ref{Agauss}) with $(x_1,y_1)=(0,0)$ for the NLS$^+$ equation.
For $A=0.1$, we chose the time step as $\Delta_t = 6*10^{-4}$, and observe that, 
as expected, the perturbed soliton under the 2d cubic NLS flow is unstable.
We show in Fig.\ref{nlseA01uts} the numerical solution at several times, it appears that the solution will 
blow up 
at $t^{*}=5.148$, where $\delta$ in (\ref{abd}) vanishes\footnote{
By vanishing, we mean here that the fitting parameter $\delta$ reaches 
the smallest distance in Fourier space which can be resolved, since we use a discrete Fourier series. It is defined by
 $m:=2\pi L_x/N_x$. No length below this threshold can
be numerically distinguished from zero, see also \cite{DSdDS}.
}, see Fig. \ref{nlseamplAs}. 
The Fourier coefficients of 
the solution at $t=5$ still decrease to machine precision $\sim 10^{-15}$, see Fig. \ref{nlseA01amplcoefs}.
That means that the system is still well resolved until $t=5$, just before the blow up time.
 At $t=t^*=5.148$, $\| u \|_{\infty}$ reaches a value of $\sim 50$, and 
$\| u_x \|_{\infty} \sim 1819$. The fitting error $p$ is of the order of $\sim 0.05$, indicating that the fitting is 
reliable, and that the blow up time is recovered with sufficient accuracy.
It is already clear from the pictures, that the blow up appears in only one spatial point, the location of the latter 
can be also identified as explained in section 3.2, here one finds $\alpha(t^*) = 7.2162$. The 
value of the numerically computed energy at $t=t^*$ is $\Delta_E \sim 10^{-15}$, indicating that the system 
is still well resolved, and that the asymptotic Fourier analysis provides also a determination of the blow up time before the 
'typical' crash of the numerical code. 
\\

\begin{figure}[htb!]
\centering
\includegraphics[width=0.7\textwidth]{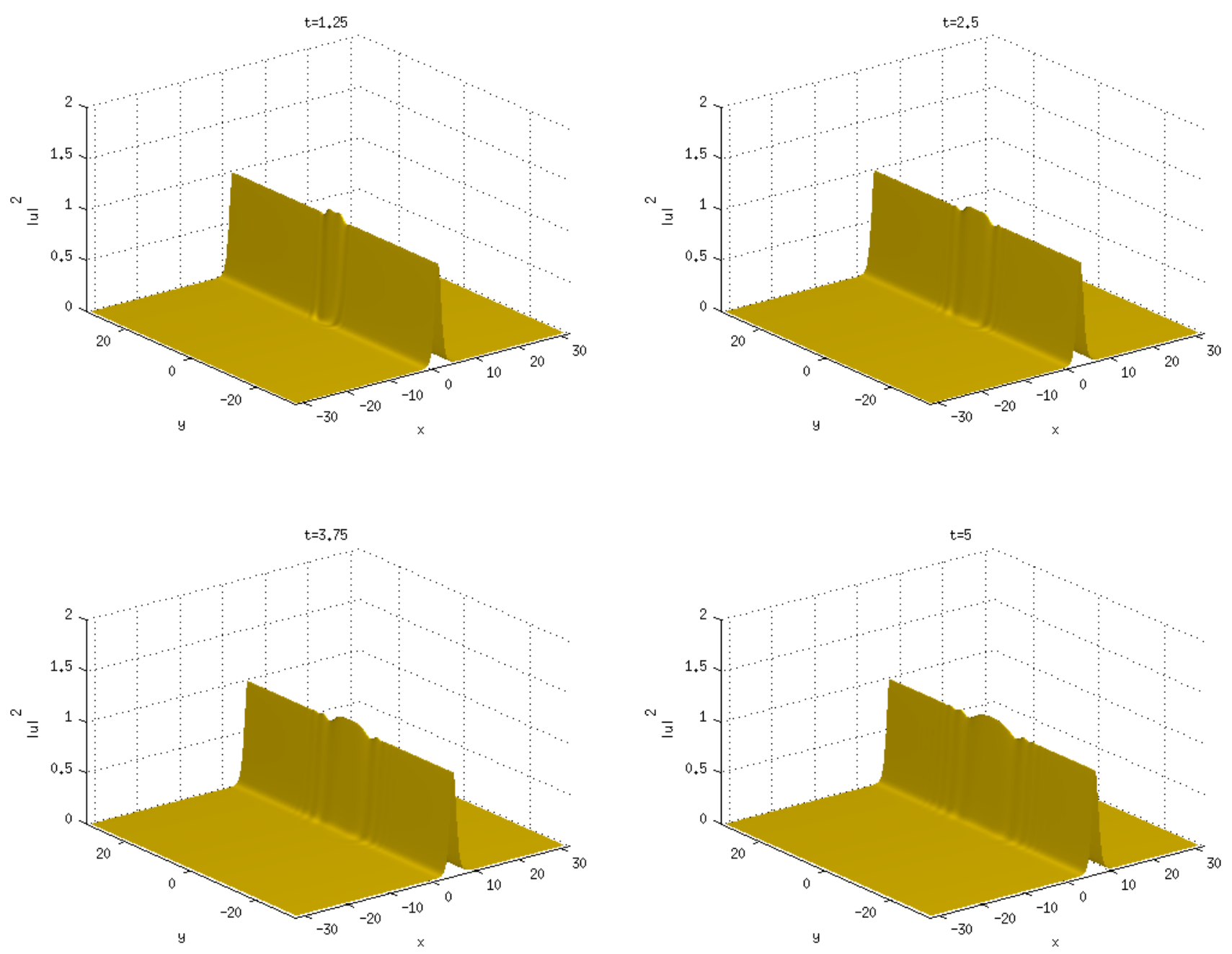} 
 \caption{Solution of the DS$^{++}$ system
 at several times for an initial data of the form (\ref{Agauss}) with $(x_1,y_1)=(0,0)$ with $A=0.1$.}
 \label{dseeA01uts}
\end{figure}

\begin{figure}[htb!]
\centering
\includegraphics[width=0.45\textwidth]{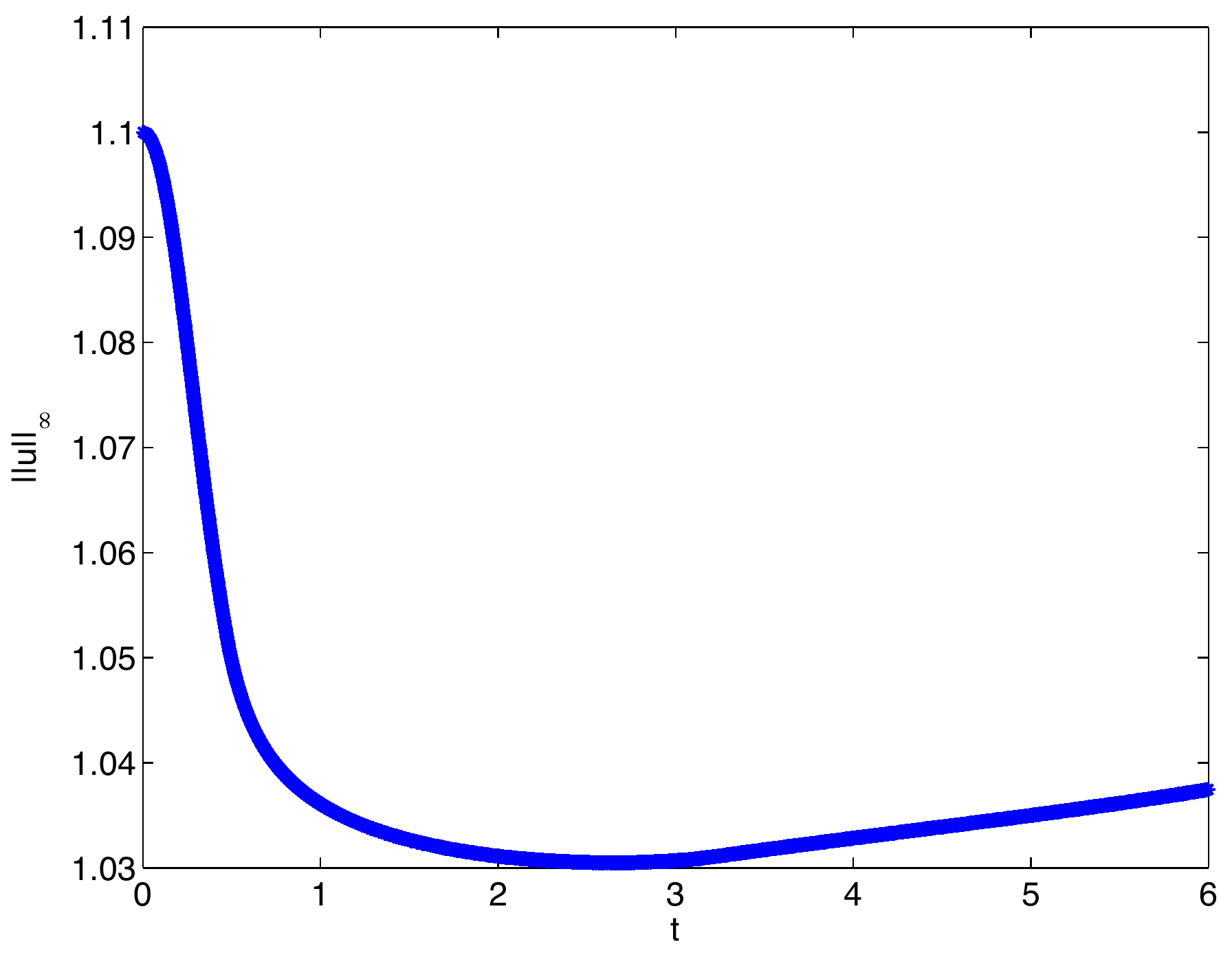} 
\includegraphics[width=0.45\textwidth]{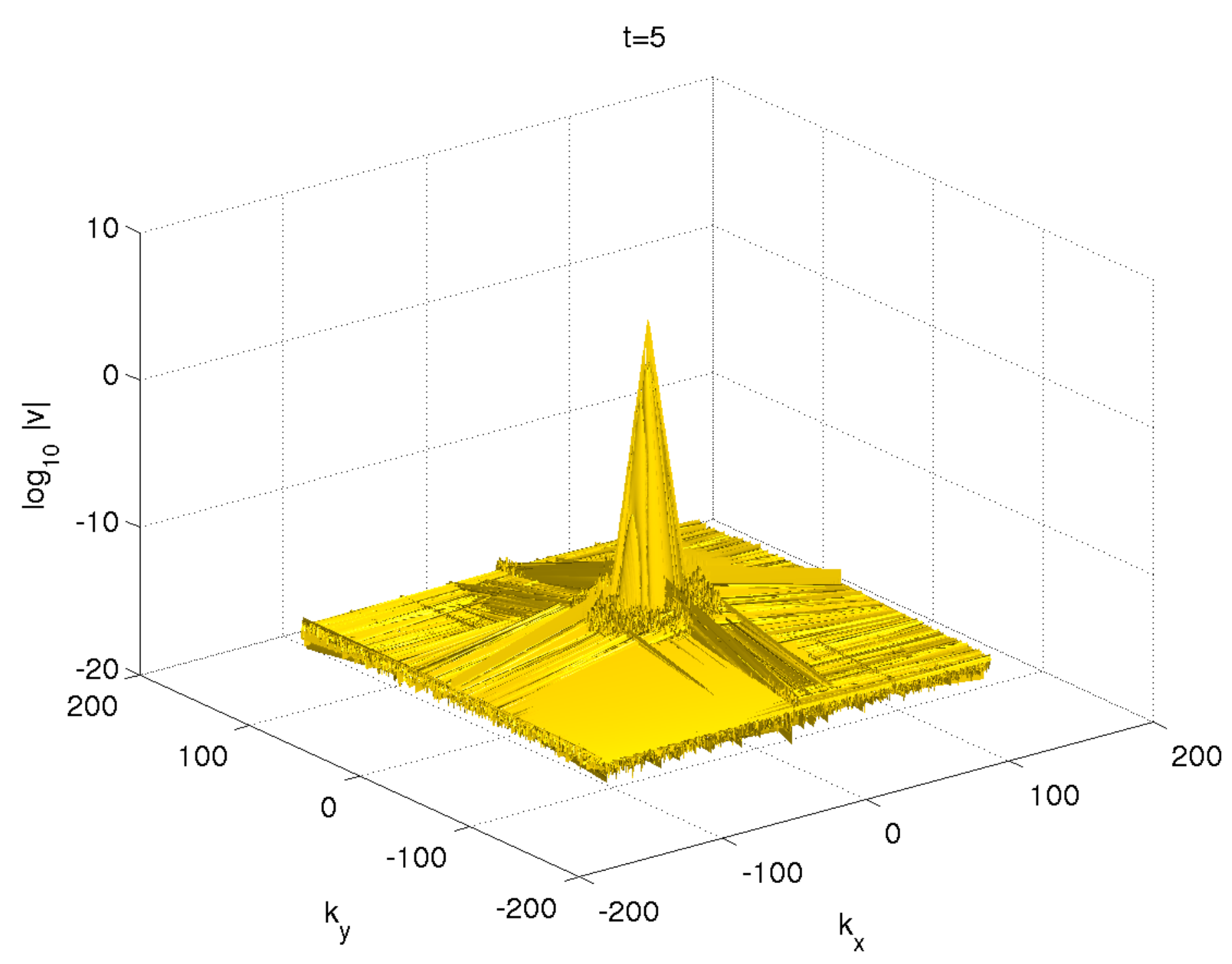} 
 \caption{Time evolution of the $L_{\infty}$-norm of the solution shown in Fig. \ref{dseeA01uts} on the left,
 and its Fourier coefficients at $t_{max}=5$ on the right.}
 \label{dseeA01linf}
\end{figure}

\begin{figure}[htb!]
\centering
\includegraphics[width=0.7\textwidth]{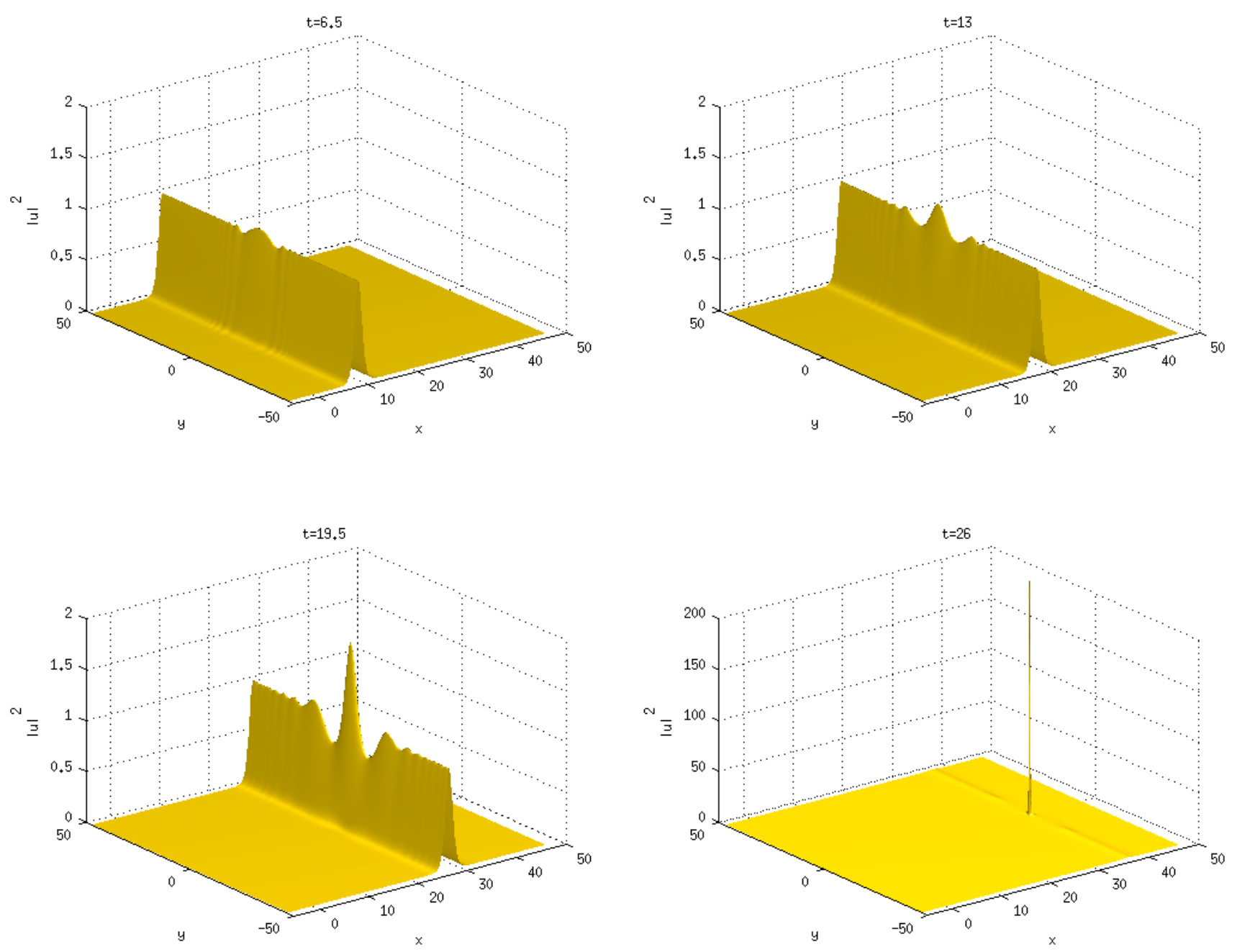} 
 \caption{Solution of the DS$^{++}$ system
 at several times for an initial data of the form (\ref{Agauss}) with $(x_1,y_1)=(0,0)$ and $A=0.1$.}
 \label{dseeA01uts2}
\end{figure}

The situation is similar for other values of $A>0$, the more we add
energy to the initial data, the earlier is the blow up time. 
We show in Fig. \ref{nlseamplAs} the time evolution of $\|u^A\|_{\infty}$ for several values of $A$, 
$u^A$ denoting the solution to the
NLS$^+$ equation for an initial data of the form (\ref{Agauss}), with $(x_1,y_1)=(0,0)$, and the time evolution of 
the corresponding fitting parameter $\delta(t)$ in (\ref{abd}).
The transverse instability of the isolated soliton (\ref{trav}) in the 2d cubic NLS$^+$ equation for localized perturbations
thus appears to be characterized by the appearance of a $L_{\infty}$ blow up of the solution at one spatial point.
\\
\\
One thus expects that a similar study for perturbations of the isolated soliton 
under the flow of the DS$^{++}$ system leads to similar results. It has been however not yet
proved, and the theory in \cite{RT08} does not seem to hold in this case due to the coupled mean field in 
the nonlinear part of the equation.
We thus consider now an initial data of the form (\ref{Agauss})  with $(x_1,y_1)=(0,0)$  for the DS$^{++}$ system.
\\
For $A=0.1$, and $\Delta_t = 5*10^{-4}$, we show the solution of the DS$^{++}$ system at several times in Fig. \ref{dseeA01uts}.
The situation differs from the previous study, due to the contribution of the 
coupled mean field $\Phi$ in the system (\ref{DSgen}). The localized perturbation is 
somehow more spread in the $y$-direction. The time evolution of the $L_{\infty}$-norm of $u$
is shown in Fig. \ref{dseeA01linf}, where we observe that after a rapid decrease, it increases again. 
The Fourier coefficients reach machine precision all along the computation, 
see for example the situation at $t=5$ in Fig. \ref{dseeA01linf}, as well as the numerically computed energy 
$\Delta_E$ typically used as an indicator of numerical accuracy, (one has $\Delta_E \sim 10^{-15} $ at $t=5$).
If the code is run for longer time, 
one finds that the numerical solution will actually blow up, as for the NLS$^+$ case.
We show in Fig. \ref{dseeA01uts2} the solution at later times.
As before, we perform the fitting of the Fourier coefficients of $u$ on the interval $10<k_x<2 max(k_x)/3$, 
following the well known de-aliasing rule. 
The vanishing of $\delta(t)$ occurs at $t=t^*=26.4$ indicating that a singularity occurs at this time.
The value of the fitting parameter $B$ at $t^*$ is of the order of $\sim 0.6$, which allows to conjecture a blow up of the solution 
in the $L_{\infty}$-norm, which can also be inferred from the value of the $L_{\infty}$-norm of $u$ at this time,
$\|u\|_{\infty}\sim 40$. The fitting error reaches a value of $p=0.3$. This indicates that the fitting is reliable, 
one typically expect an error not higher than $0.5$ especially for 
blow up situations we are looking at, see also \cite{DSdDS}.
The determination of the location of the singularity as in (\ref{phi}) yields $\alpha(t^*)=36.79$, and $\Delta_E \sim{10^{-14}}$ at $t=t^*$.

\begin{figure}[htb!]
\centering
\includegraphics[width=0.45\textwidth]{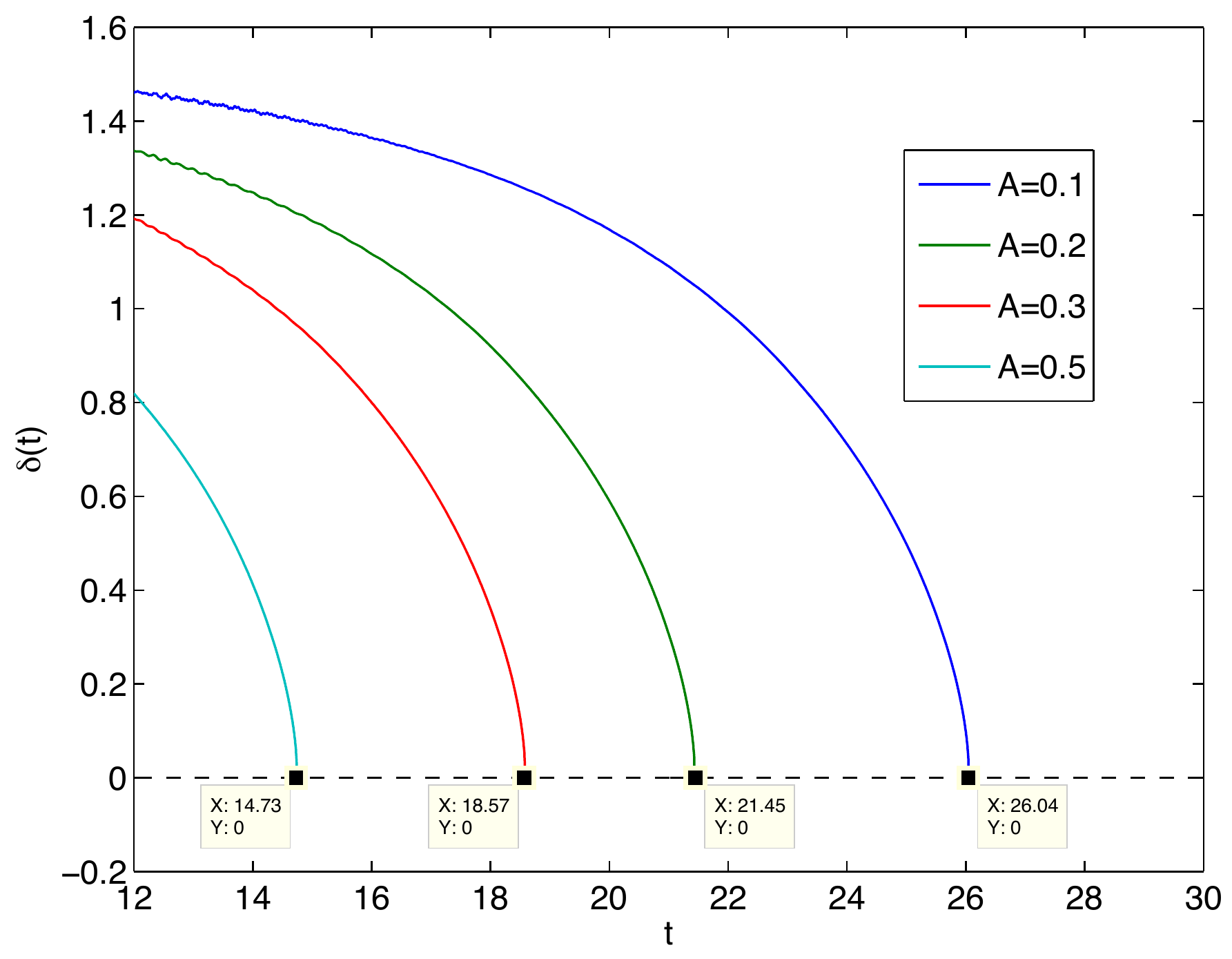} 
\includegraphics[width=0.45\textwidth]{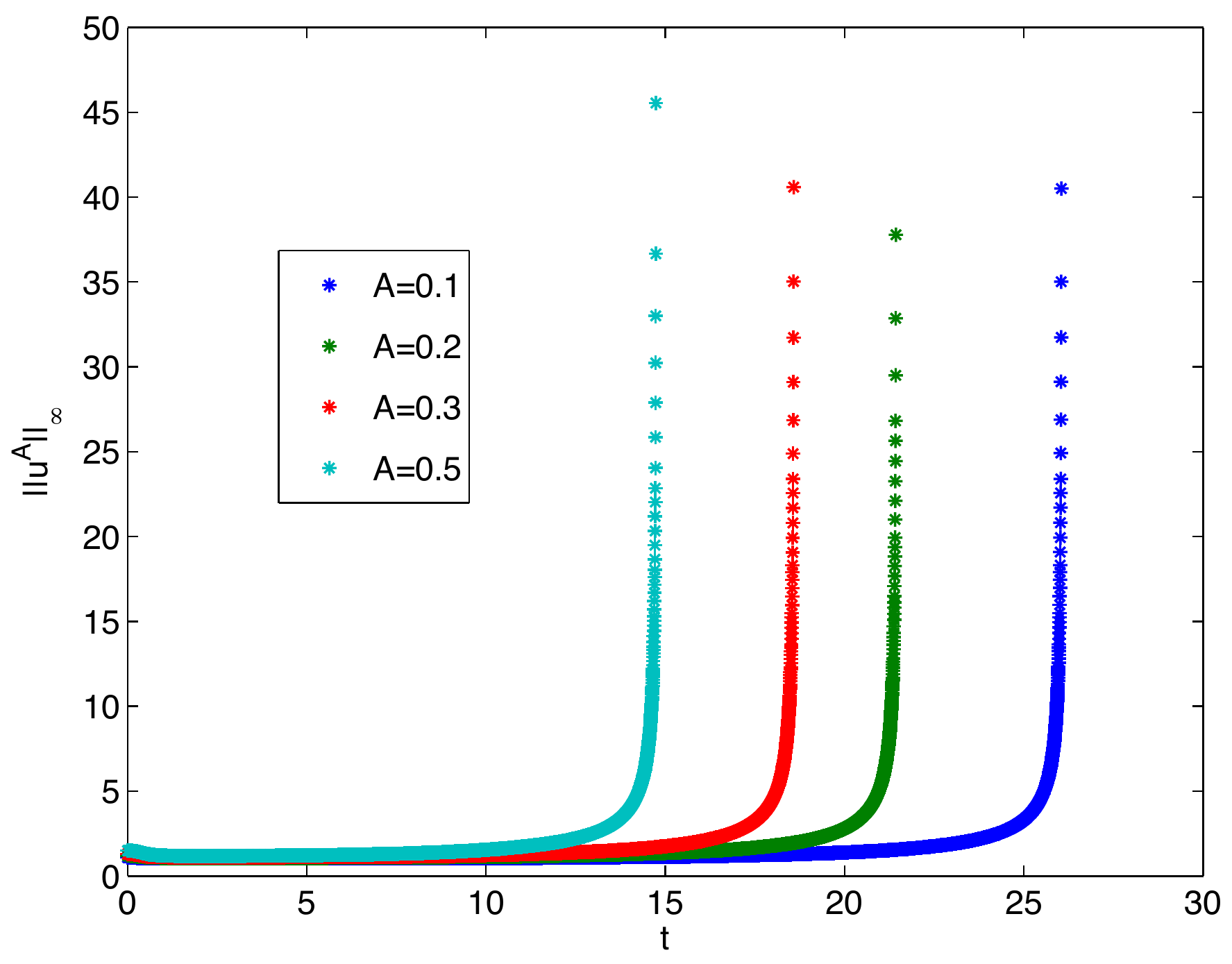} 
 \caption{Time evolution of the fitting parameter $\delta(t)$ on the left and of the $L_{\infty}$-norm of the solutions
 $u^A$ (on the right) to the
DS$^{++}$ equation for an initial data of the form (\ref{Agauss}) and $(x_1,y_1)=(0,0)$ for several values of $A$.}
 \label{dseeamplAs}
\end{figure}

The situation is similar for higher values of $A$, and we show in Fig. \ref{dseeamplAs}
the $L_{\infty}$-norm of the solutions $u^A$ to the
DS$^{++}$ equation for an initial data of the form (\ref{Agauss}) with $(x_1,y_1)=(0,0)$
for several values of $A$ on the right, and
the time evolution of the corresponding $\delta(t)$ in (\ref{abd}) on the left.
The blow up times are also reported in Table \ref{nlseDSeetblows} for the NLS$^+$ and DS$^{++}$ equations. 
\begin{table}
 \centering
 \begin{tabular}{|c|cccc|}
  \hline
  $A$ & $0.1$ & $0.2$ & $0.3$ & $0.5$ \\
  \hline
  $t^{*}_{NLS^{+}}$   & $5.16$ & $3.89$ & $3.10$ &  $2.06$  \\
  $t^{*}_{DS^{++}}$   & $26.04$ & $21.45$ & $18.57$ & $14.73$ \\
  \hline
   \end{tabular}
\label{nlseDSeetblows}
\caption{Blow up times of the solutions of the NLS$^+$ ($t^{*}_{NLS^{+}}$) and of the DS$^{++}$ 
($t^{*}_{DS^{++}}$) equation for an initial data of the form (\ref{Agauss}) with $(x_1,y_1)=(0,0)$ for several values of $A$}
\end{table}
\\

In this section, we illustrated numerically the instability of the isolated soliton (\ref{trav})
under the flow of the 2d elliptic NLS equation, and observed that this instability is characterized by the appearance 
of a blow up in one spatial point. 
Moreover, we performed the same study for the DS$^{++}$ system. 
Due to its elliptic character and the parameter chosen, a similar behavior was expected
and actually observed. Localized perturbations to the isolated soliton lead also
to a blow up here, at a later time though, 
because of the contribution of the 
the coupled elliptic equation for the mean field $\Phi$, which tends to extend the perturbation along the $y$-axis.
As expected, the isolated soliton is thus found to be nonlinearly unstable 
under the flow of the DS$^{++}$ system. 

\subsection{Hyperbolic NLS Equations}

In this section, we investigate the transverse (in)-stability of the 
isolated soliton (\ref{trav}) under the flow of hyperbolic NLS equations
by studying again Gaussian perturbations of the form (\ref{Agauss}).
We first perform simulations for the 2d hyperbolic cubic NLS (NLS$^-$), and then for the focusing DS II equation.
It is found that we the perturbed isolated soliton 
is unstable under the flow of the 2d hyperbolic NLS equation, and that this instability occurs via dispersion, 
and it appears to be orbitrarily stable under the flow of DS II for
Gaussian perturbations of the form \ref{Agauss} with $0<A<1$.
\\

\begin{figure}[htb!]
\centering
\includegraphics[width=0.7\textwidth]{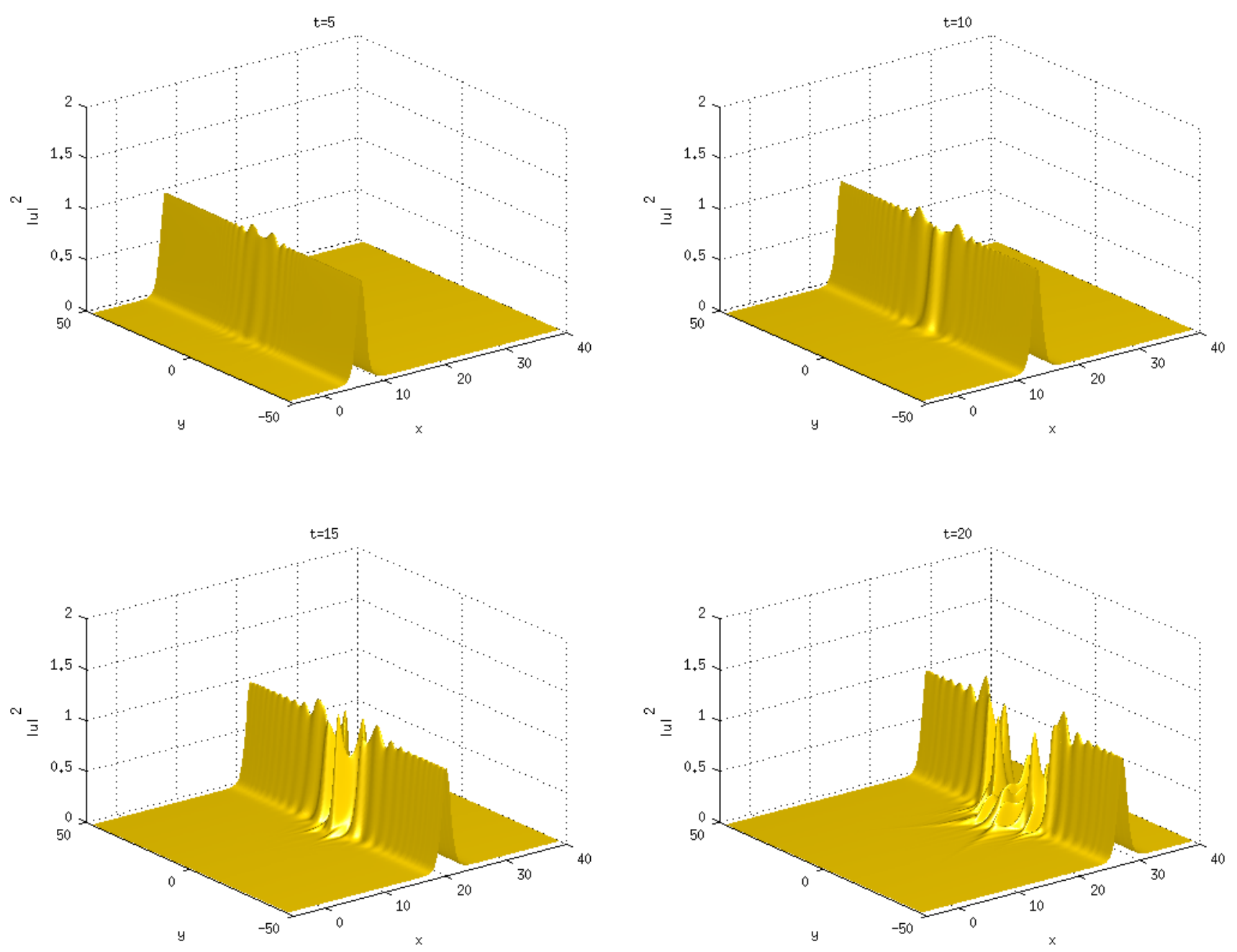} 
 \caption{Solution of the 2d cubic NLS$^-$ equation
 at several times for an initial data of the form (\ref{Agauss}) with $A=0.1$ and $(x_1,y_1)=(0,0)$.}
 \label{nlshA01uts}
\end{figure}

\begin{figure}[htb!]
\centering
\includegraphics[width=0.7\textwidth]{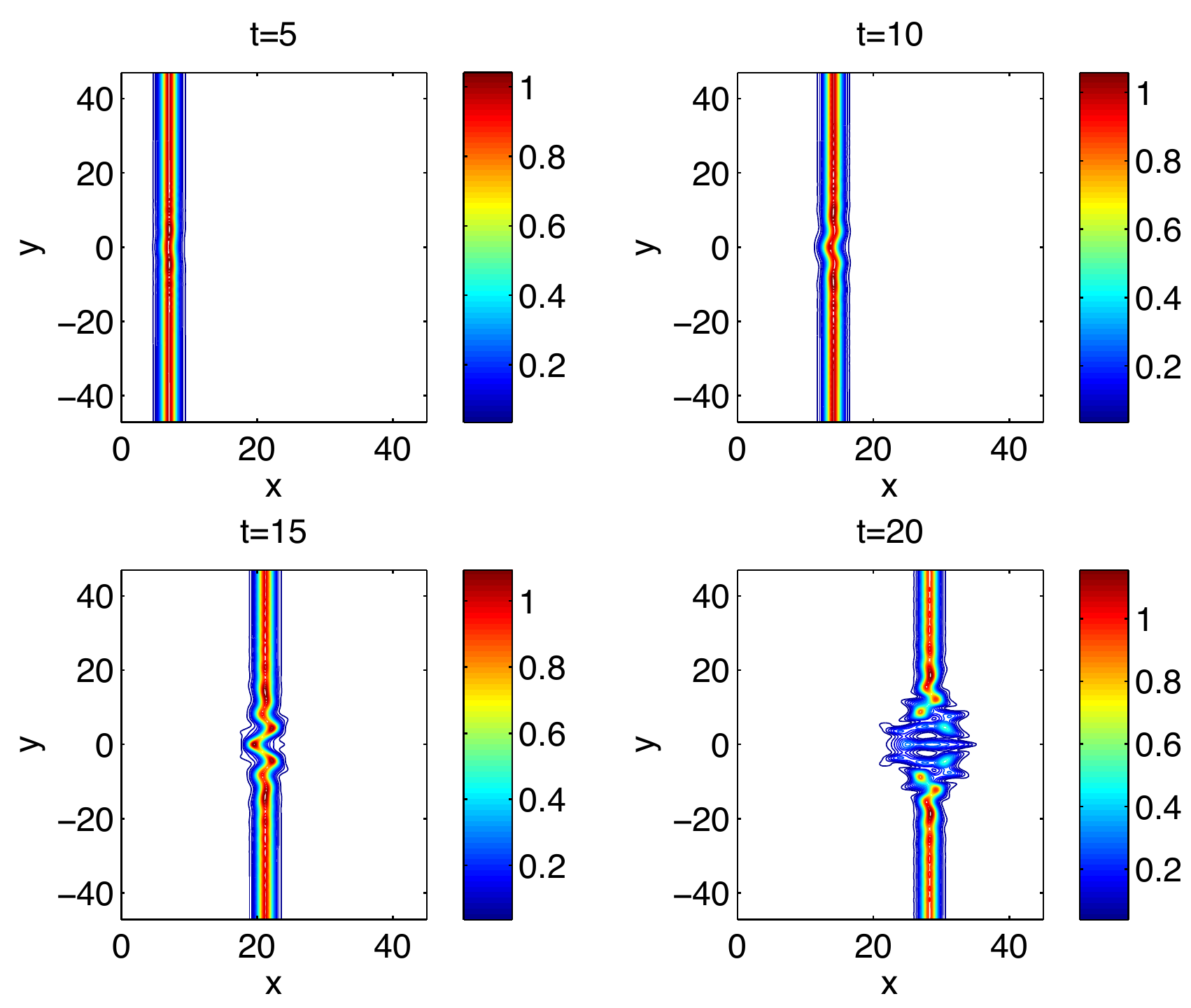} 
 \caption{Contour plots of the solution shown in Fig . \ref{nlshA01uts}}
 \label{nlshA01contts}
\end{figure}

\begin{figure}[htb!]
\centering
\includegraphics[width=0.7\textwidth]{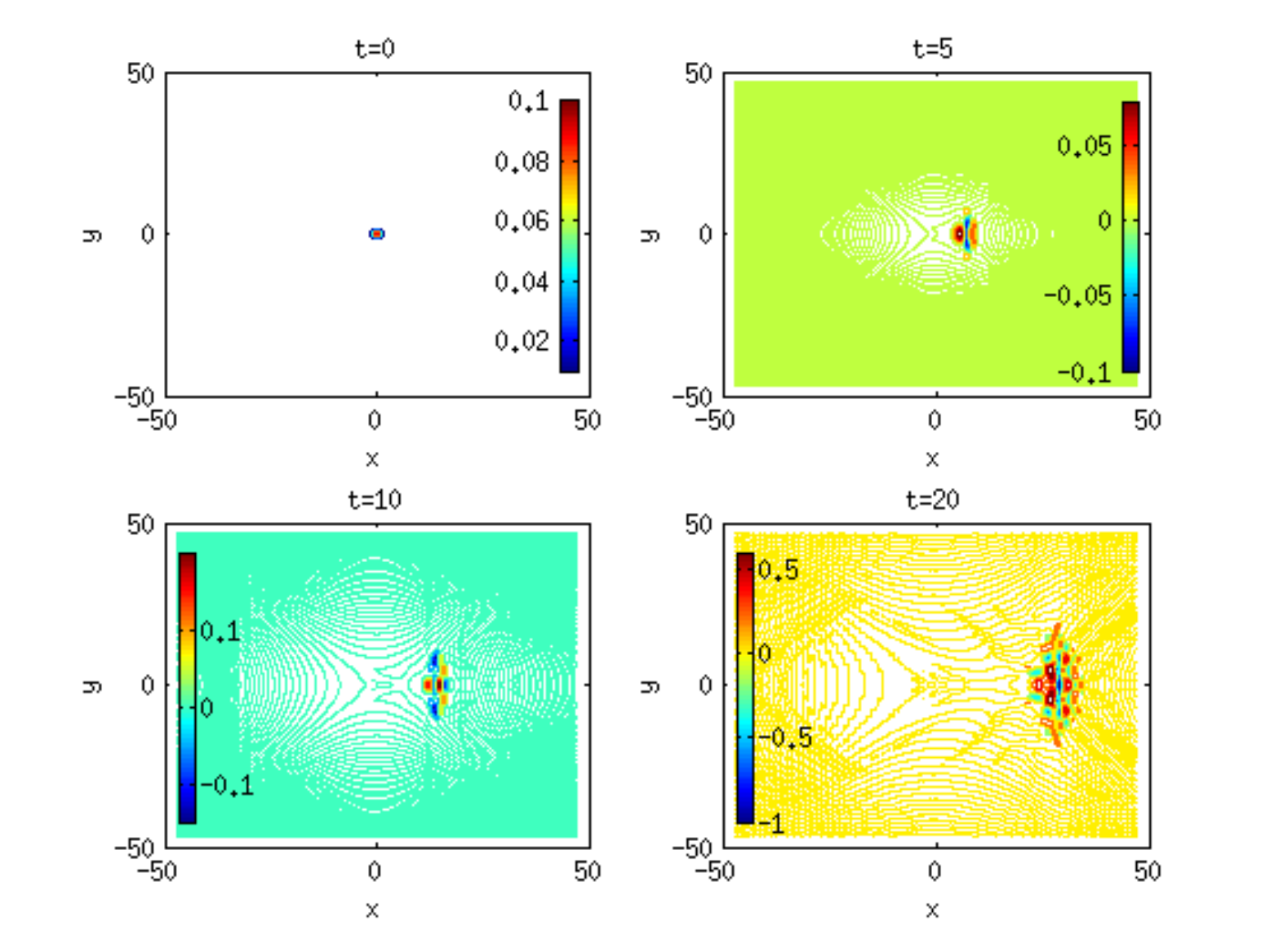} 
 \caption{Contour plots of the difference 
between the numerical solution shown in Fig. \ref{nlshA01uts} and the original soliton at several times}
 \label{nlshA01diffcontts}
\end{figure}

For $A=0.1$, we choose the time step as $\Delta_t = 2*10^{-3}$, and observe that, 
as expected, the perturbed soliton under the flow of NLS$^-$ flow is unstable. Recall that in this context only linear analysis 
is available in the literature.
We show in Fig.\ref{nlshA01uts} the numerical solution at several times, 
it appears that the solution becomes  
chaotic and totally looses the shape and typical features of the original soliton.
The situation is even clearer in Fig. \ref{nlshA01contts}, where we show the corresponding contour 
plots of the solution.
It seems the original soliton will totally disperse for longer times.
In fact, as we can infer from Fig. \ref{nlshA01diffcontts}, where we show the contour plot of the difference 
between the numerical solution and the original soliton, the perturbation travels at the same velocity as the 
isolated soliton and simply disperses away. The dispersion of the perturbation yields then the dispersion of the full numerical solution.

\begin{figure}[htb!]
\centering
\includegraphics[width=0.45\textwidth]{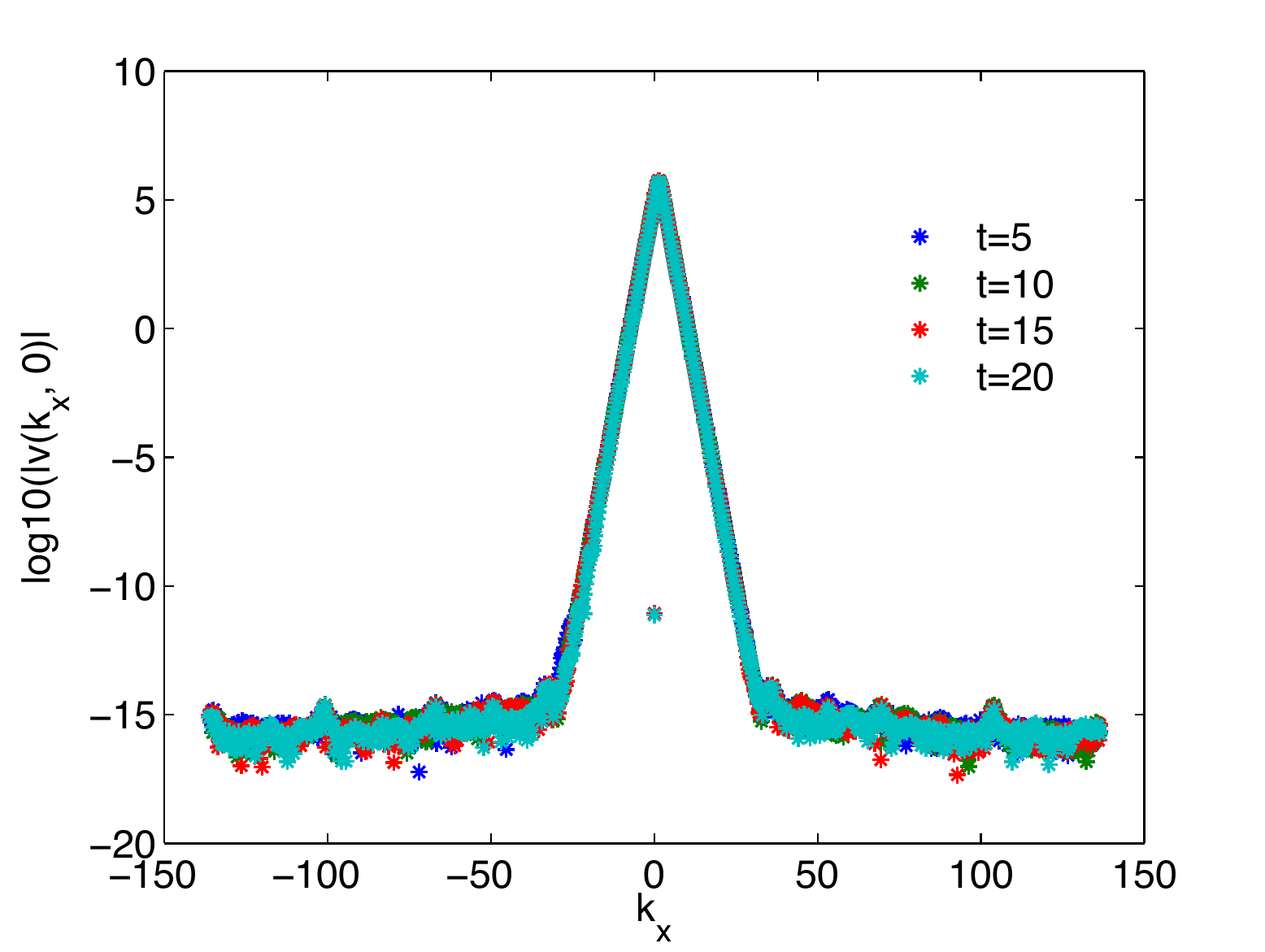} 
\includegraphics[width=0.45\textwidth]{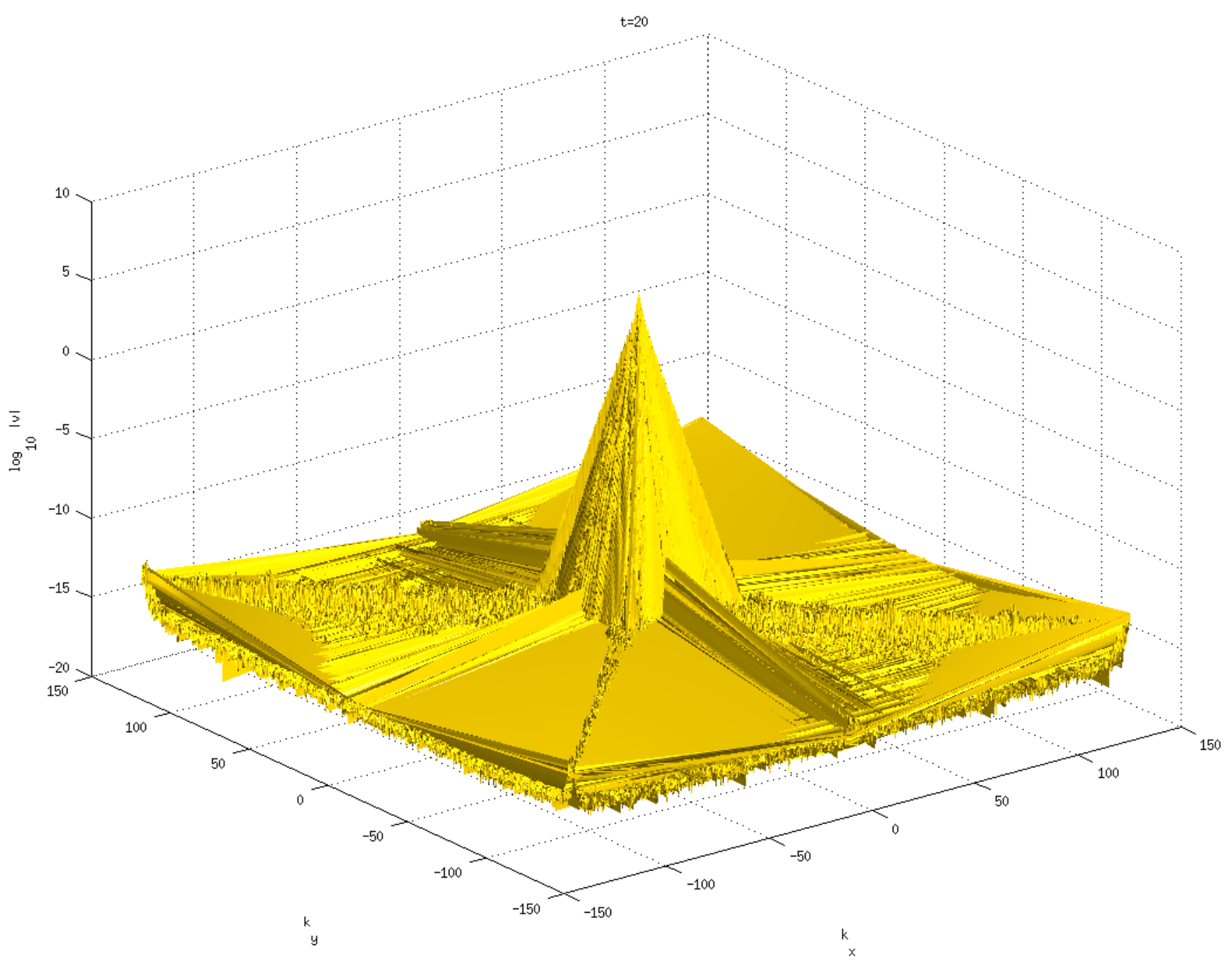} 
 \caption{Fourier coefficients of 
the solution shown in Fig. \ref{nlshA01uts} at several times plotted on the $k_x$-axis 
 on the left, and at $t=20$ on the right (plotted in both directions)}
 \label{nlshA01vts}
\end{figure}

\begin{figure}[htb!]
\centering
\includegraphics[width=0.7\textwidth]{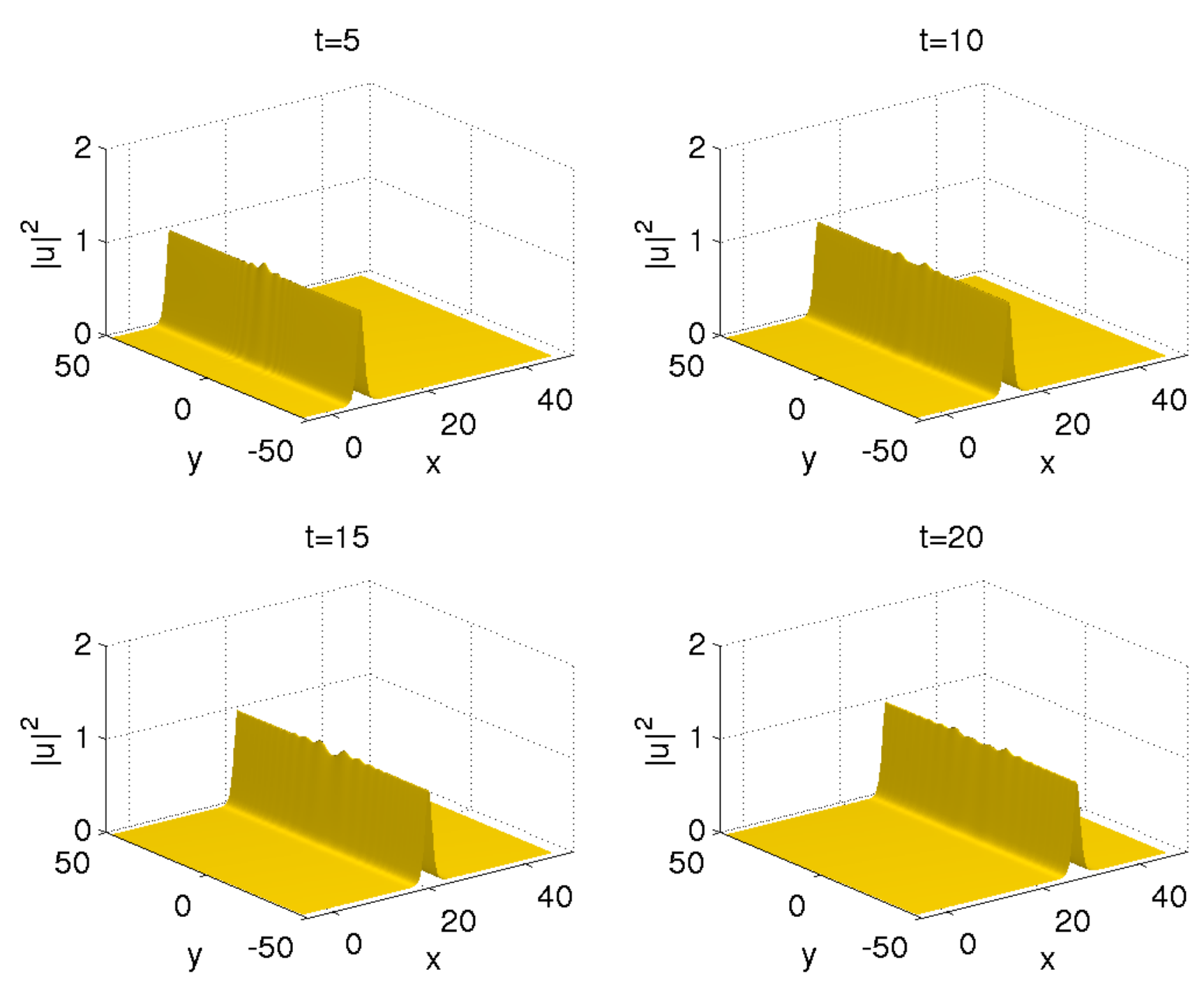} 
 \caption{Solution to the focusing DS II equation for an initial condition 
 of the form (\ref{Agauss}), with $A=0.1$ and $(x_1,y_1)=(0,0)$, at several times.}
 \label{ds2isopg01u}
\end{figure}

\begin{figure}[htb!]
\centering
\includegraphics[width=0.5\textwidth]{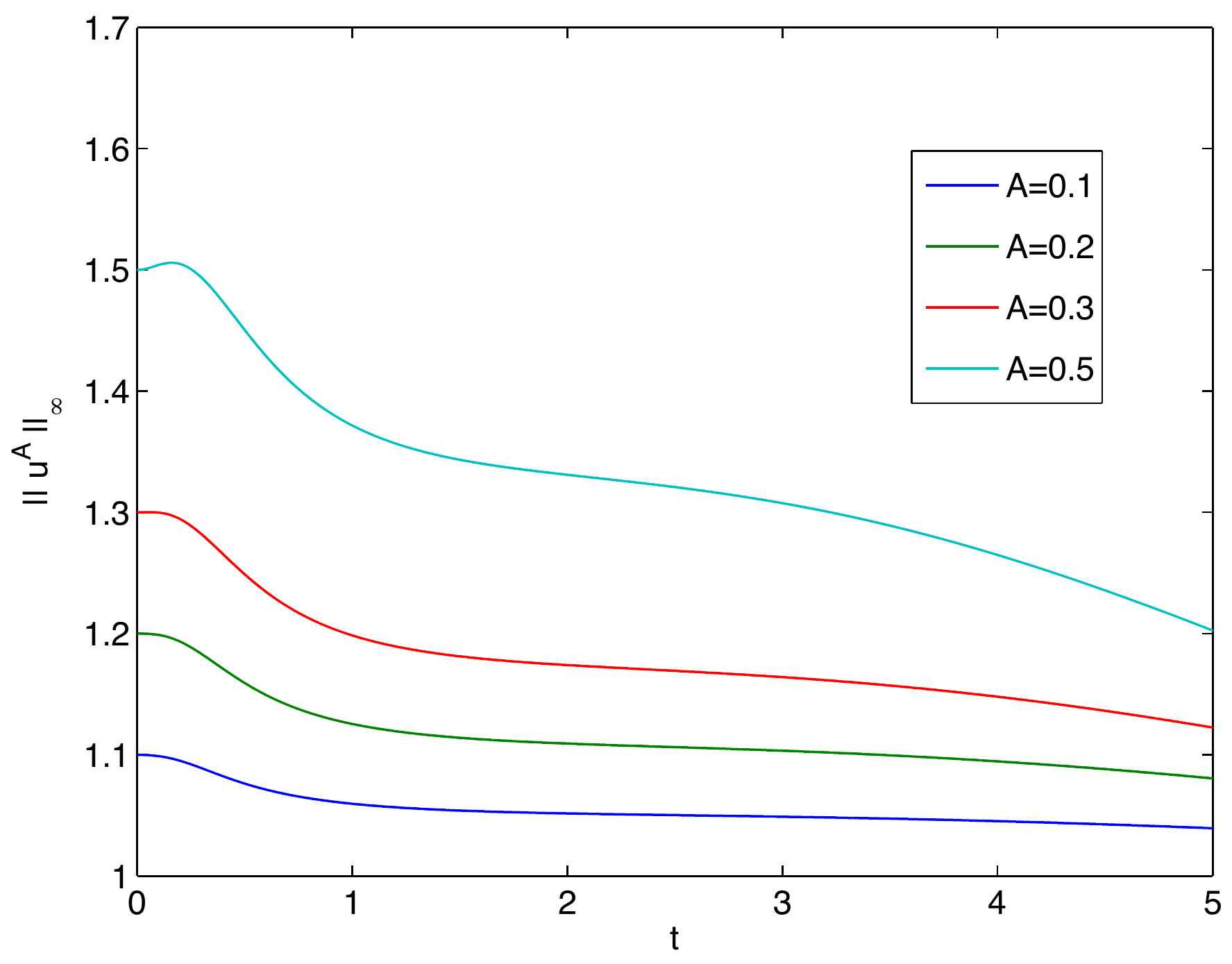}
 \caption{Time evolution of the $L_{\infty}$-norm of the solutions $u^A$ to the 
 focusing DS II equation for an initial condition of the form 
 (\ref{Agauss}) with $(x_1,y_1)=(0,0)$, for several values of  $A$.}
 \label{ds2isopg01a}
\end{figure}

We ensure that the system is well resolved by checking the good decay of the Fourier 
coefficients all along the computation, they decrease to machine precision ($10^{-15}$)
at all times considered. We show in Fig. \ref{nlshA01vts} the Fourier coefficients of 
the solution at several times plotted on the $k_x$-axis 
 on the left, and plotted in both spatial directions at $t=20$ on the right.
The same behavior was observed for higher values of $A$.
The instability of the isolated soliton thus appears to occur though/via the dispersion of the solution
as $t$ tends to infinity.
\\
\\
We now perform the same study for the DS II equation. 
We show the numerical solution at several times in Fig .\ref{ds2isopg01u}.
In this case the perturbation seems to be somehow distributed 
along the isolated soliton, in the $y$-direction. 
The $L_{\infty}$-norm of $u$ decreases as $t\to \infty$, 
see Fig. \ref{ds2isopg01a}, where we show also the situation for other values 
of $0<A<1$ until $t=5$. 
It seems it will stabilise for later times.
The fitting of the Fourier coefficients appears to be reliable, with a fitting error of the order of $p\sim 0.1$ all along the computation.
One finds that
$\delta(t)$ is almost constant $\delta(t) \sim 1.5$,
for all times  $t \leq 20$ studied, 
indicating the regularity of the solution.

\begin{figure}[htb!]
\centering
\includegraphics[width=0.45\textwidth]{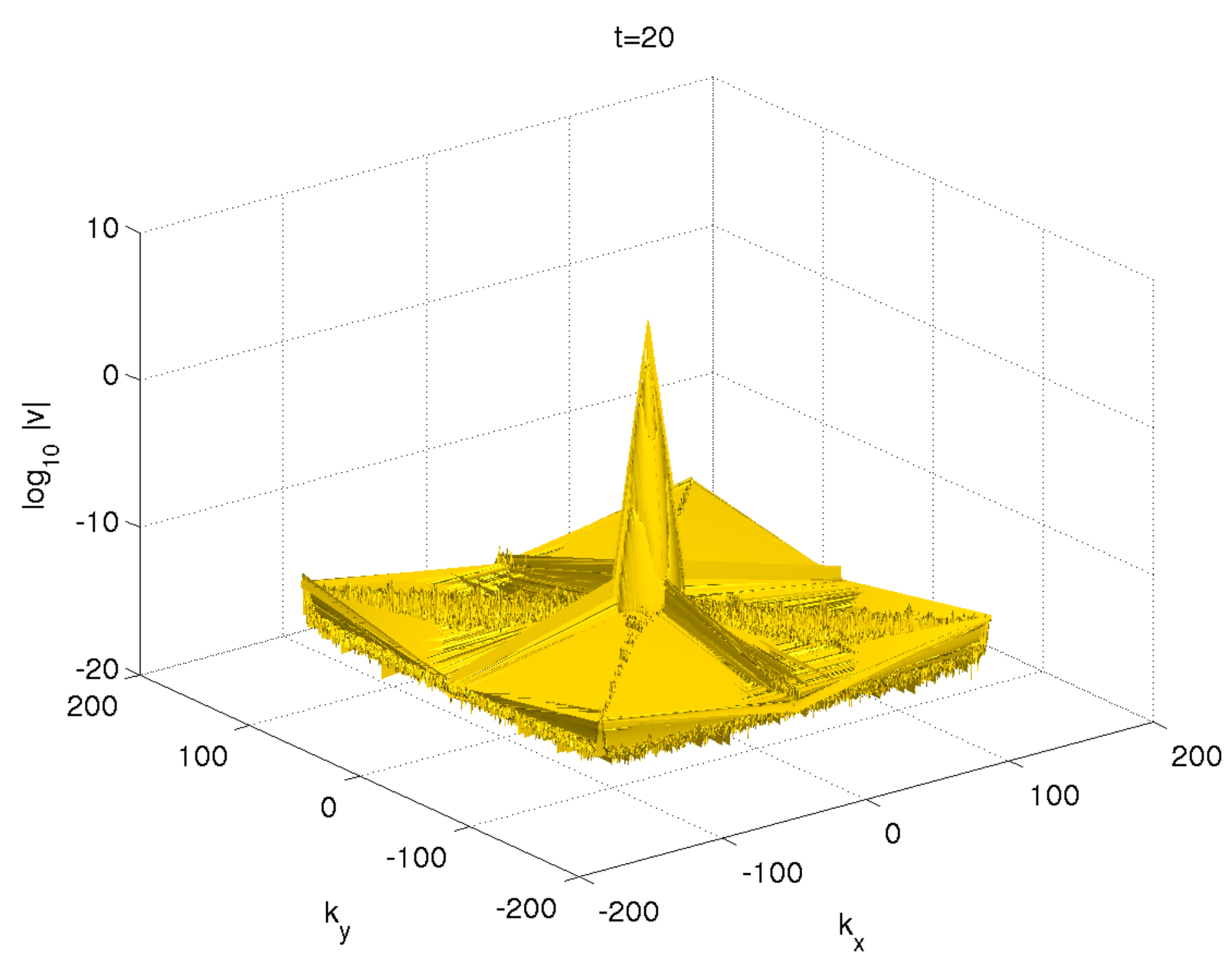} 
\includegraphics[width=0.45\textwidth]{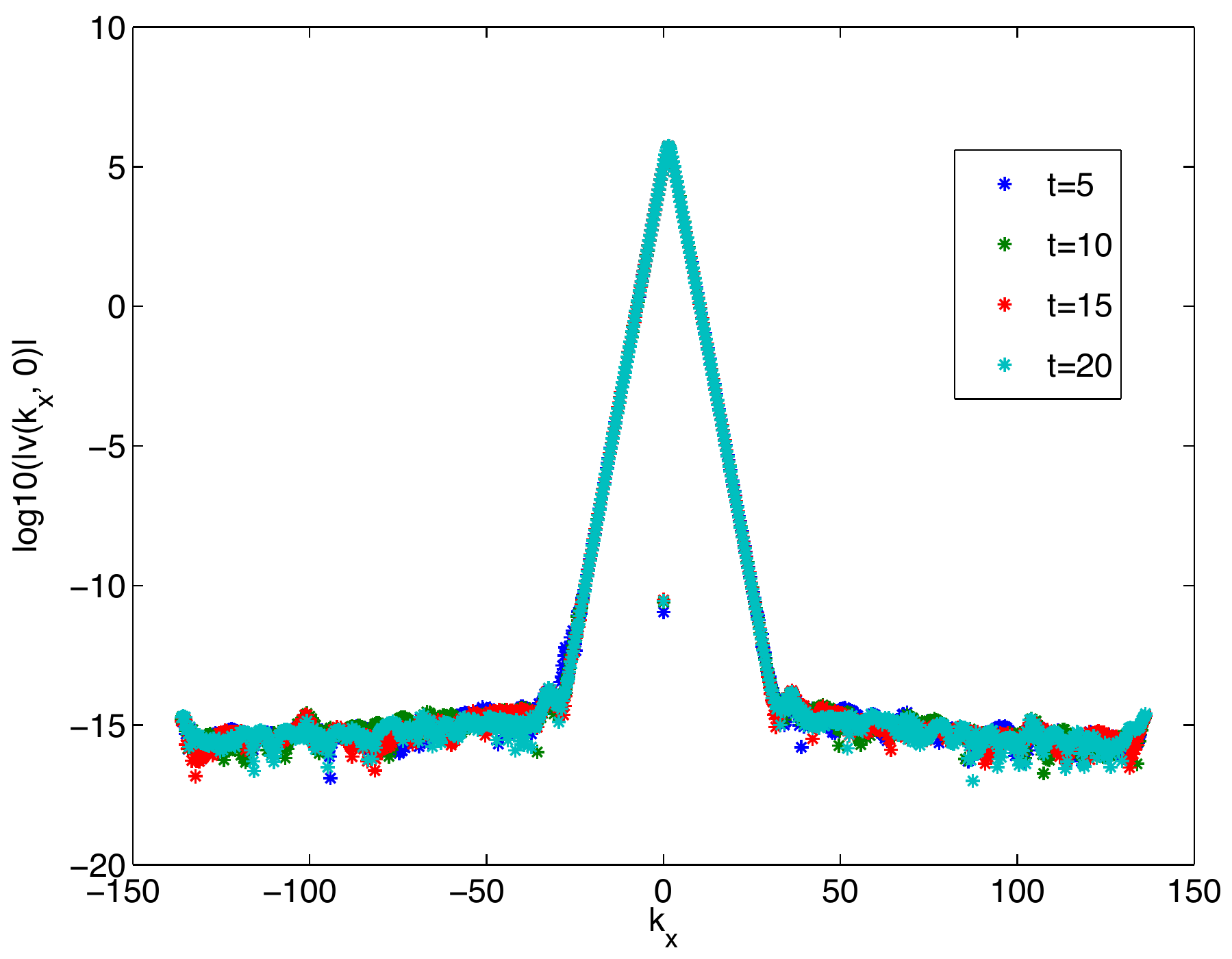} 
\caption{Fourier coefficients of the solution to the focusing DS II equation for an initial 
condition of the form (\ref{Agauss}), with $A=0.1$ and $(x_1,y_1)=(0,0)$, at $t=20$ on the left, and at several times (plotted 
on the $k_x$-axis on the right).}
\label{ds2isopg01co}
\end{figure}

\begin{figure}[htb!]
\centering
\includegraphics[width=0.45\textwidth]{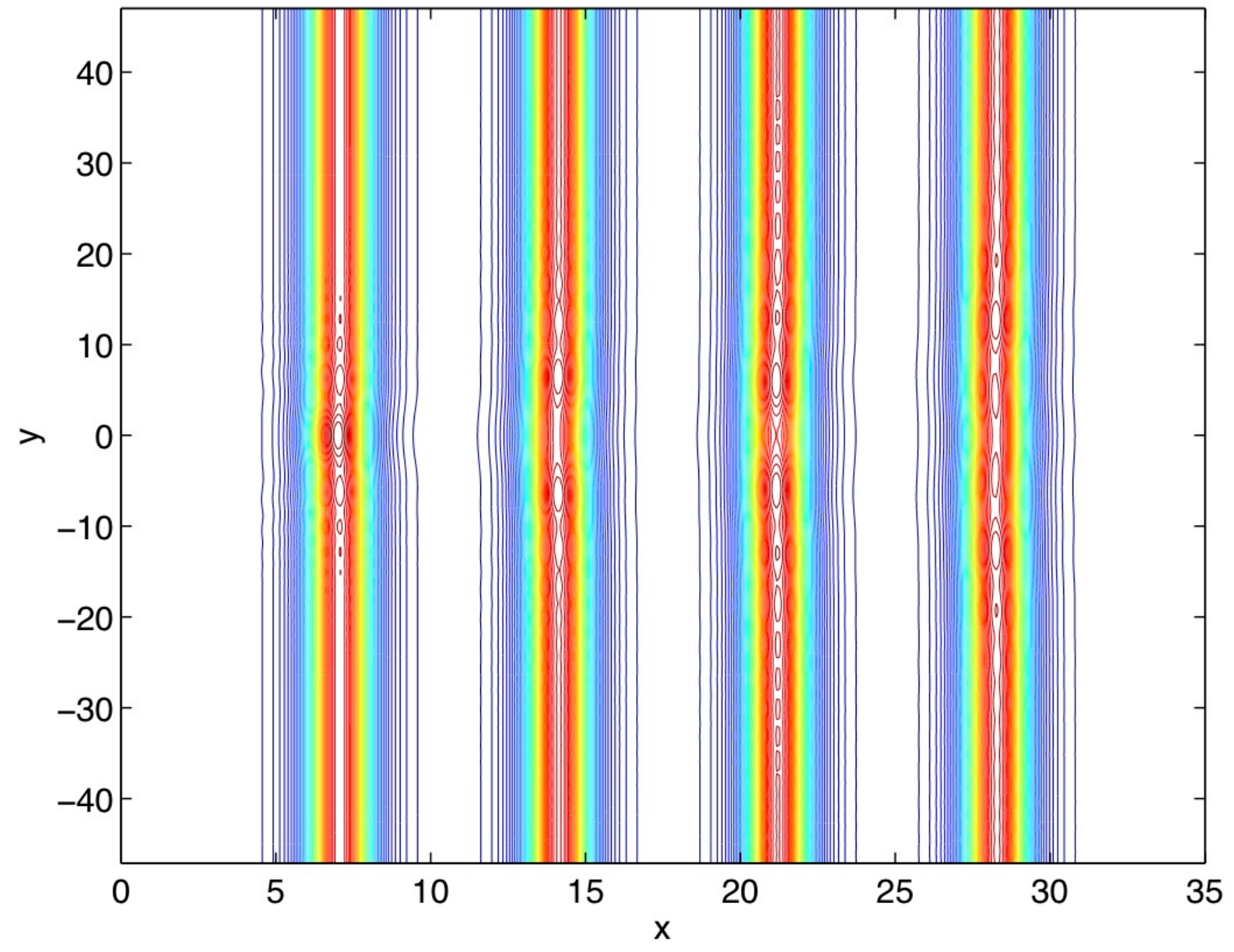} 
 \caption{Contour plots of the solution shown in Fig. \ref{ds2isopg01u} at several times, 
 from the left to the right: $t=5, 10, 15, 20$.}
 \label{ds2isopg01utscontt20}
\end{figure}

\begin{figure}[htb!]
\centering
\includegraphics[width=0.7\textwidth]{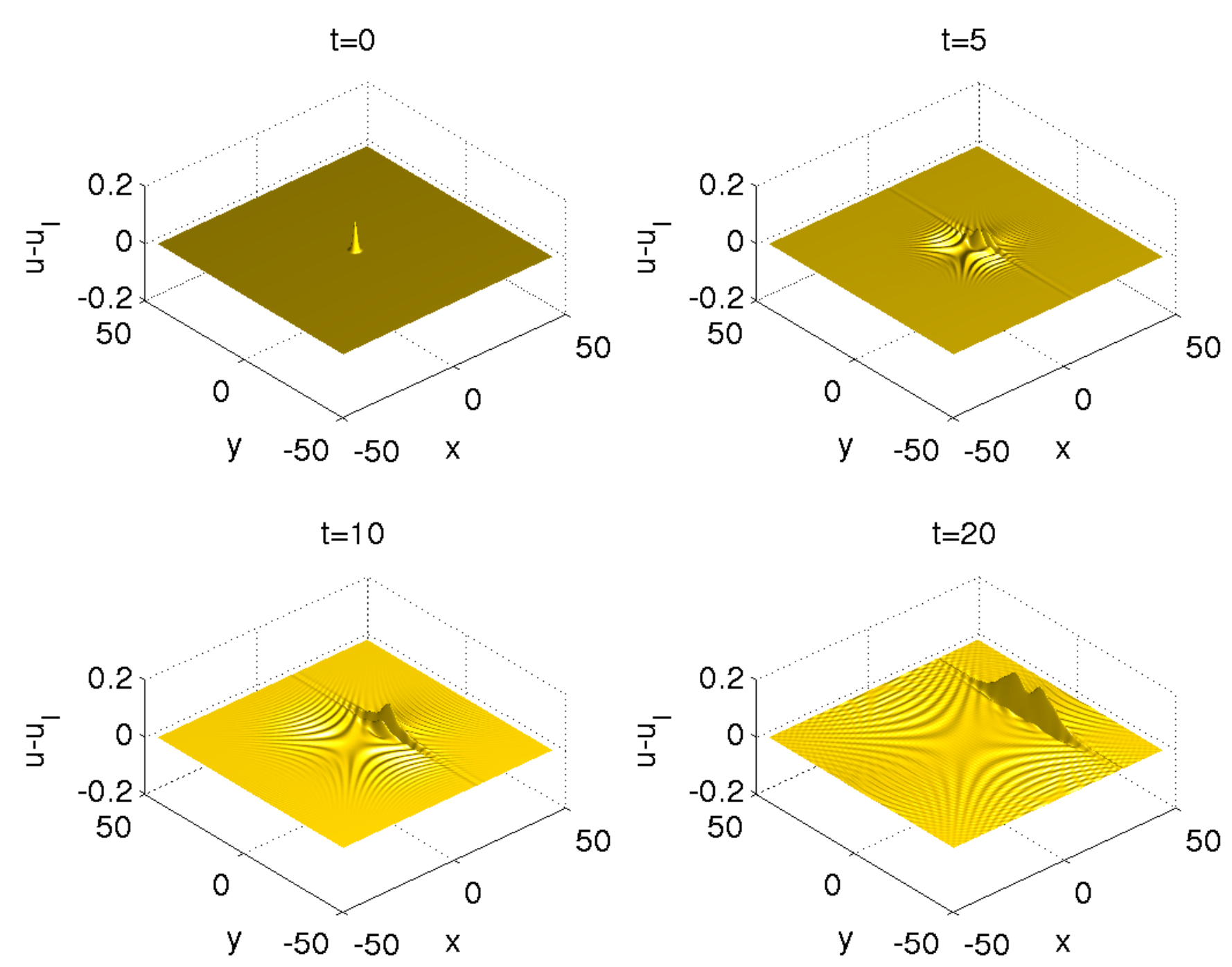} 
 \caption{Difference between the solution of the DS II equation for an initial data of the form
 (\ref{Agauss}) with $(x_1,y_1)=(0,0)$ and the original soliton $u_I$ at several times.}
 \label{ds2A01diffts}
\end{figure}

The numerical accuracy is ensured by the decay to machine precision ($\sim 10^{-15}$) of the Fourier 
coefficients, , see Fig. \ref{ds2isopg01co},
 and the conservation of the numerically computed energy, $\Delta_E$, which reaches the same order 
as the latter ($\Delta_E \sim 10^{-15}$) all along the computation.
\\
\\
In this case, the solution preserves the shape of the original soliton, see 
Fig. \ref{ds2isopg01utscontt20} for the contour plots of the solution shown in Fig. \ref{ds2isopg01u},
and the perturbation itself takes the form of a soliton, see Fig. \ref{ds2A01diffts}. 

\begin{figure}[htb!]
\centering
\includegraphics[width=0.7\textwidth]{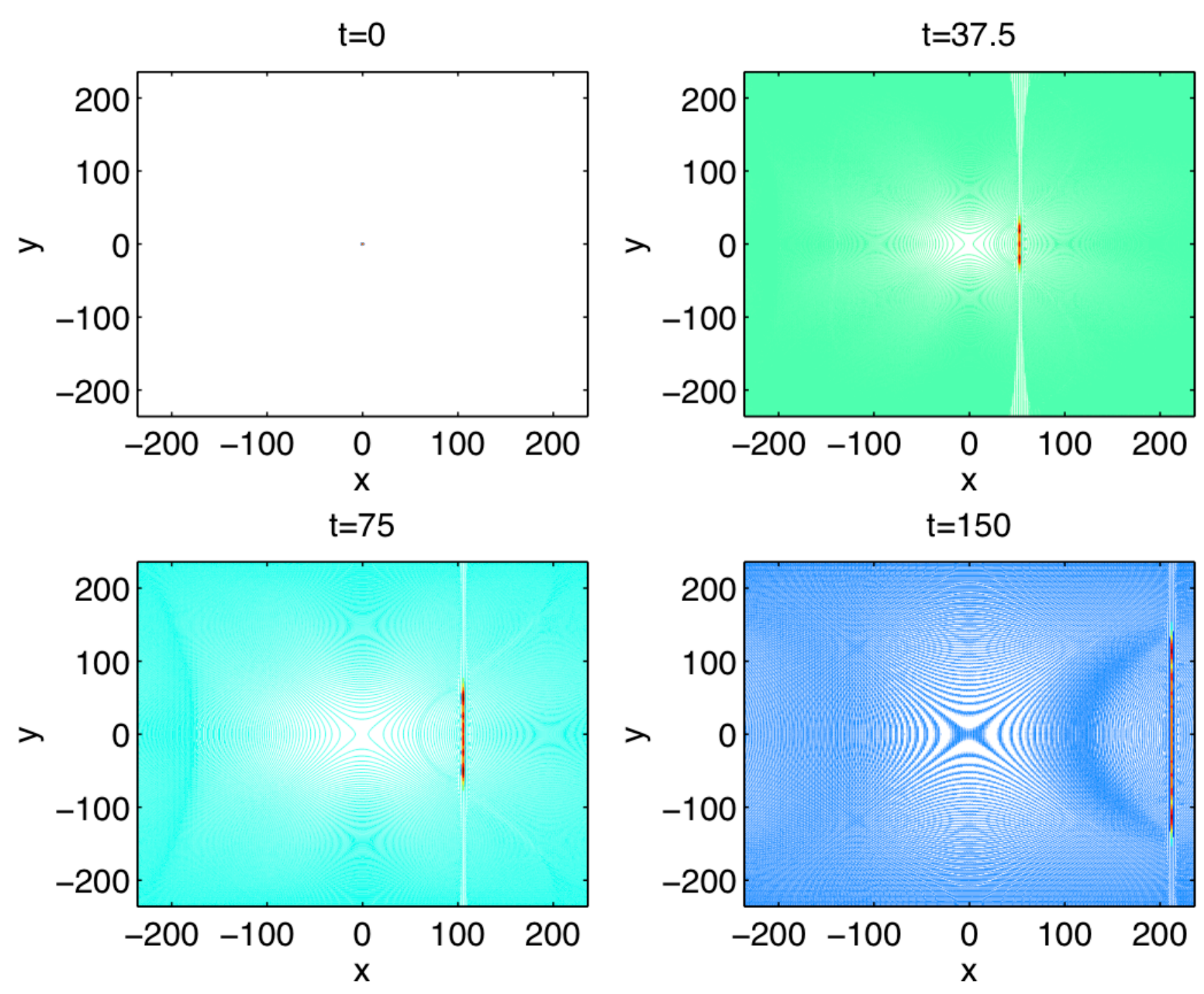} 
 \caption{Contour plots of the 
 difference between the solution of the DS II equation for an initial data of the form
 (\ref{Agauss}) with $(x_1,y_1)=(0,0)$ and the original soliton $u_I$ at several times}
 \label{ds2A01diffcontts}
\end{figure}

\begin{figure}[htb!]
\centering
\includegraphics[width=0.45\textwidth]{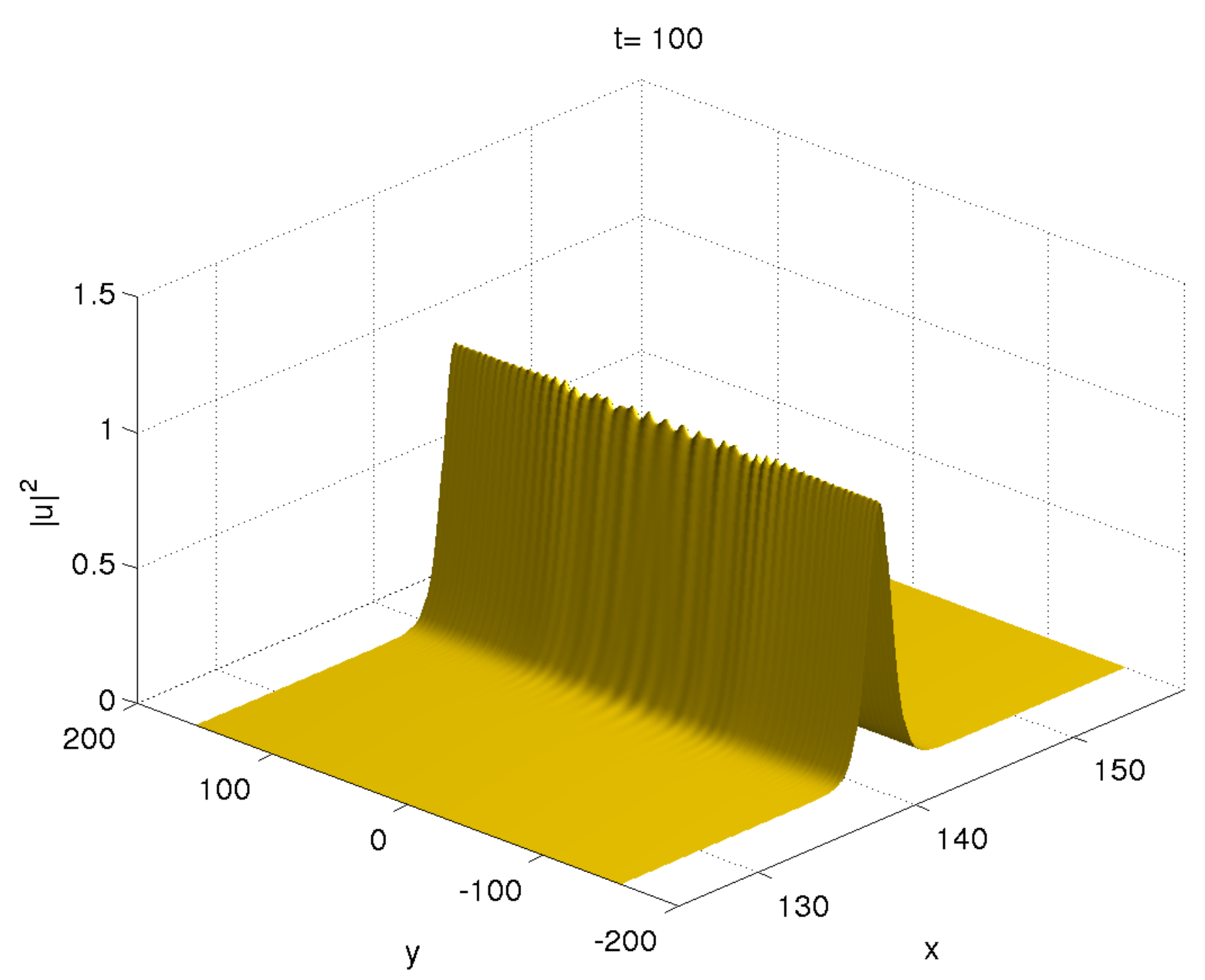} 
\includegraphics[width=0.45\textwidth]{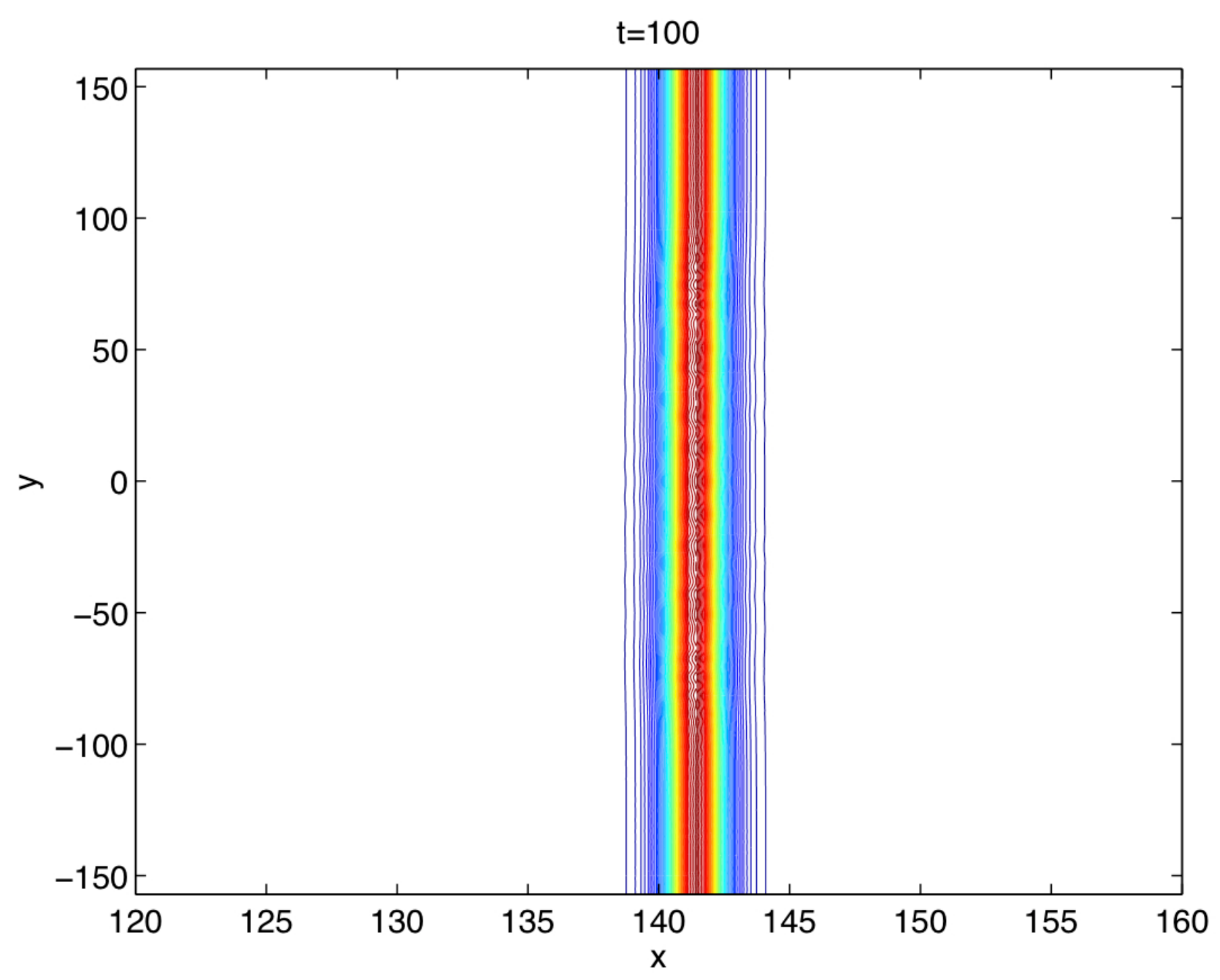} 
\label{isopg01ut10}
 \caption{Solution of the DS II equation for an initial data of the form (\ref{Agauss}) 
 with $A=0.1$, at $t=100$ on the left, and contour plot on the right.}
 \end{figure}

It is clearer when  
 the code is run for longer times on a bigger domain of computation. One finds that
the initially localized perturbation is indeed 
spread in the $y$-direction, and take finally itself the shape of a soliton, see 
see Fig. \ref{ds2A01diffcontts}, where we show
the contour plot of the difference 
between the numerical solution and the original soliton at several times.
The solution asymptotically preserves the soliton's shape, see Fig. \ref{isopg01ut10} for the situation at $t=100$
and the contour plot in the same picture.
It appears that the perturbation travels with the soliton and leads also to oscillations around it.
Notice that the speed of the soliton is not affected, indicating that no new soliton is formed here.
The propagation of the perturbation is still present for large times, so it is difficult to decide if they will finally disappear.
\\

\begin{figure}[htb!]
\centering
\includegraphics[width=0.45\textwidth]{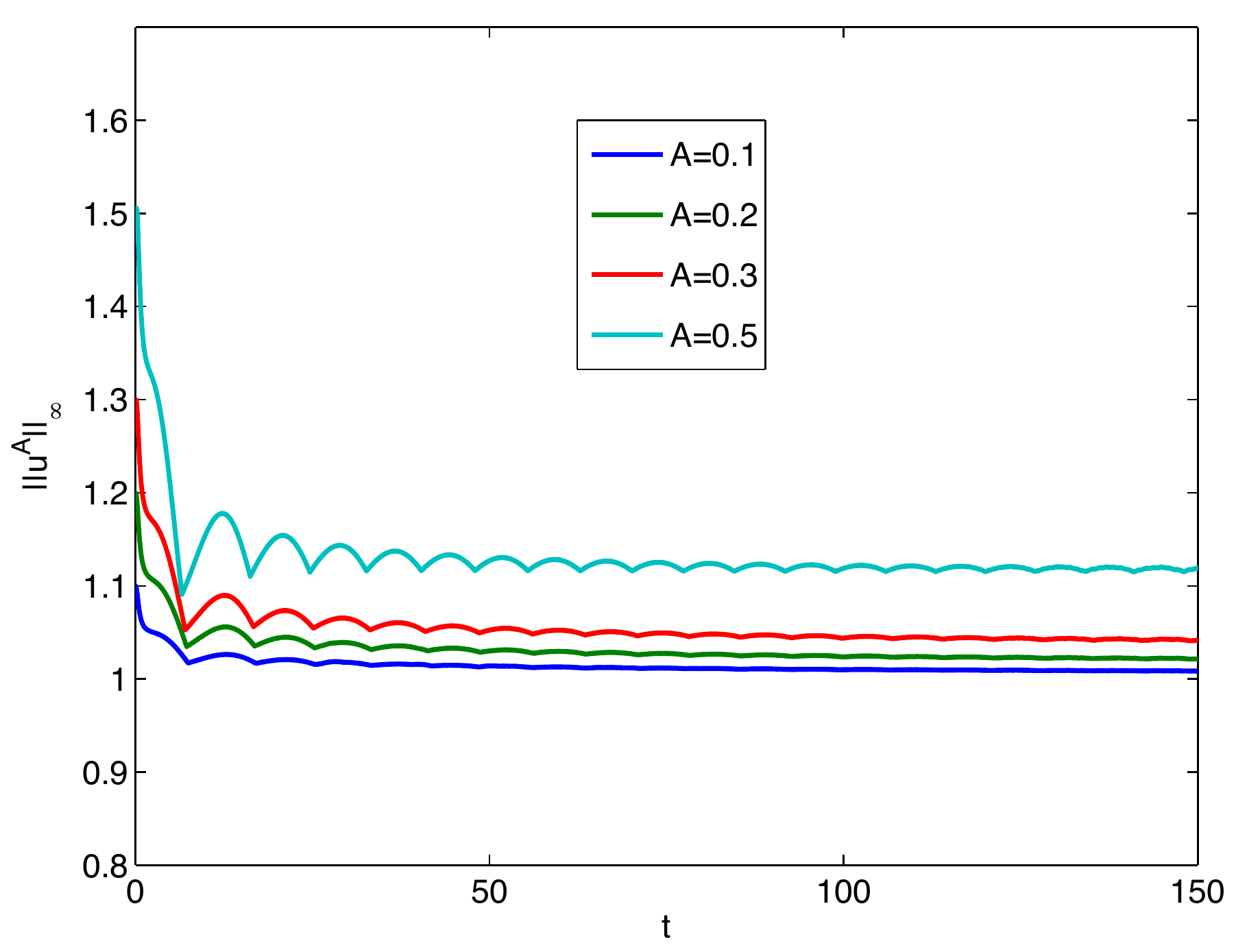} 
 \caption{Time evolution of the $L_{\infty}$-norm of the solutions $u^A$ to the focusing DS II 
 equation for an initial condition of the form (\ref{Agauss}) with $(x_1,y_1)=(0,0)$, for several values of  $A$, until $t=150$.}
 \label{isopg01linft150}
\end{figure}

The situation is similar for all values of $A$ studied, and we show the $L_{\infty}$-norm of $u^A$ in Fig. \ref{isopg01linft150},
it decreases before stabilizing for longer times in all cases.

Until now we considered perturbations centered as the same location as the original soliton, 
i.e., $x_0=0$ in (\ref{trav}) and $u_{pert}=\exp(-(x-x_1)^2 - (y-y_1)^2)$ with $(x_1,y_1)=(0,0)$.
If instead, one considers a de-centered Gaussian perturbation, i.e., either $(x_1,y_1)=(a,0)$ or
$(x_1,y_1)=(a,b)$, with $(a,b) \in [-L_x\pi, L_x\pi]\times[-L_y\pi, L_y\pi]$, one observes that the perturbation simply 
disperses as $t$ goes to infinity, and has no real impact on the soliton behavior.

\begin{figure}[htb!]
\centering
\includegraphics[width=0.7\textwidth]{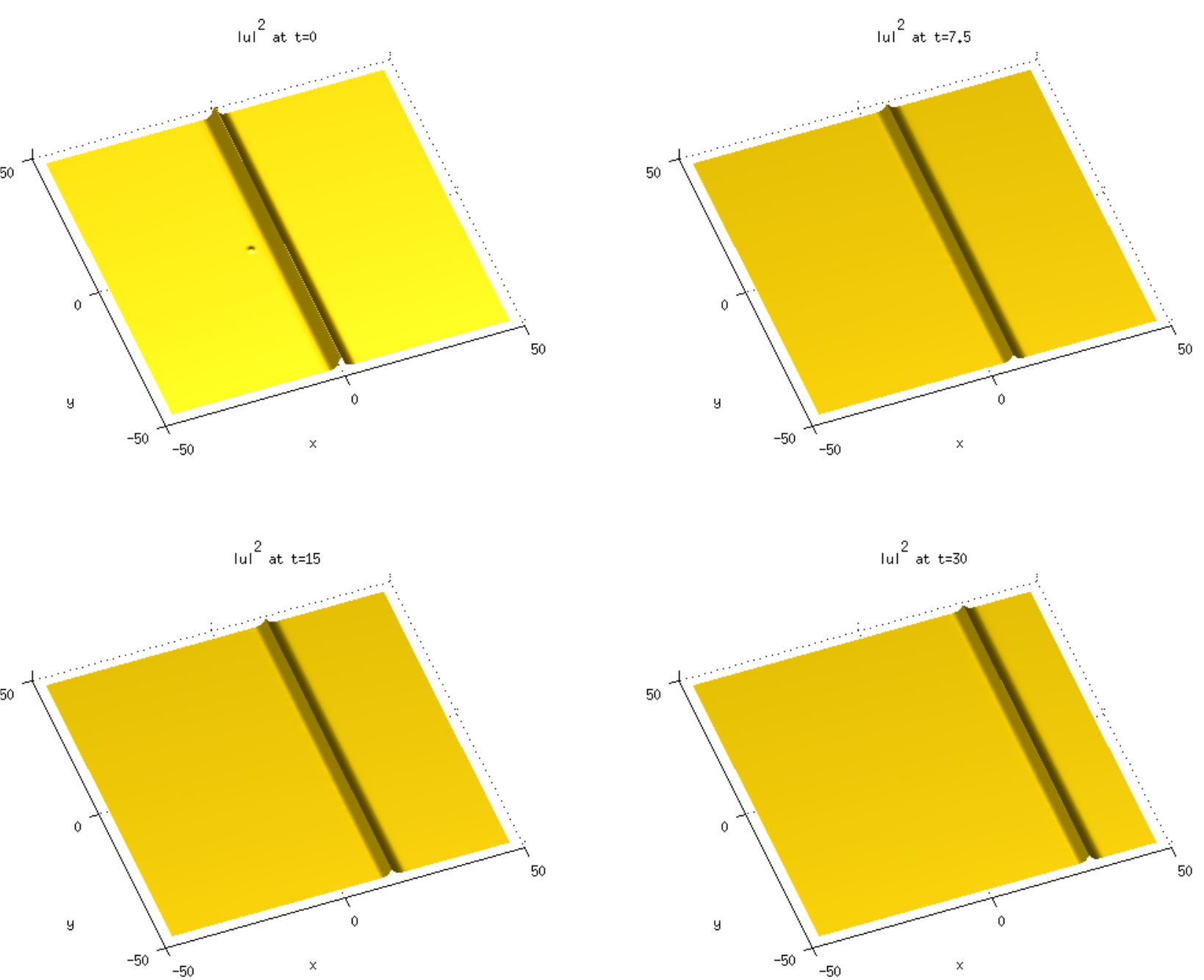} 
 \caption{Solution to the focusing DS II equation for an initial condition 
 of the form (\ref{Agauss}), with $A=0.1$ and $(x_1,y_1)=(-L_x/2,0)$, at several times.}
 \label{ds2dcxuts}
\end{figure}

\begin{figure}[htb!]
\centering
\includegraphics[width=0.7\textwidth]{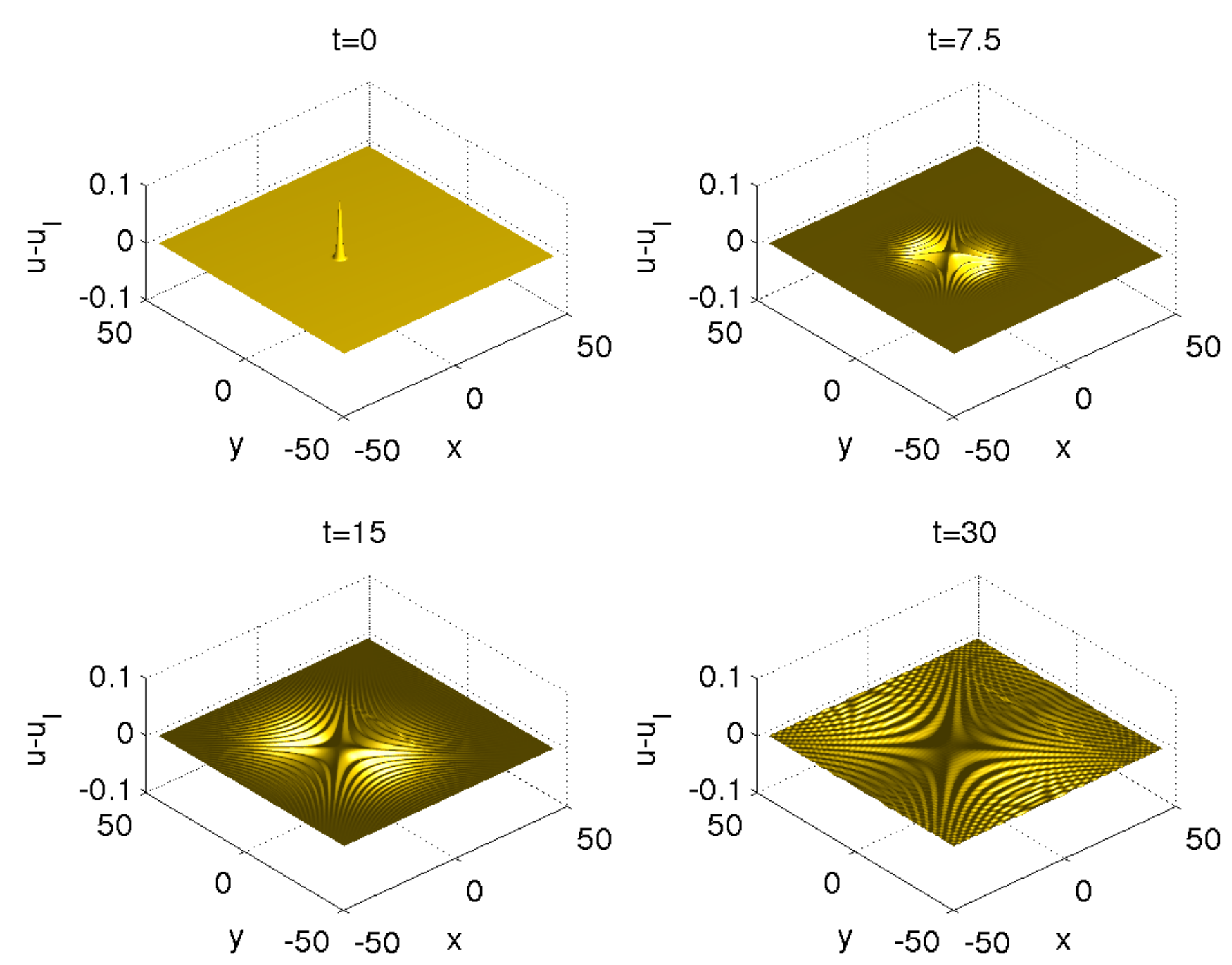} 
 \caption{Difference between the solution of the DS II equation for an initial data of the form
 (\ref{Agauss}) with $(x_1,y_1)=(L_x/2,0)$ and the original soliton $u_I$ at several times}
 \label{ds2dcxdiffts}
\end{figure}

\begin{figure}[htb!]
\centering
\includegraphics[width=0.7\textwidth]{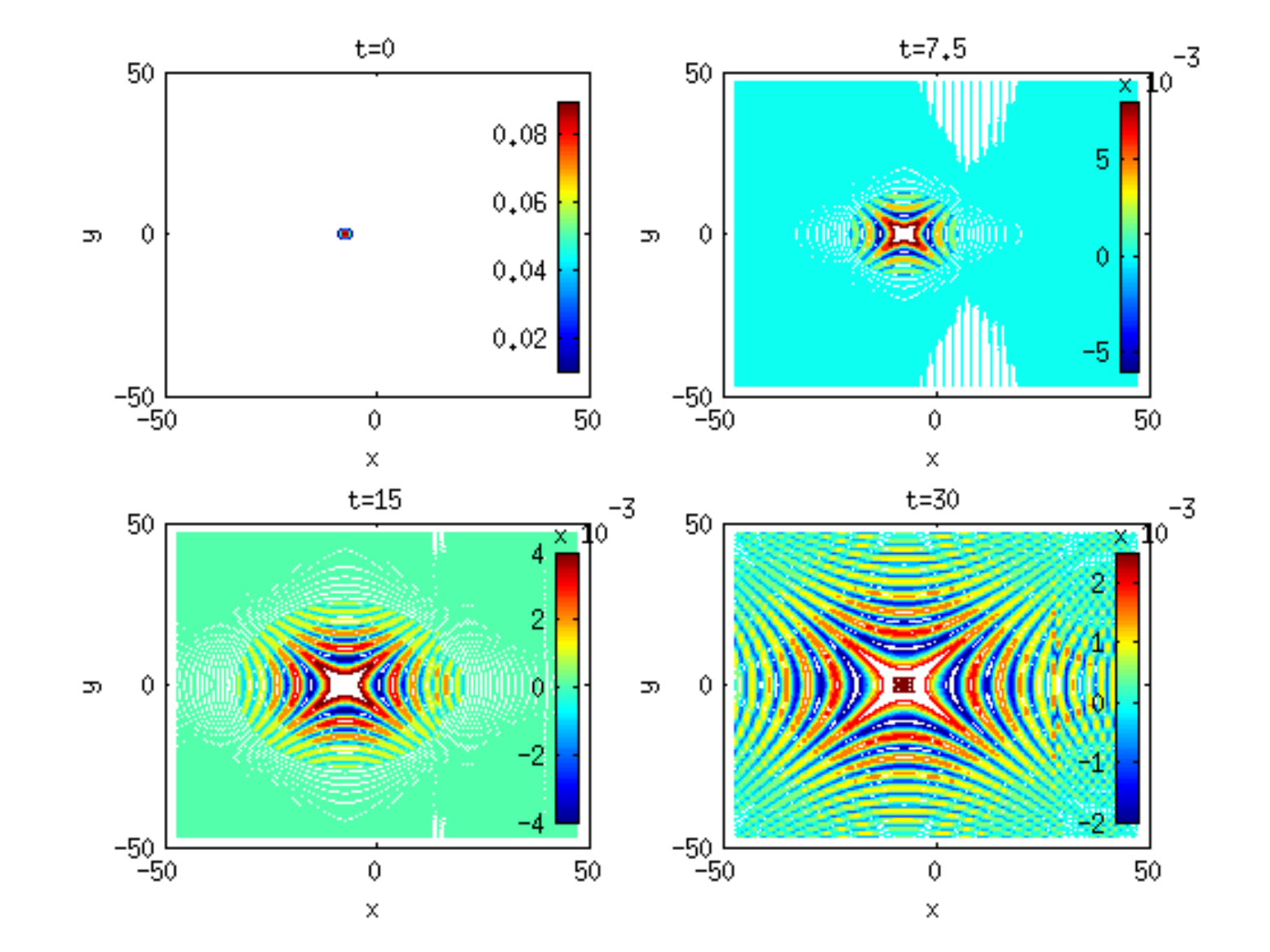} 
 \caption{Contour plots of the solution shown in Fig . \ref{ds2dcxdiffts}}
 \label{ds2dcxdiffcontts}
\end{figure}

To illustrate this, we consider an initial data of the form (\ref{Agauss}), with $(x_1,y_1)=(-L_x/2,0)$.
The computation is carried out with $2^{13}\times 2^{13}$ points for
$ x \times y \in [-15 \pi, 15 \pi] \times [-15 \pi, 15 \pi]$ and $\Delta_t=3*10^{-3}$. 
We show in Fig. \ref{ds2dcxuts} the resulting numerical solution of DS II at several times.
The solution travels with the original velocity, and its shape is preserved.
The difference between the solution and the original soliton $u_I$ 
are shown in Fig. \ref{ds2dcxdiffts}. One can see that the perturbation simply disperses away in the form of tails to infinity. 
It is even clearer in the contour plots shown in Fig. \ref{ds2dcxdiffcontts}.

\begin{figure}[htb!]
\centering
\includegraphics[width=0.45\textwidth]{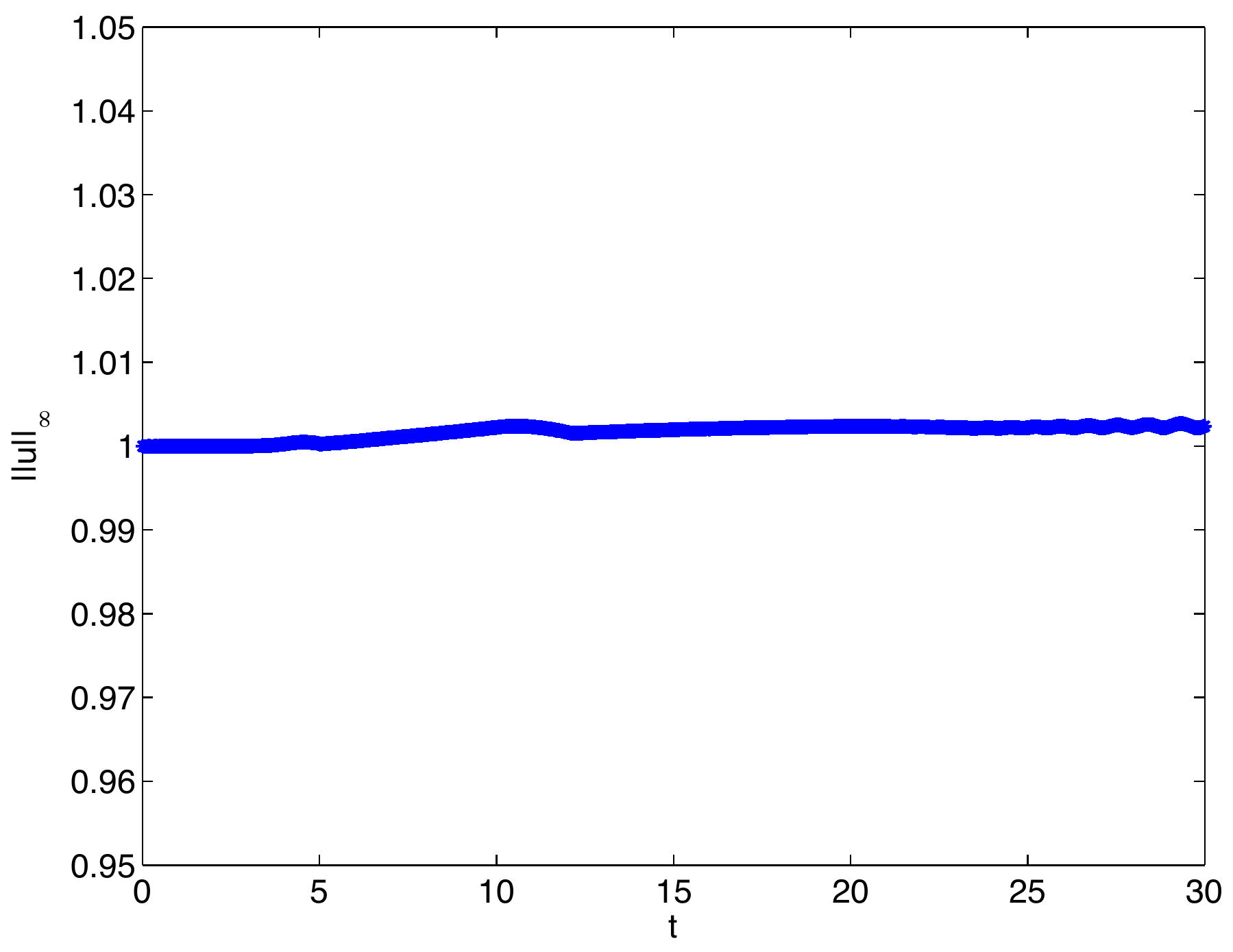} 
\includegraphics[width=0.45\textwidth]{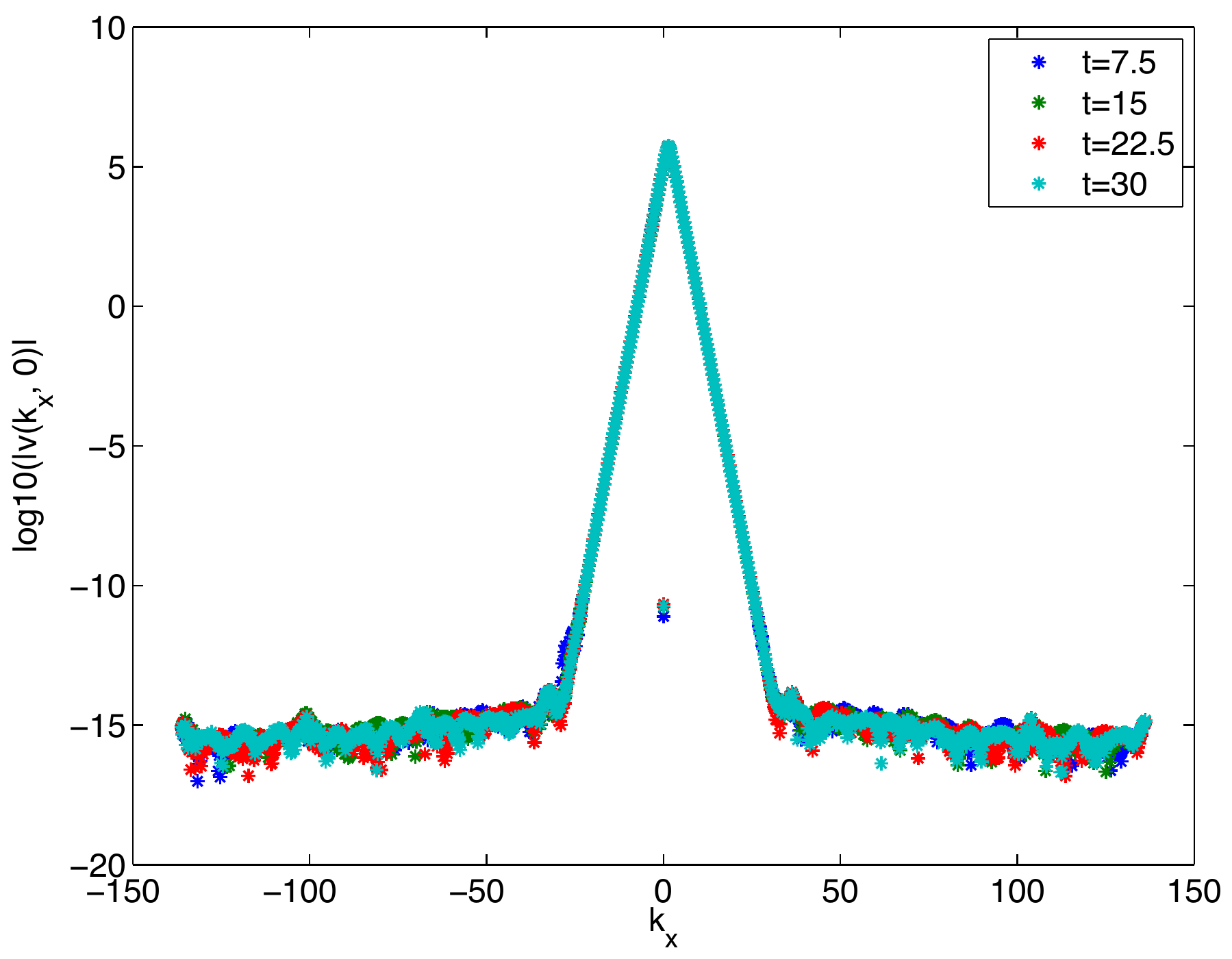} 
\caption{Time evolution of the $L_{\infty}$-norm of the solution shown in Fig. \ref{ds2dcxuts} on the left, and its Fourier coefficients at several times on the right}
 \label{ds2dcxamplvts}
\end{figure}

The $L_{\infty}$-norm of the solution is shown in Fig. \ref{ds2dcxamplvts} together  with the Fourier coefficients at several times. The latter 
reach machine precision all along the computation, indicating sufficient accuracy, and the $L_{\infty}$-norm of $u$ stays almost constant.
The soliton appears to be unaffected by the perturbation which appears to smear out in the background of the soliton.
The same behavior was observed for initial data of the form (\ref{Agauss}) with $y_1$ non equal to $0$.
\\

For localized perturbations, we found that the perturbed isolated soliton 
is unstable under the flow of the 2d hyperbolic NLS equation, and that this instability occurs via dispersion.
The study of (\ref{trav}) under the flow of DS II however indicates that numerical solutions issued from 
Gaussian perturbations travel with unchanged speed, and
 have a profile very close to the original one, in particular the $L_{\infty}$-norm of $u$  
  becomes constant as $t \to \infty$. The perturbed solution recovers the features of a soliton for all cases studied there, including 
  de-centered Gaussian perturbations.
The isolated soliton appears in this sense orbitrarily stable under 
Gaussian perturbations of the form \ref{Agauss} with $0<A<1$.
 This a noticeable difference with the results for the 2d (both hyperbolic and elliptic) NLS equations, in which the isolated soliton is unstable.

\section{Periodic deformations} 
We now consider periodic deformations of the isolated soliton (\ref{trav}). 
More precisely, we consider initial data of the form
\begin{equation}
u(x,y,0) = u_{I}(x,0)(1 + \epsilon \cos(\gamma y/L_y)), \epsilon \ll 1, \gamma \in \mathbb{R}
\label{perio}
\end{equation} 
for the four models we are considering in this paper.
The analytical results in \cite{RT09} provide also here the instability of (\ref{trav}) for the 2d elliptic NLS equation under this kind of perturbations. 
We will see that in this case, this instability occurs via a blow up in multiple spatial points. 
For the NLS equation, such behavior has been studied by Merle \cite{ Merlekblow}, and we find in addition that 
such phenomena occur also in the case of DS$^{++}$, for which however no theoretical study is available in this context.
The isolated soliton (\ref{trav}) appears to be unstable under periodic perturbations of the form (\ref{perio}) for all cases here. 

\subsection{Elliptic NLS Equations}

\begin{figure}[htb!]
\centering
\includegraphics[width=0.32\textwidth]{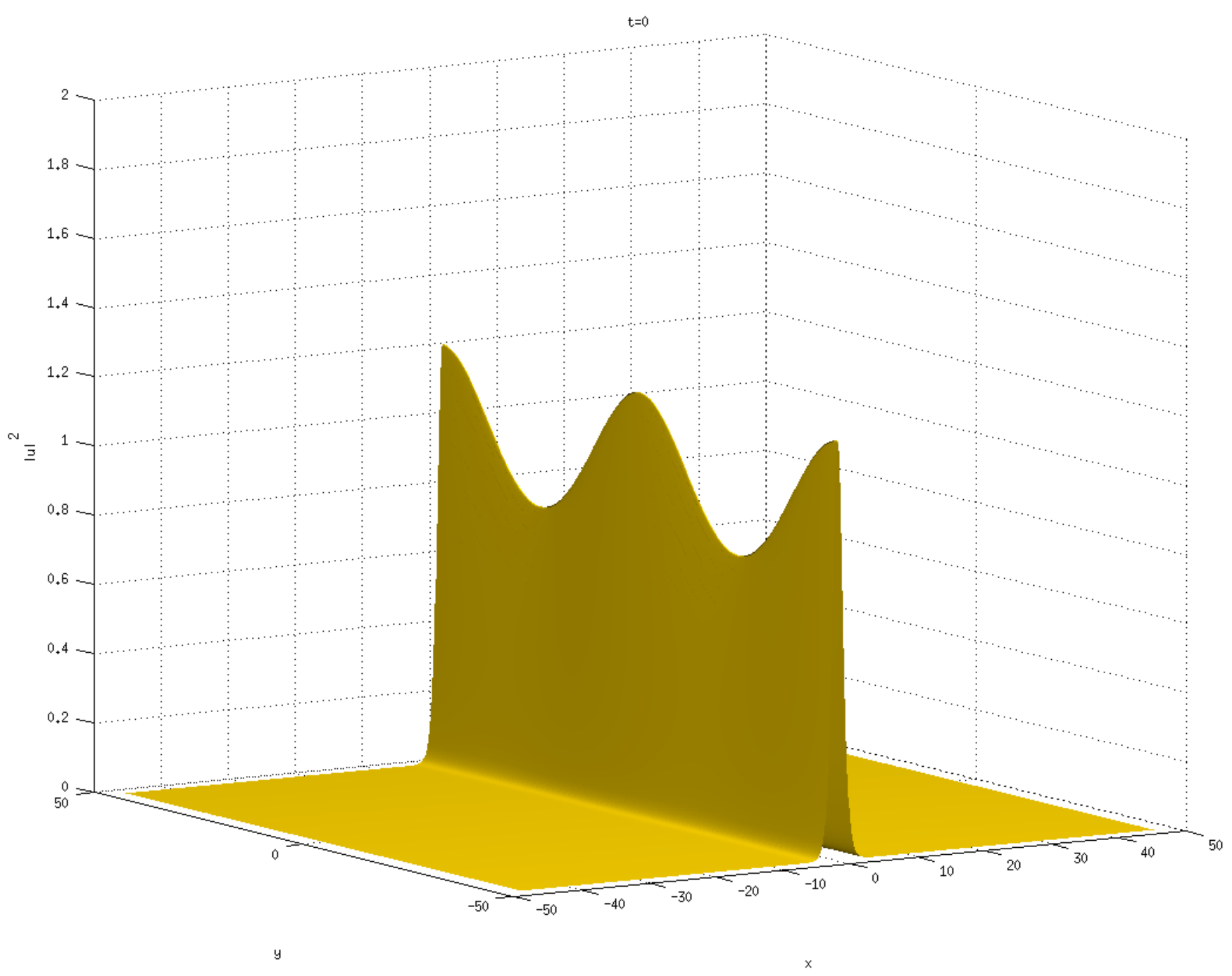} 
\includegraphics[width=0.32\textwidth]{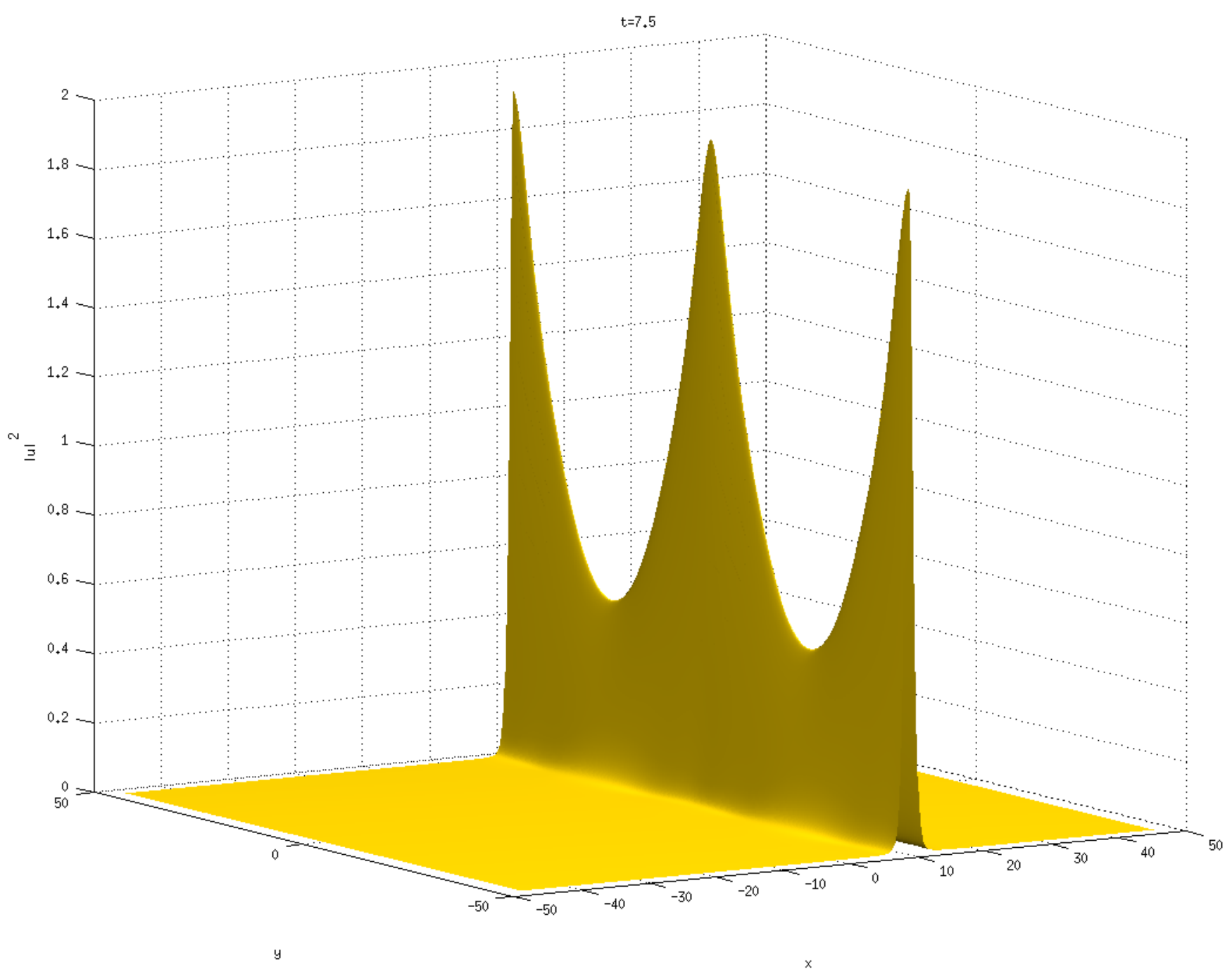} 
\includegraphics[width=0.32\textwidth]{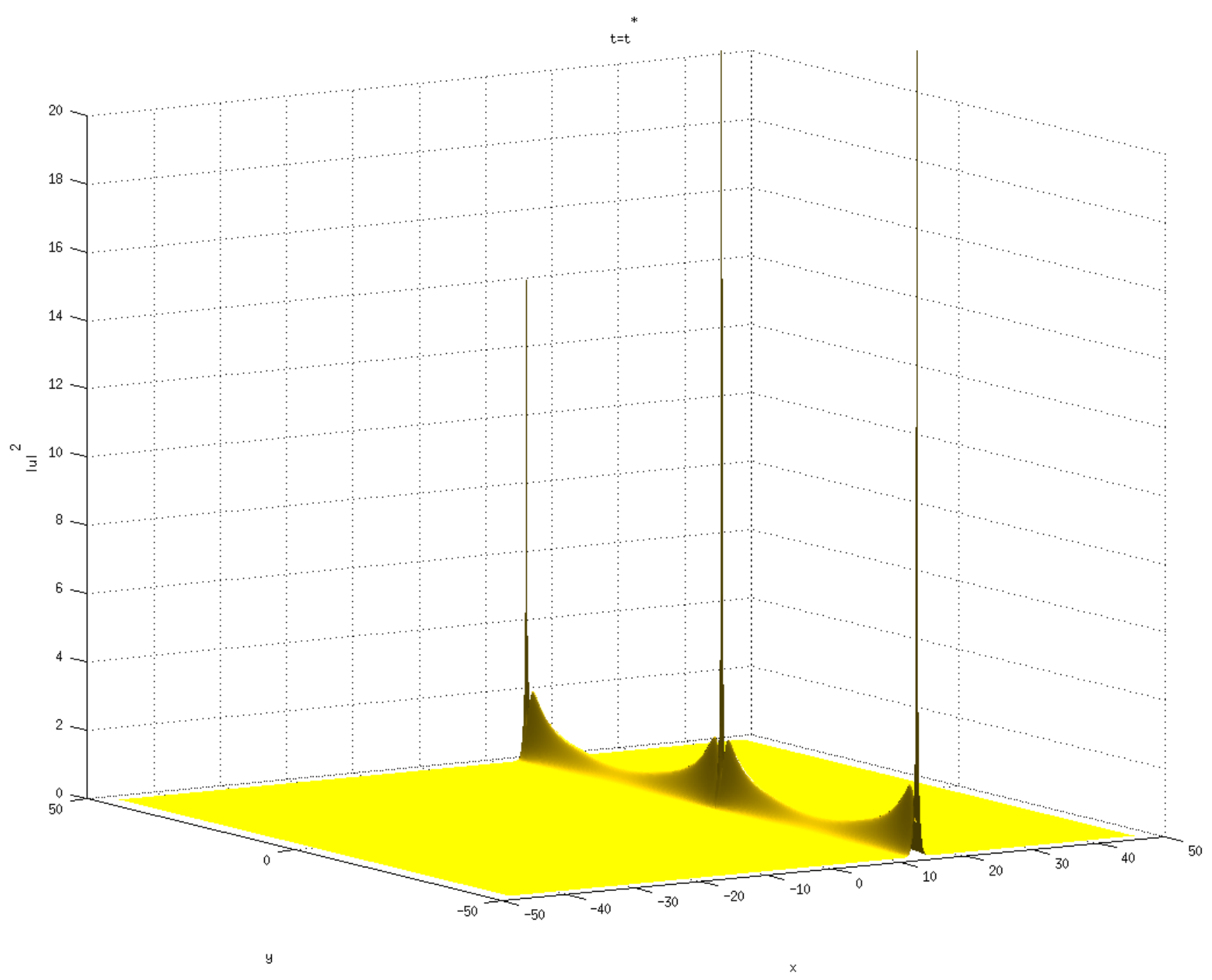} 
\caption{Solution to NLS$^+$ for an initial 
condition of the form (\ref{perio}), with $\epsilon=0.1$, $b=2$ at several times}
\label{nlseperiouts}
\end{figure}

\begin{figure}[htb!]
\centering
\includegraphics[width=0.45\textwidth]{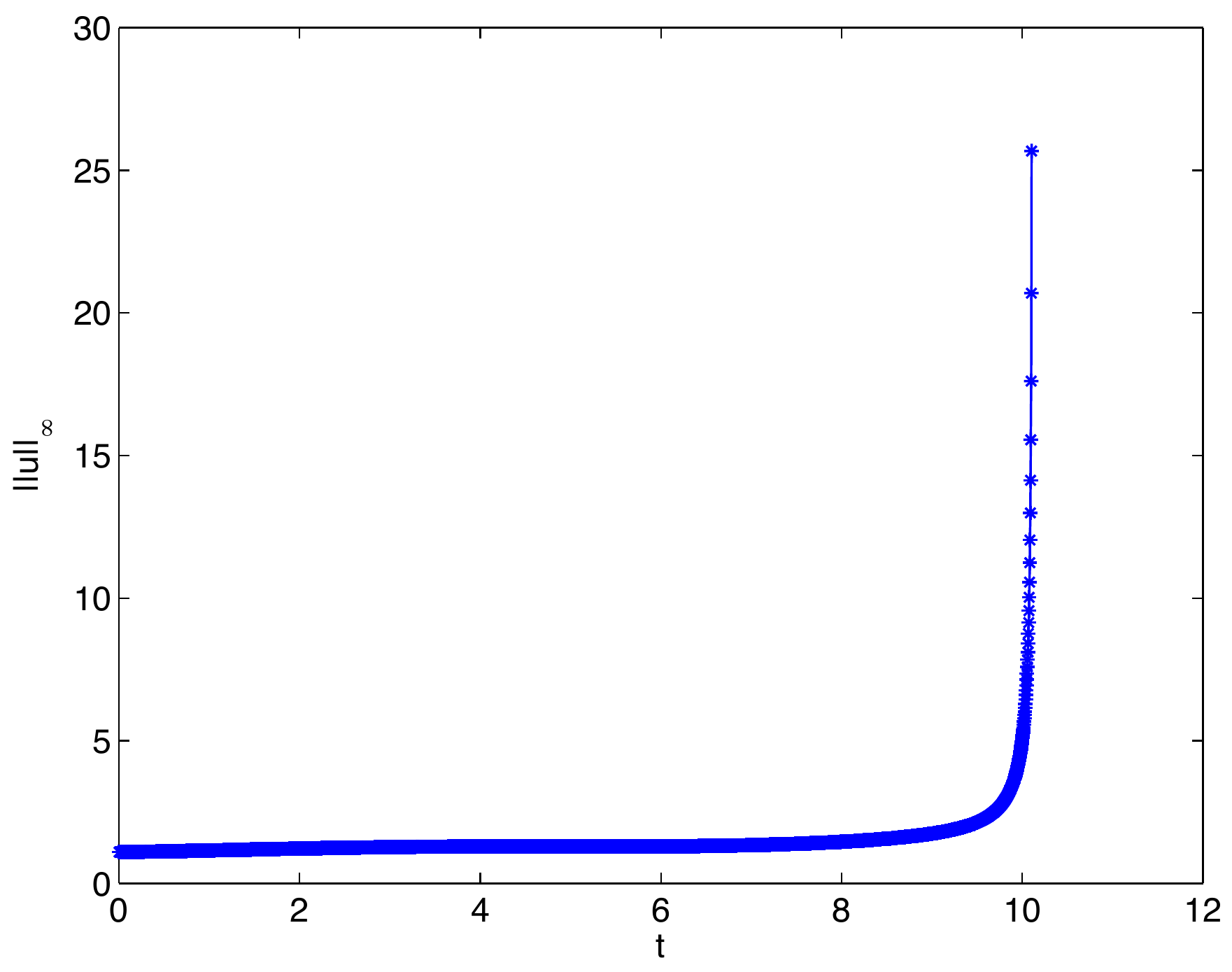} 
\includegraphics[width=0.45\textwidth]{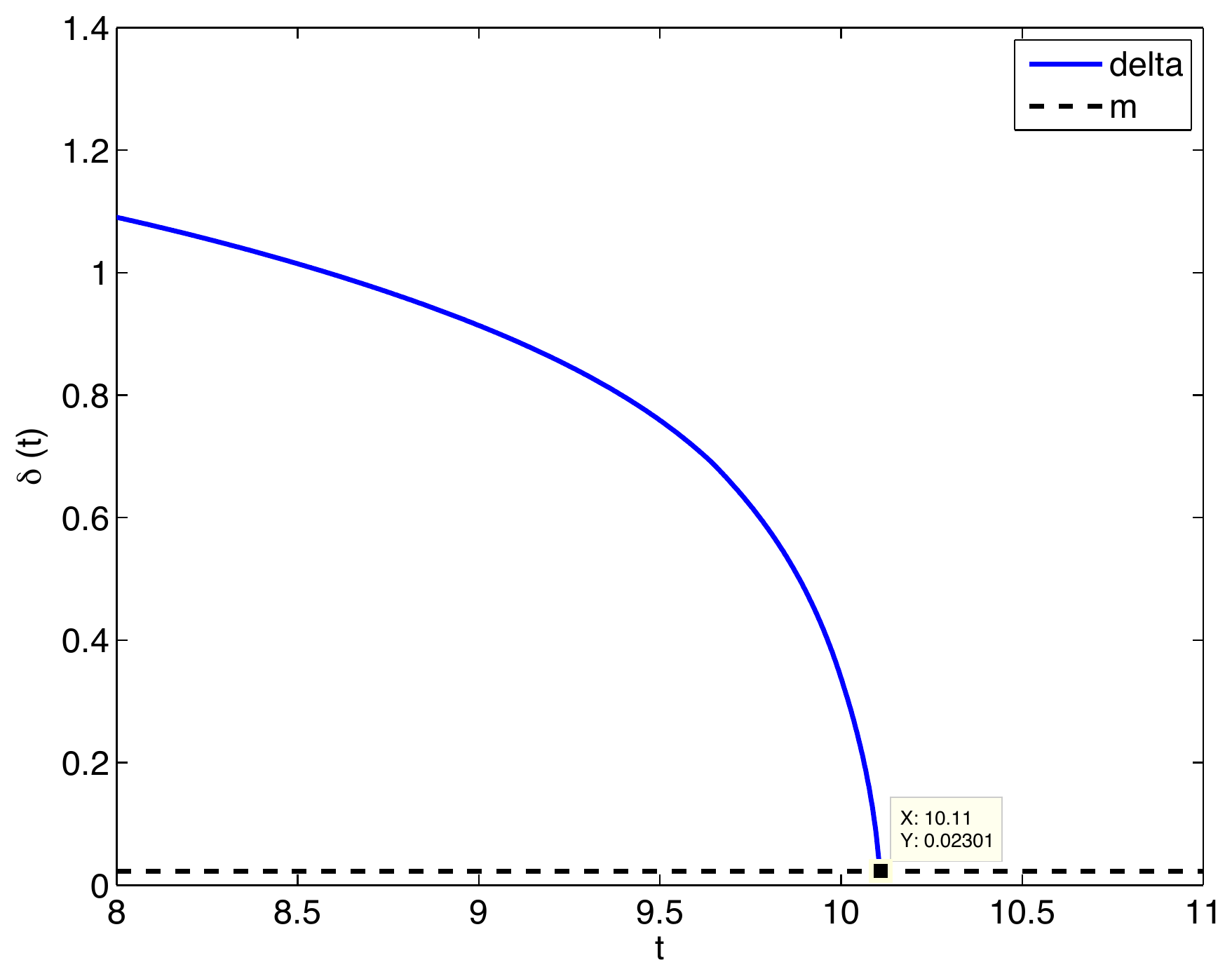} 
\caption{Time evolution of the $L_{\infty}$-norm of the solution shown in Fig. \ref{nlseperiouts} 
on the left,
and the corresponding fitting parameter $\delta(t)$ in (\ref{abd}) on the right.}
\label{nlseperioampldel}
\end{figure}

\begin{figure}[htb!]
\centering
\includegraphics[width=0.45\textwidth]{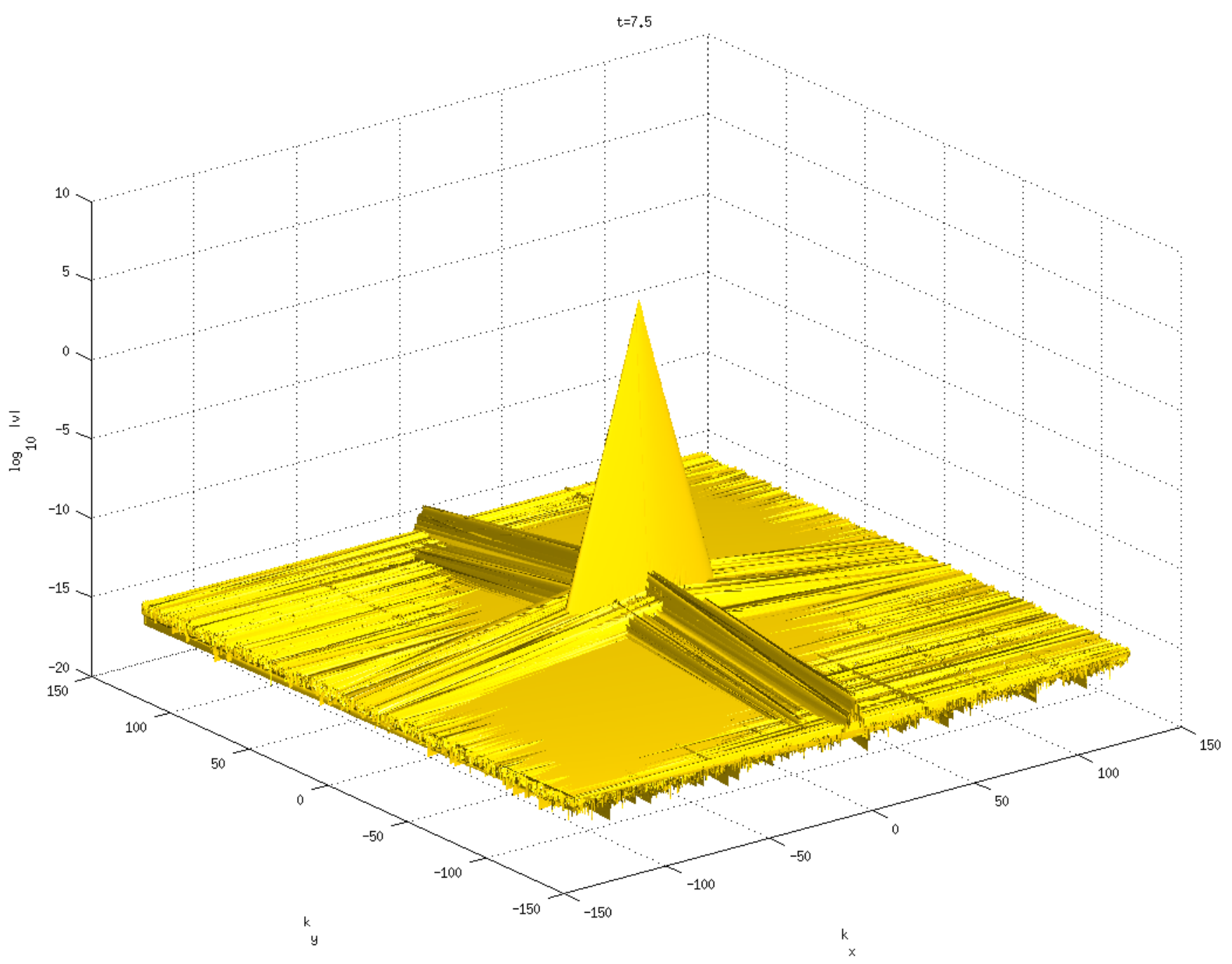} 
\includegraphics[width=0.45\textwidth]{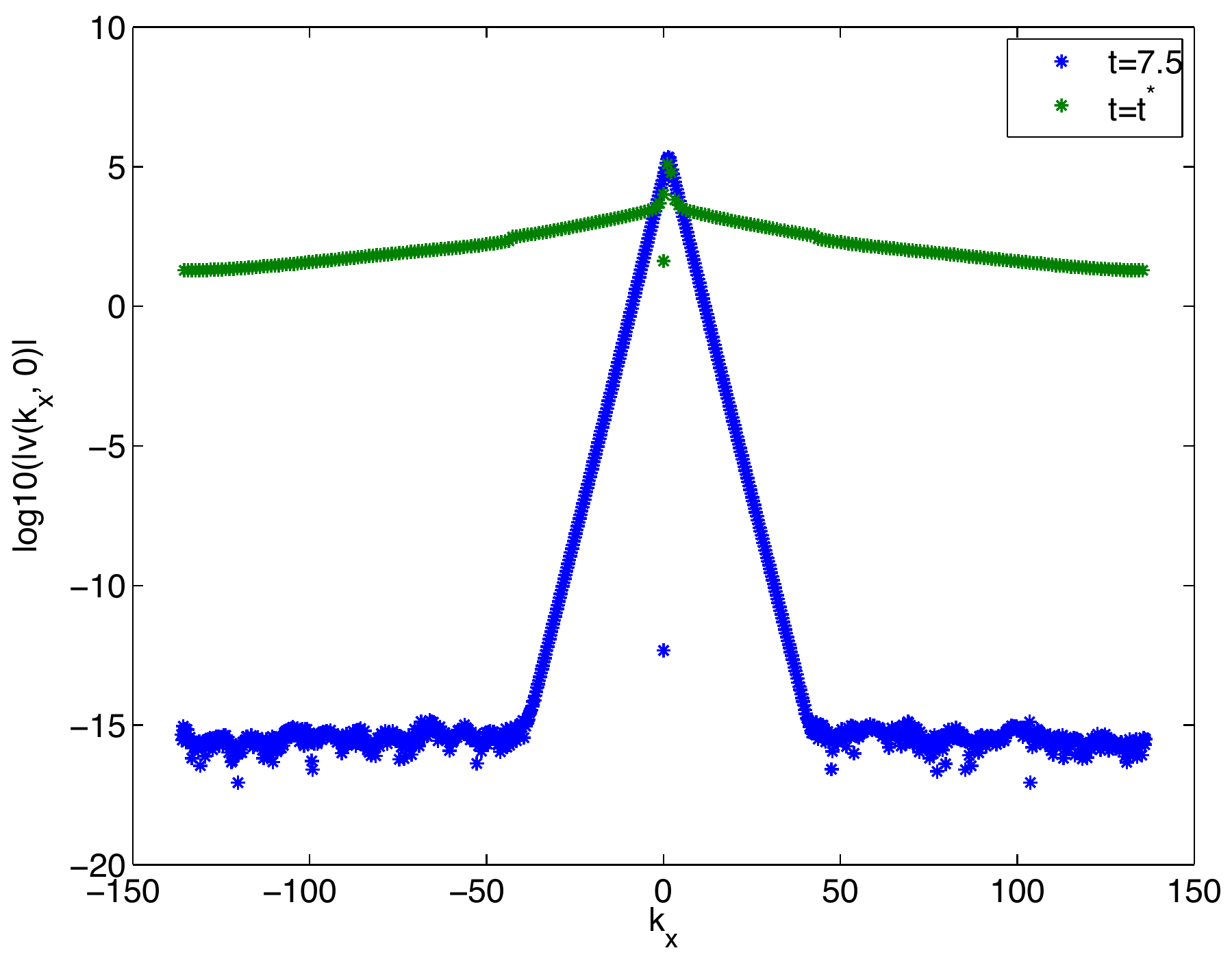} 
\caption{Fourier coefficients of the solution shown in Fig. \ref{nlseperiouts}}
\label{nlseperiovts}
\end{figure}

We first consider an initial data of the form (\ref{perio}) with $\epsilon=0.1$ and $b=2$ for the 2d elliptic NLS equation (NLS$^{+}$).
The computations are carried out with $2^{13} \times 2^{13}$ points for 
$x \times y \in  [-15\pi, 15\pi]\times[-15\pi, 15\pi] $ and $\Delta_t=6*10^{-4}$.
As in the previous section, we study the asymptotics of the Fourier coefficients all along the computation, 
and choose the following range of wavenumbers for the fitting, $10<k_x<2 \max(k_x)/3$.
For the situation here, we found that a multiple blow up occurs at $t^*=10.107$, where $\delta$ as in (\ref{abd}) vanishes. 
We show in Fig. \ref{nlseperiouts}  the numerical 
solution at several times, including $t^*$.
The time evolution of $\delta$ as in (\ref{abd}) and of the $L_{\infty}$-norm of $u$ are shown in Fig. \ref{nlseperioampldel}.
At $t^*$, $\|u \|_{\infty} \sim 25$.
The fitting error is of the order of $p\sim 0.13$ at $t=t^*$, and $B$ in \ref{abd} reaches a value of 
$B(t^*)=0.4852$ indicating clearly a $L_{\infty}$ blow up. Notice that, in this configuration, (multiple blow up points) 
it was not clear how well the parameter $B$ could reach such value. Some other tests in this context are needed to be able to decide 
if the asymptotics of the Fourier coefficients can identify or not with such good precision the kind of the singularity 
via the parameter $B$.
Notice that even for one spatial point blow up, the parameter B is not always reliable, as pointed out in \cite{DSdDS}.
The numerically computed energy is of the order of $\Delta_E \sim 10^{-14}$, indicating that the system is 
well resolved until the singularity formation.
In this case, the blow up of the solution occurs in multiple spatial points, located on 
the same $x$ locations, that we can identify via the formula (\ref{phi}). One finds here that $\alpha(t^*)=14.2739$. 
The $y$-locations correspond to the maxima of the solution, i.e. the values of $y$ such 
that $\sin(\gamma y/L_y)=0$, i.e., $\gamma y/L_y=0 \,\, \mbox{modulo} \,\, \pi$. For the example here, 
we thus get three spatial blow up points.
The Fourier coefficients are shown in Fig. \ref{nlseperiovts} at several times.

\begin{figure}[htb!]
\centering
\includegraphics[width=0.45\textwidth]{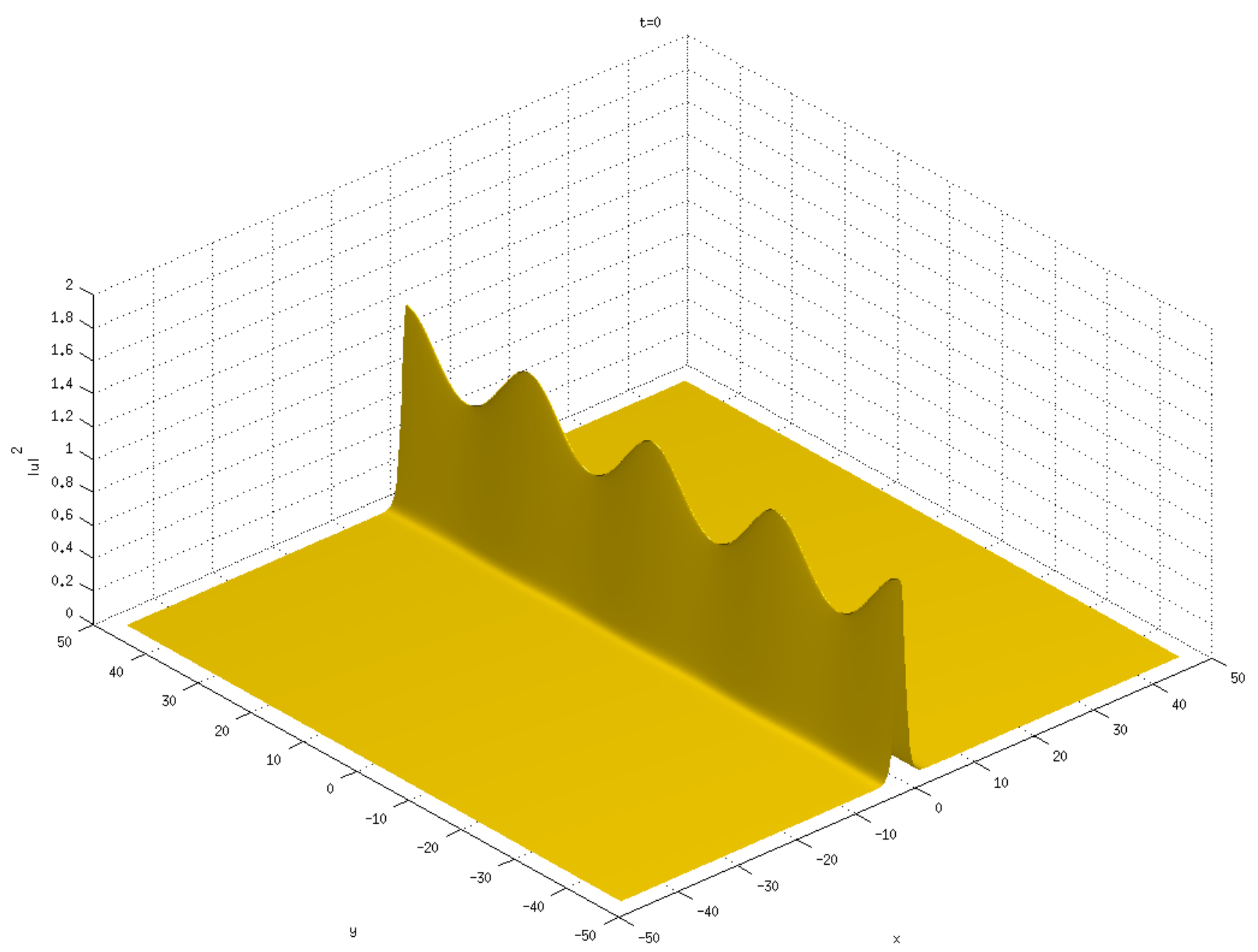} 
\includegraphics[width=0.45\textwidth]{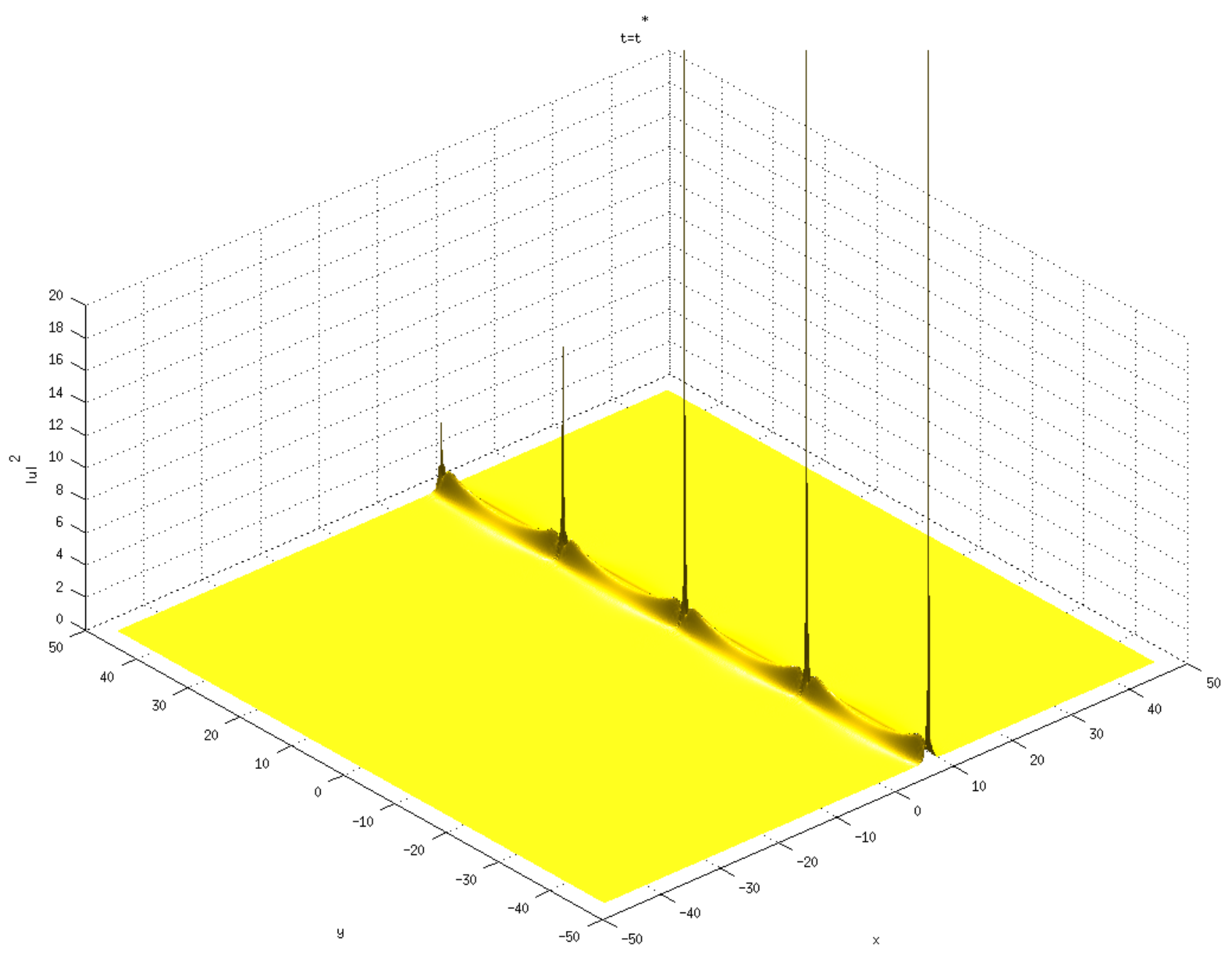} 
\caption{Solution to NLS$^+$ for an initial 
condition of the form (\ref{perio}), with $\epsilon=0.1$, $b=4$ at several times}
\label{nlseperio2uts}
\end{figure}

If one varies $\gamma$, i.e., the period, one observes that the number of blow up points corresponds to the number of maximum of the solution, for example for $\gamma=4$, one gets 5 blow up space locations, see Fig .\ref{nlseperio2uts}.  
In this case, the blow up occurs at $t^*=5.871$, where $B=0.5435$ and $p\sim 0.04$, with $\|u \|_{\infty} \sim 32$, $\Delta_E \sim 10^{-14}$ and $\alpha(t^*)=8.2983$.

\begin{figure}[htb!]
\centering
\includegraphics[width=0.32\textwidth]{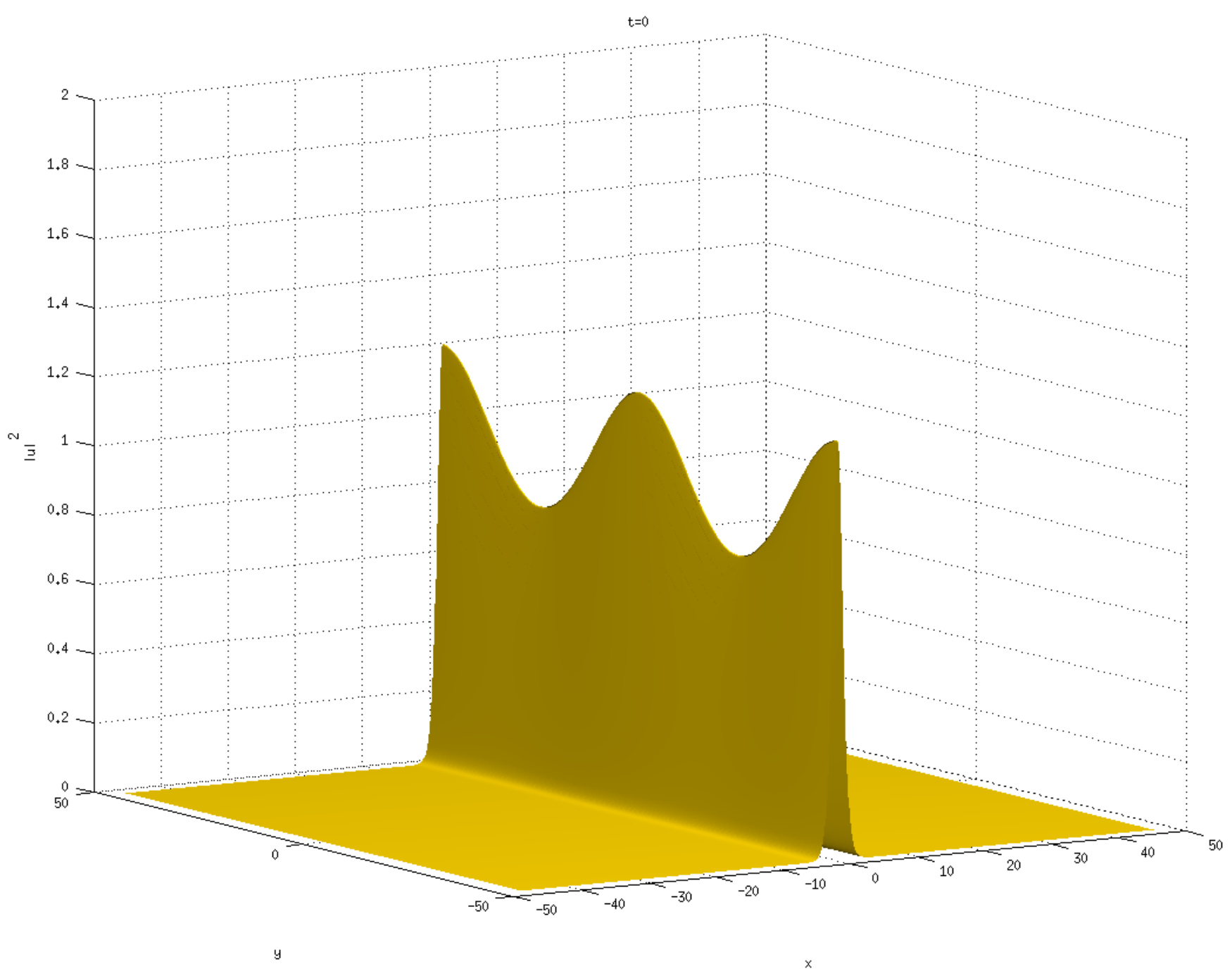} 
\includegraphics[width=0.32\textwidth]{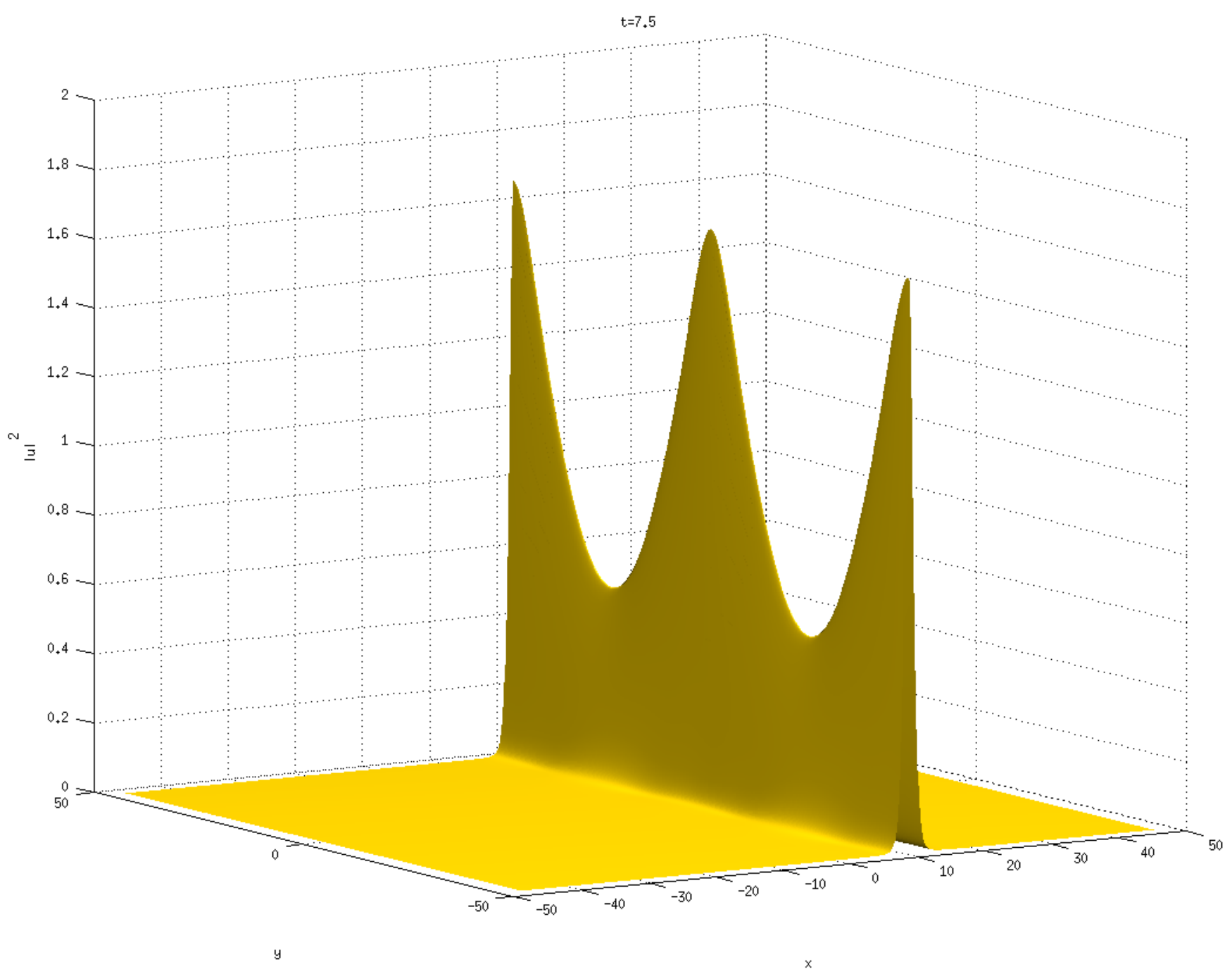} 
\includegraphics[width=0.32\textwidth]{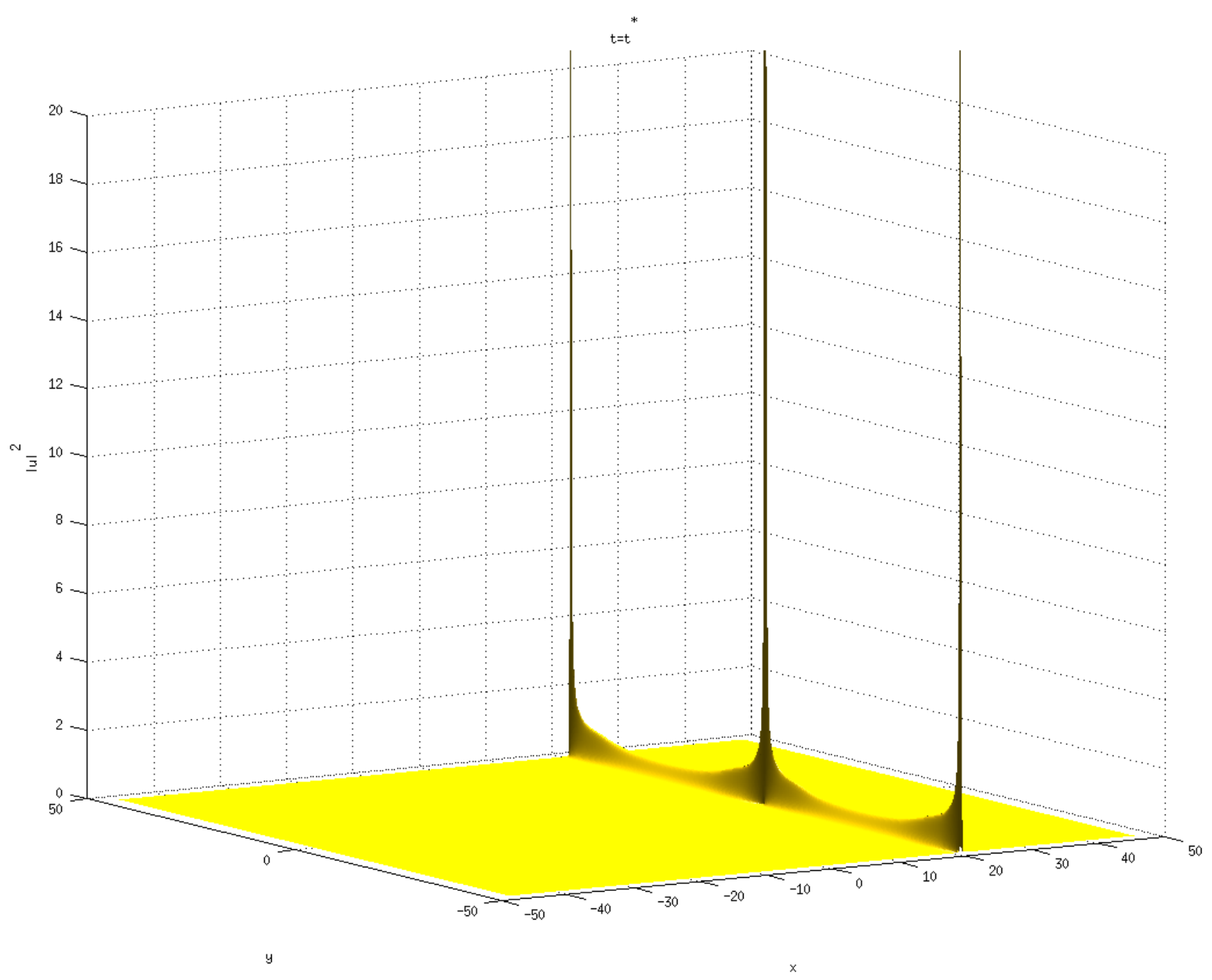} 
\caption{Solution to DS$^++$ for an initial 
condition of the form (\ref{perio}), with $\epsilon=0.1$, $b=2$ at several times}
\label{dseeperiouts}
\end{figure}

\begin{figure}[htb!]
\centering
\includegraphics[width=0.45\textwidth]{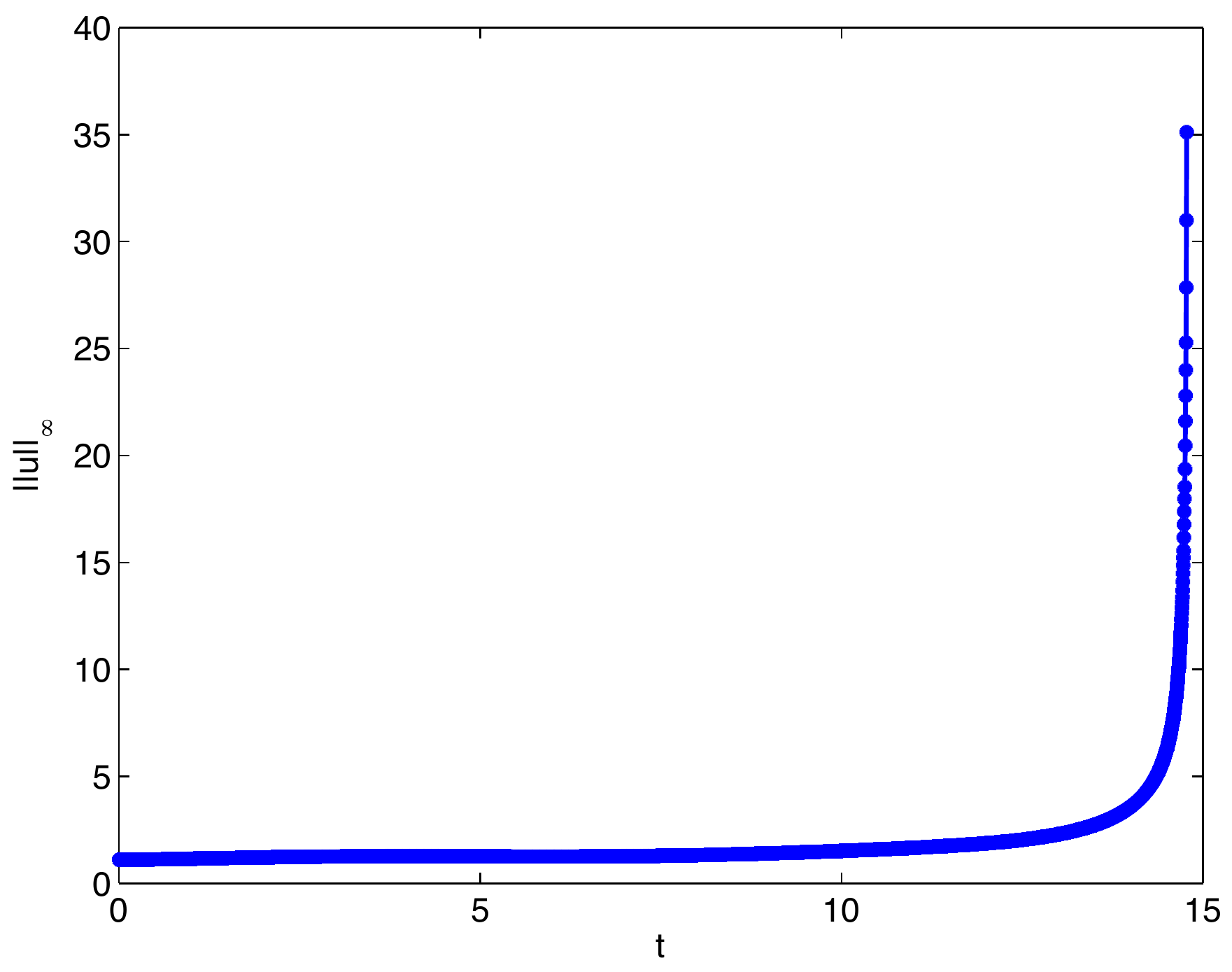} 
\includegraphics[width=0.45\textwidth]{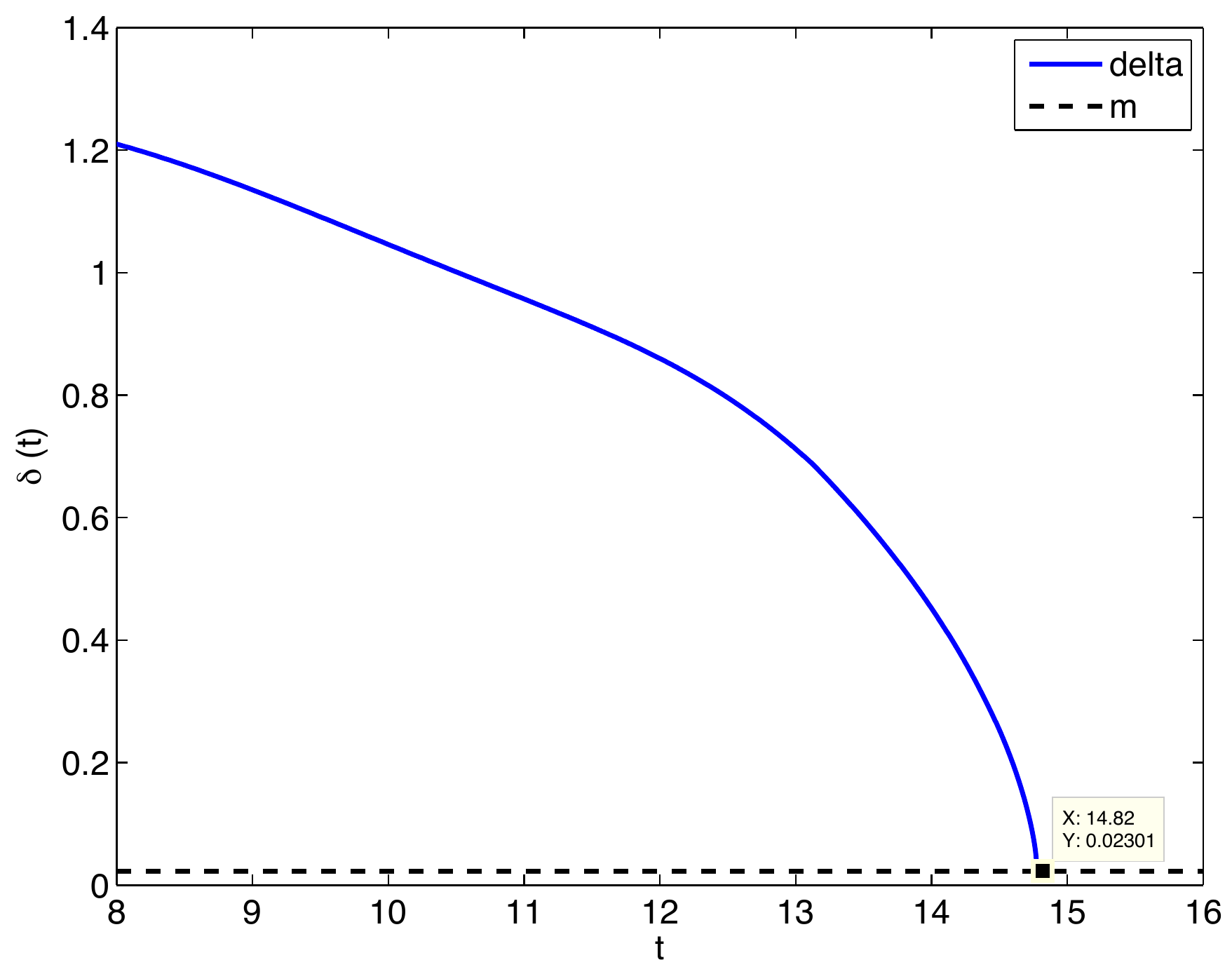} 
\caption{Time evolution of the $L_{\infty}$-norm of the solution shown in Fig. \ref{dseeperiouts} 
on the left,
and the corresponding fitting parameter $\delta(t)$ in (\ref{abd}) on the 
right.}
\label{dseeperioampldelv}
\end{figure}

The same phenomena are observed if one considers an initial data of the form (\ref{perio}) for the DS$^{++}$ equation.
We show in Fig. \ref{dseeperiouts} the solution of DS$^{++}$ for an initial data of the form (\ref{perio}) with $\gamma=2$ and 
$\epsilon=0.1$ at several times. 
In this case, the vanishing of $\delta$ (as in (\ref{abd})) occurs at $t^*=14.778$, with $\|u \|_{\infty} \sim 35$, see Fig. \ref{dseeperioampldelv}, 
where we show the time evolution of these two quantities.
One finds also $B=0.5036$ and $p\sim 0.3$, $\Delta_E \sim 10^{-14}$ and $\alpha(t^*)=20.8731$.
We recover in this case also a value of $B$ in \ref{abd} really convincing. 
The situation here is really similar to the previous case, and we see 
that the instability of 
the isolated soliton \ref{trav} under periodic perturbations for elliptic models studied there occurs via a multiple blow up point.
Recall once again that this instability was known for the NLS$^+$ but not yet for DS$^{++}$, and that the features of this instability 
were up to now not known. The phenomena of multiple blowing-up solutions in DS$^{++}$ has been also not studied (as far as we know), and it would be interesting to see if the theory as by Merle in \cite{Merlekblow} can be applied there.

\subsection{Hyperbolic NLS Equations}

We now perform the same study, but for the hyperbolic NLS equations, NLS$^-$ and the DS II equation.
The computations are carried out with $2^{13} \times 2^{13}$ points for 
$x \times y \in  [-15\pi, 15\pi]\times[-15\pi, 15\pi] $ and $\Delta_t=2*10^{-3}$.

\begin{figure}[htb!]
\centering
\includegraphics[width=0.55\textwidth]{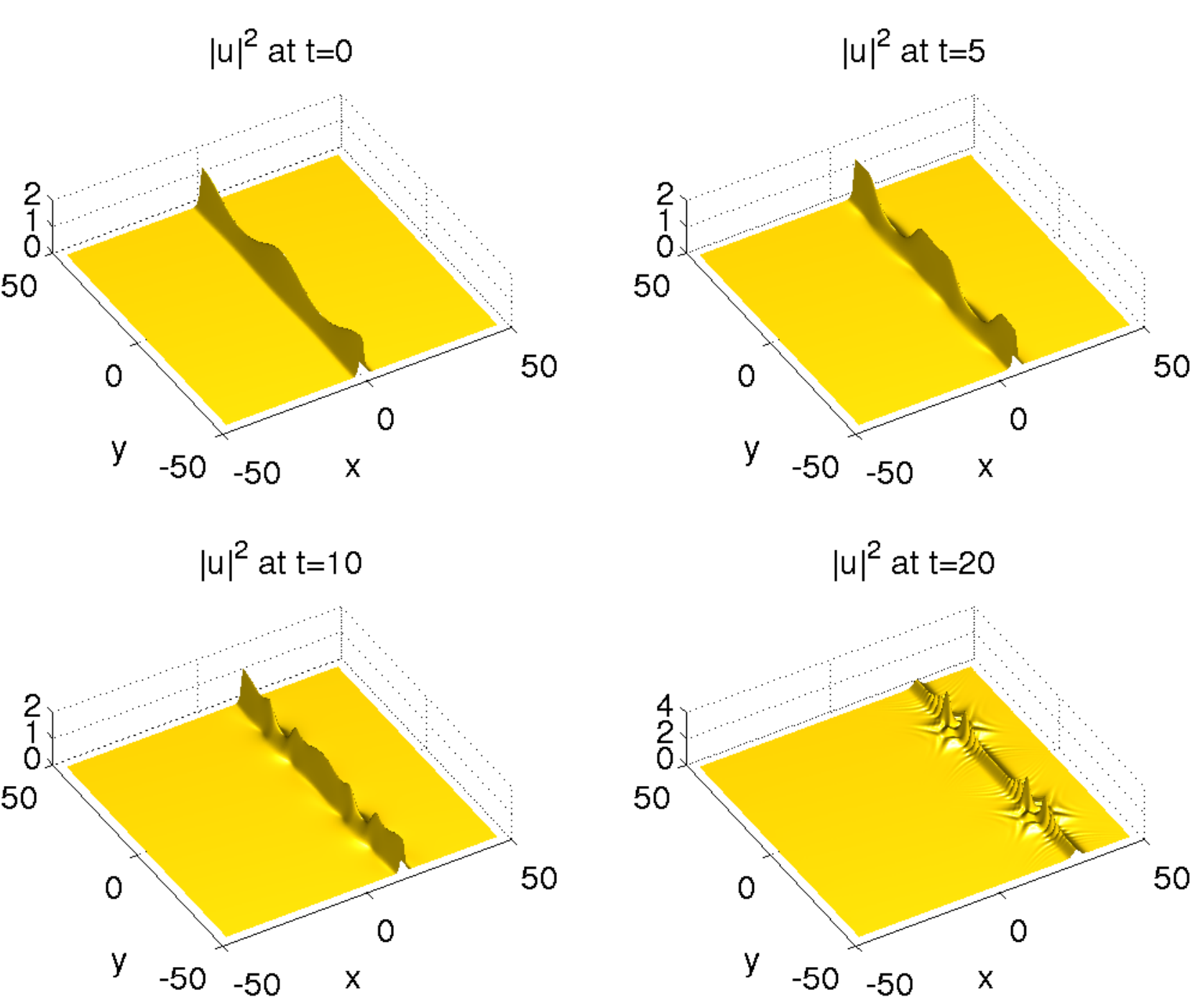}\includegraphics[width=0.55\textwidth]{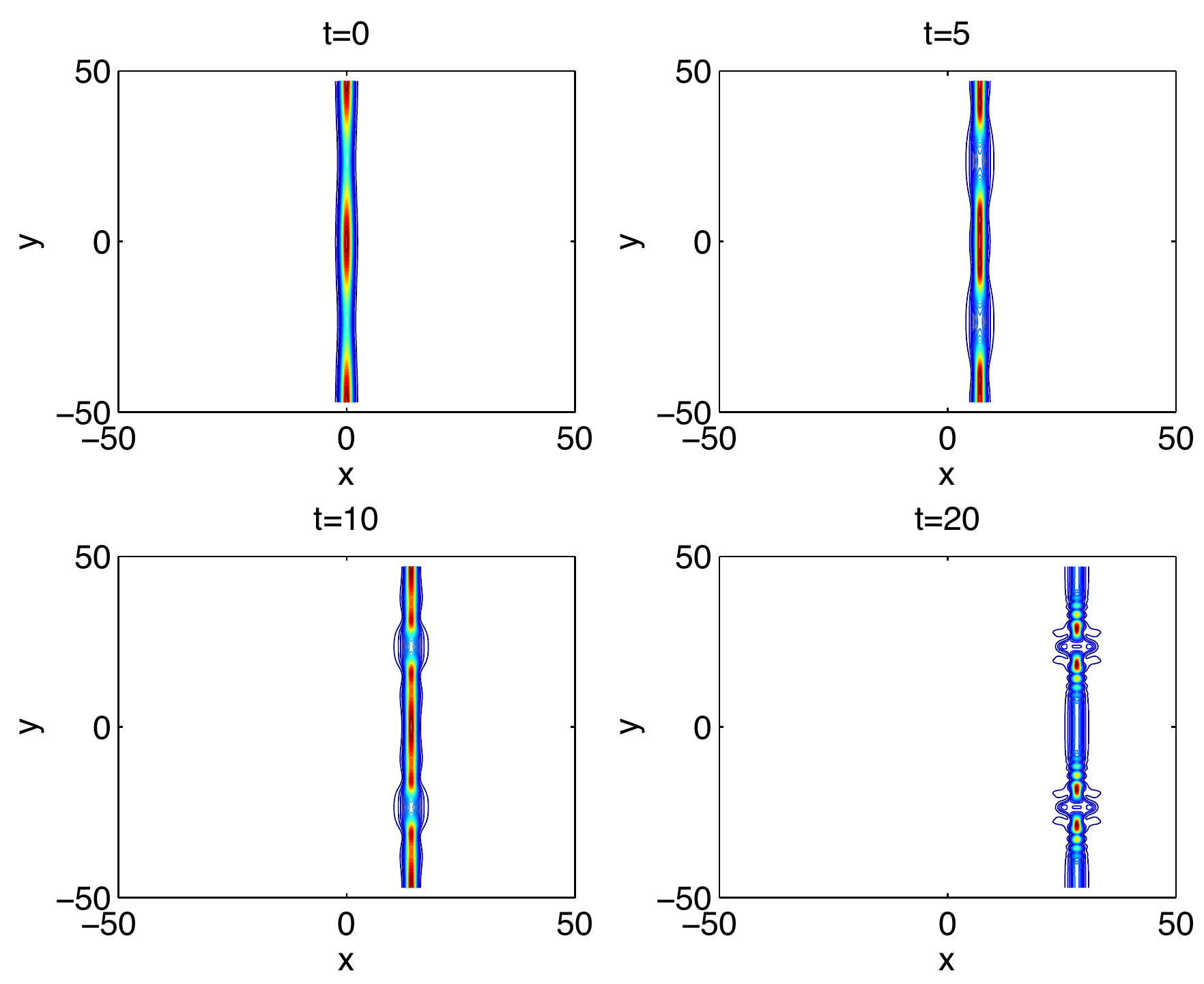} 
\caption{Solution to NLS$^-$ for an initial 
condition of the form (\ref{perio}), with $\epsilon=0.2$, $b=2$ at several times}
\label{nlshperiouts}
\end{figure}

For the NLS$^-$ equation we consider an initial data of the form (\ref{perio}) with $b=2$ and $\epsilon=0.2$, and show in Fig. \ref{nlshperiouts} the solution at several times.
We can see that the solution spreads in the $y$-direction, 
and tends to recover the initial shape of the soliton, before 
being decomposed as in the case of localized perturbations studied in the previous section.

\begin{figure}[htb!]
\centering
\includegraphics[width=0.55\textwidth]{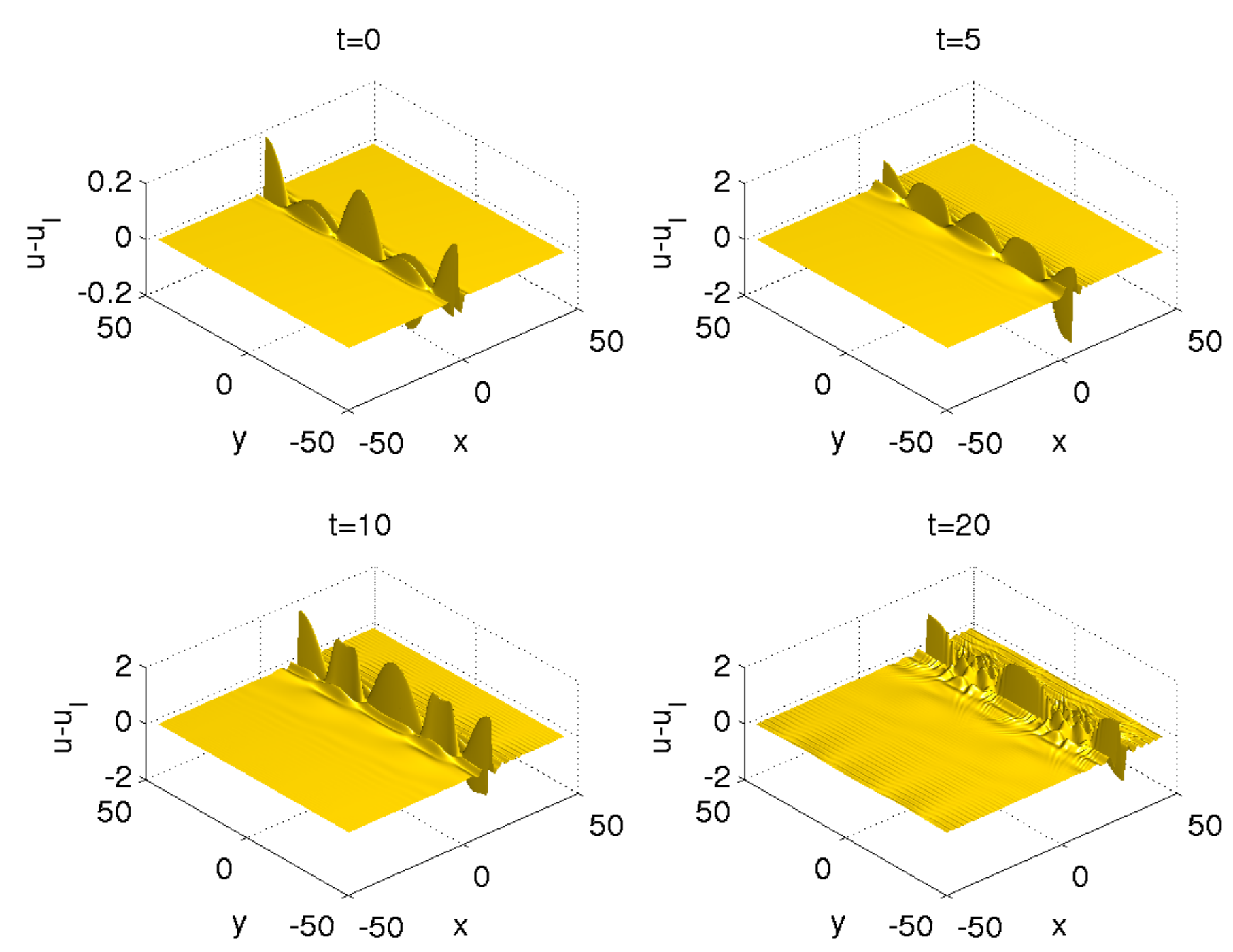}\includegraphics[width=0.55\textwidth]{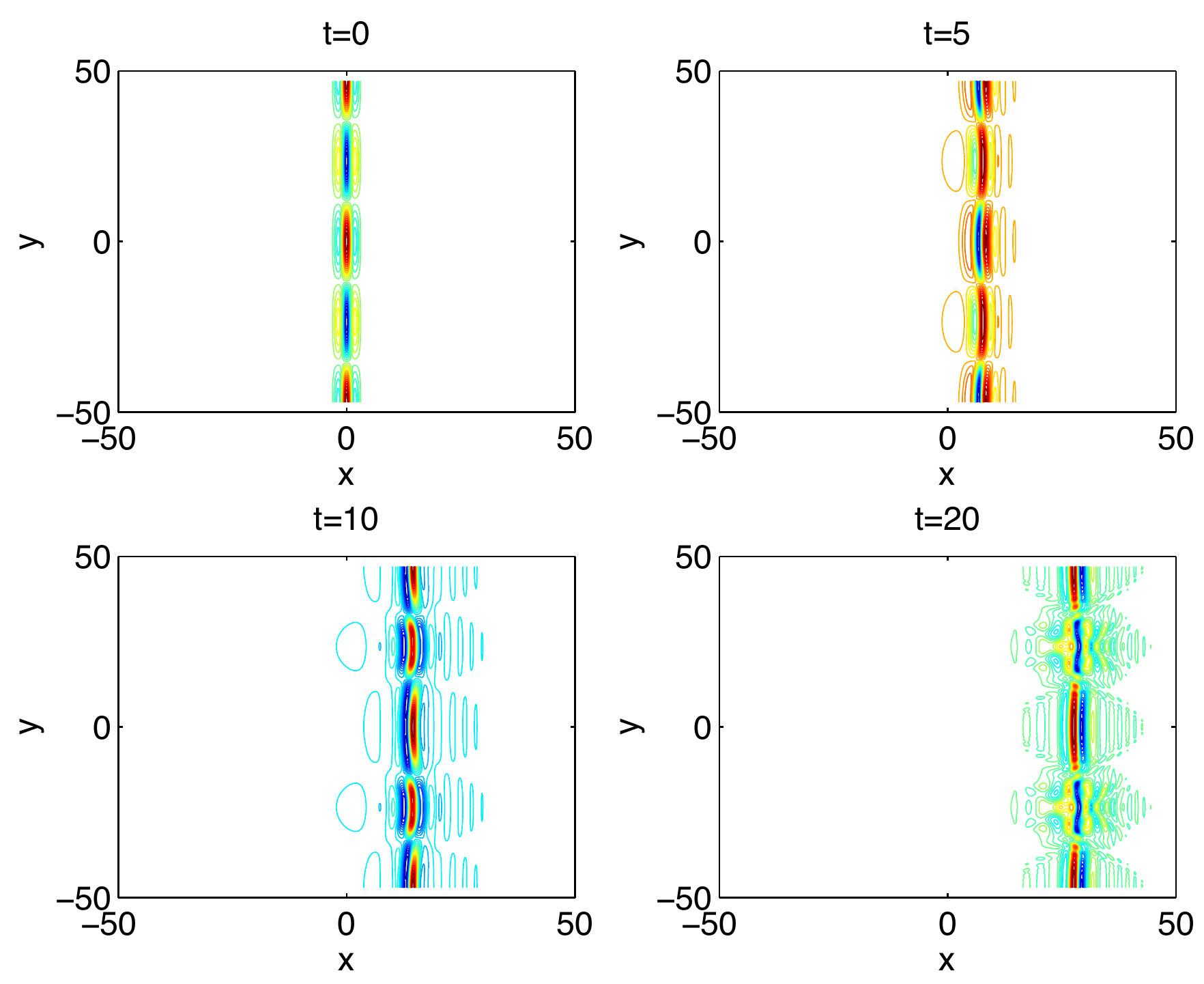} 
\caption{Difference between the solution to NLS$^-$ for an initial 
condition of the form (\ref{perio}), with $\epsilon=0.2$, $b=2$ and the original soliton $u_I$ at several times}
\label{nlshperiodiffts}
\end{figure}

The difference between the solution to NLS$^-$ for an initial 
condition of the form (\ref{perio}), with $\epsilon=0.2$, $b=2$ and the original soliton $u_I$ is shown 
in Fig. \ref{nlshperiodiffts} at several times.

\begin{figure}[htb!]
\centering
\includegraphics[width=0.45\textwidth]{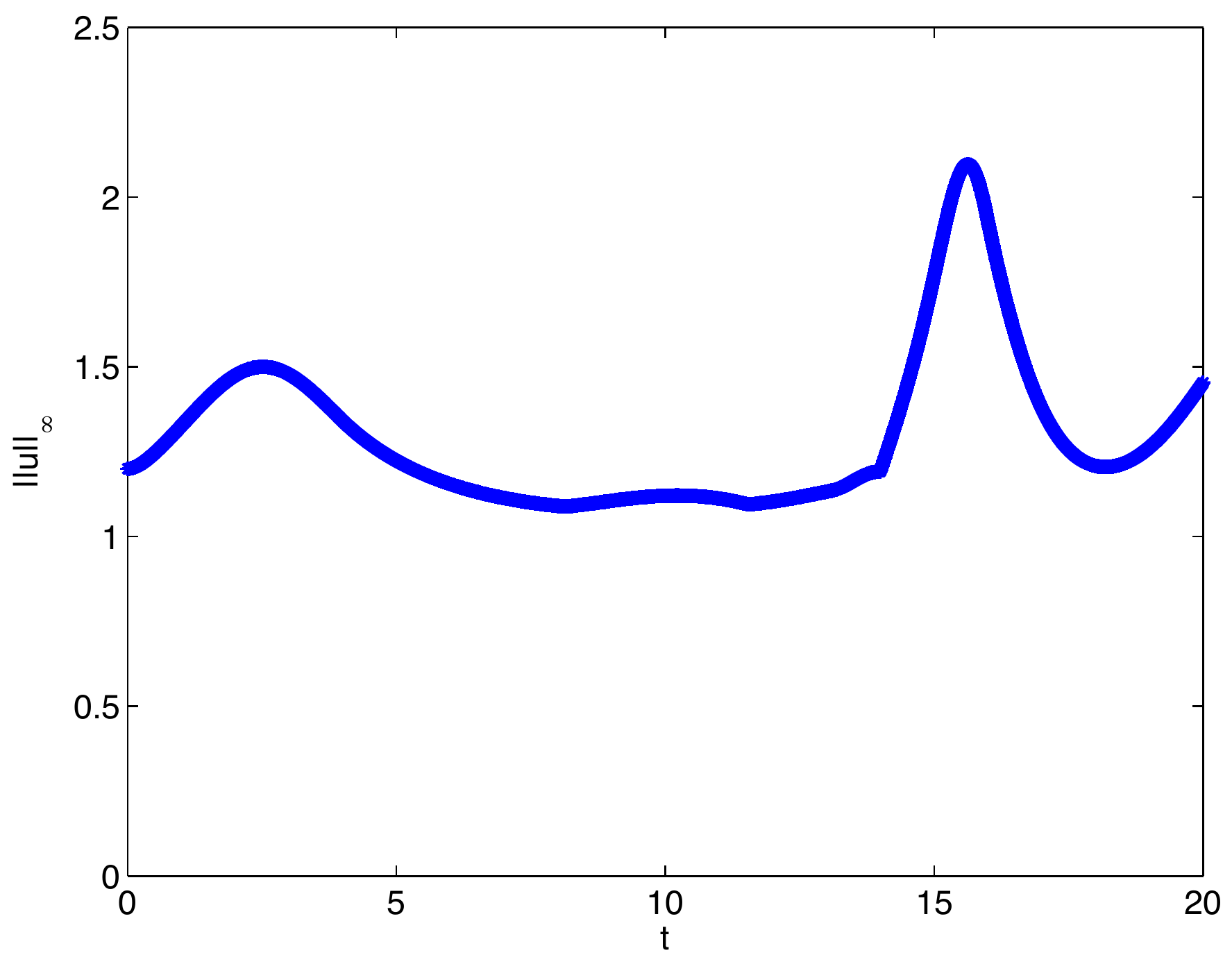} 
\includegraphics[width=0.45\textwidth]{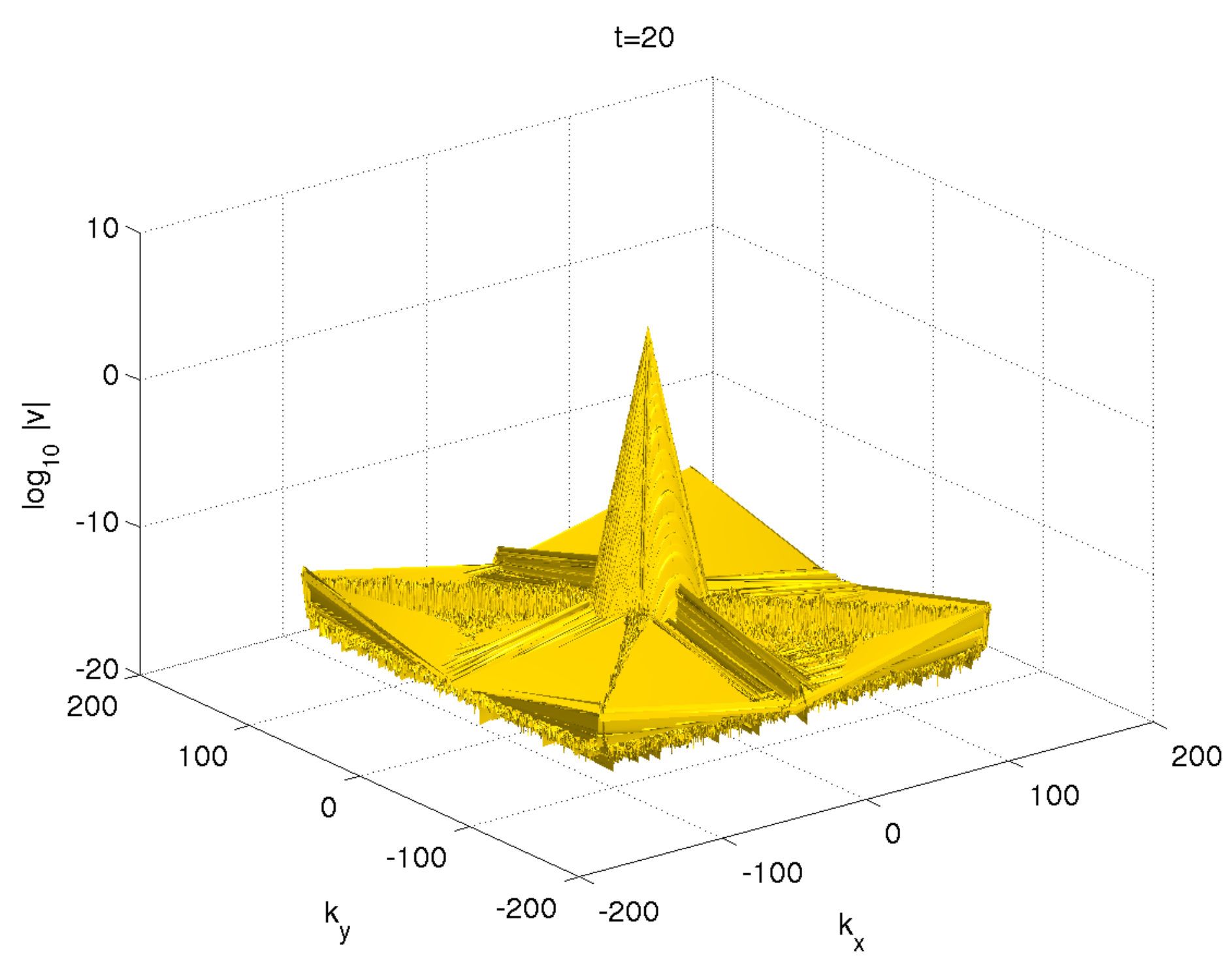} 
\caption{Time evolution of the $L_{\infty}$-norm of the solution shown in Fig. \ref{nlshperiouts} 
on the left, and Fourier coefficients of the solution at $t=t_{max}$ on the right}
\label{nlshperioampl}
\end{figure}

The $L_{\infty}$-norm of the solution is shown in Fig. \ref{nlshperioampl}, together with the Fourier coefficients 
at the maximal time of computation. The latter decrease to machine precision ($10^{-15}$) all along the computation, and $\Delta_E \sim 10^{-14} $ at the end of the computation.
As in the case of localized perturbations, the isolated soliton appears to be unstable under periodic perturbations here in 
the hyperbolic 2d NLS equation, and this instability occurs via the dispersion of the solution.

\begin{figure}[htb!]
\centering
\includegraphics[width=0.55\textwidth]{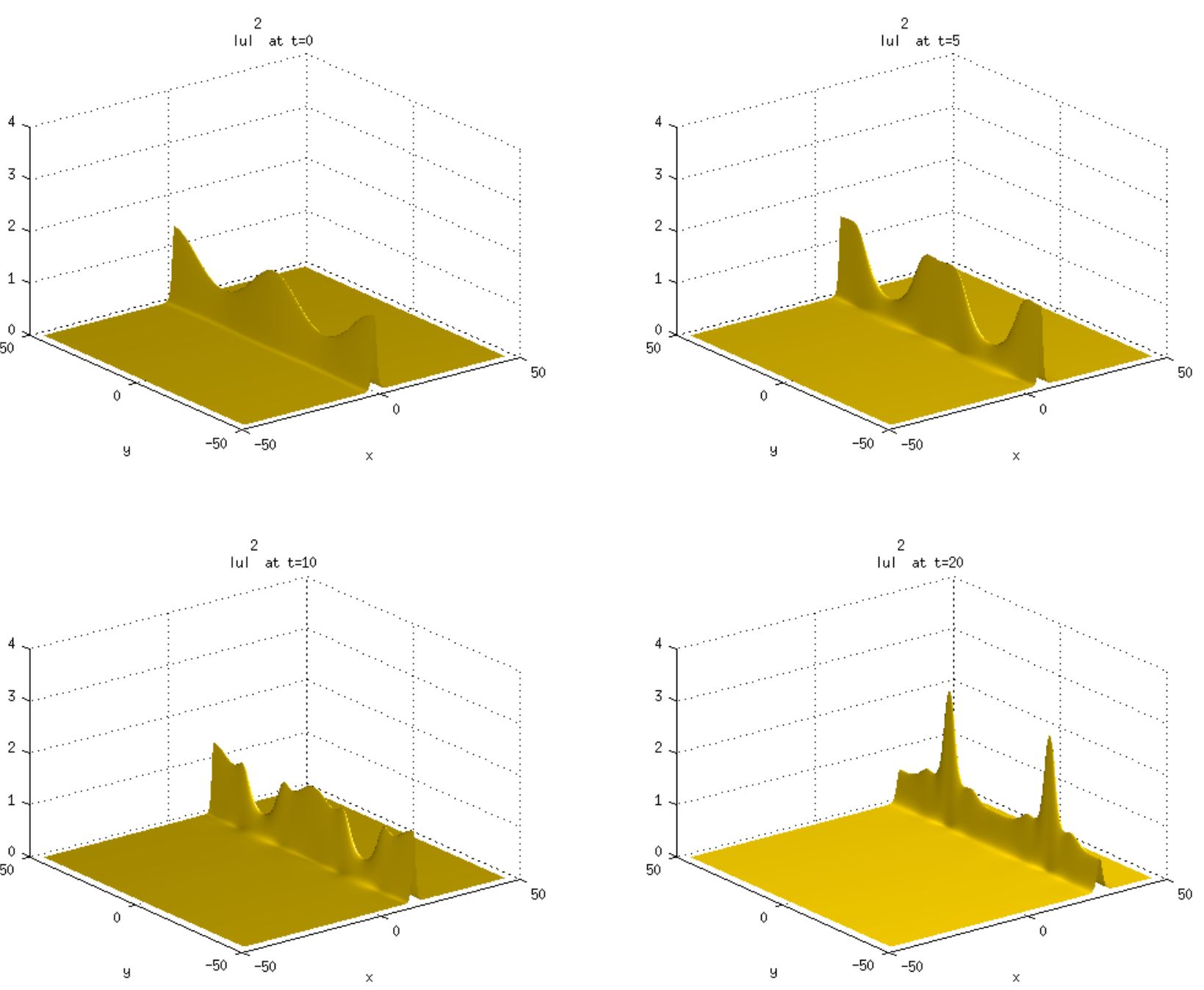}\includegraphics[width=0.55\textwidth]{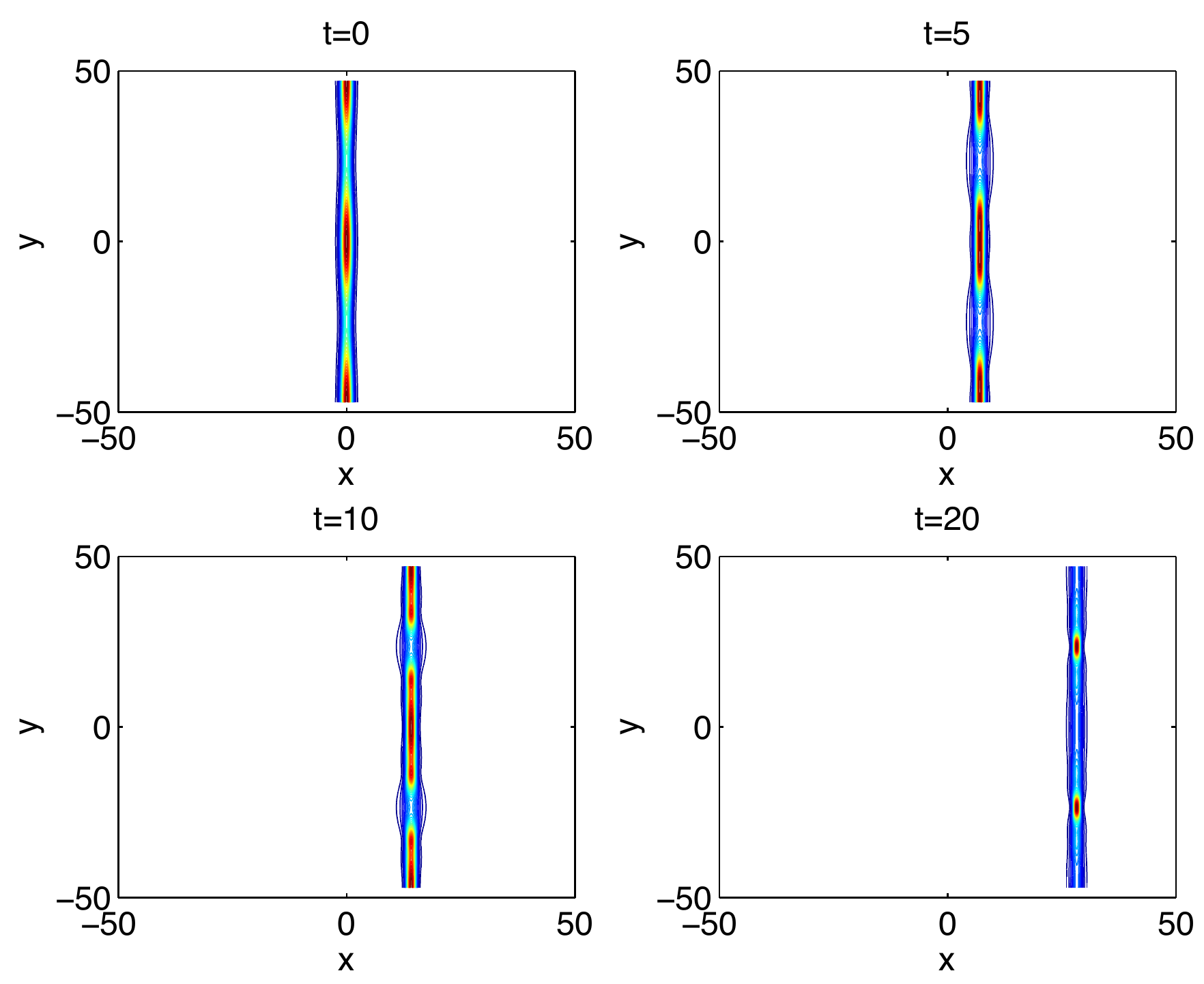} 
\caption{Solution to DS II  for an initial 
condition of the form (\ref{perio}), with $\epsilon=0.2$, $b=2$ at several times}
\label{ds2periouts}
\end{figure}

\begin{figure}[htb!]
\centering
\includegraphics[width=0.55\textwidth]{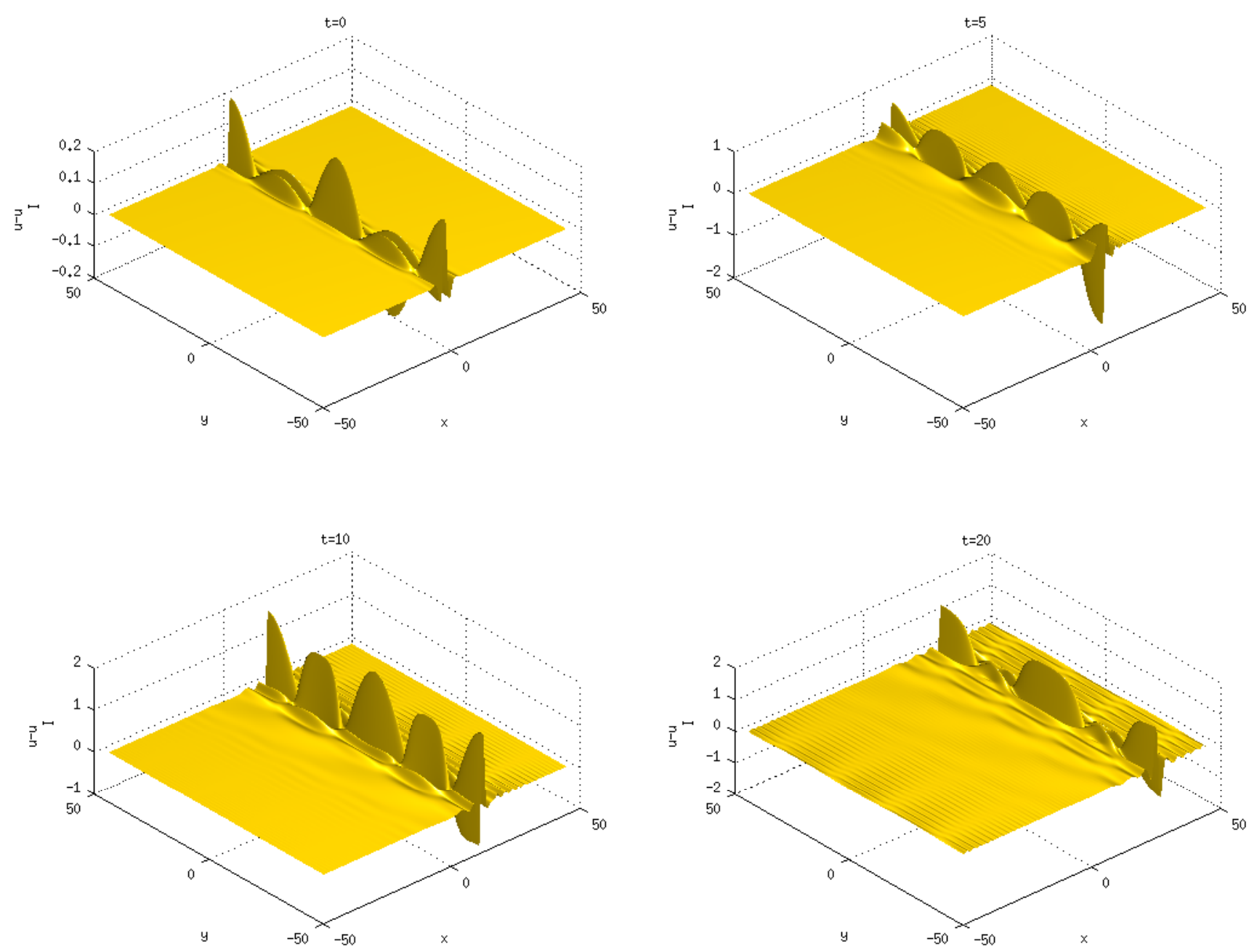}\includegraphics[width=0.55\textwidth]{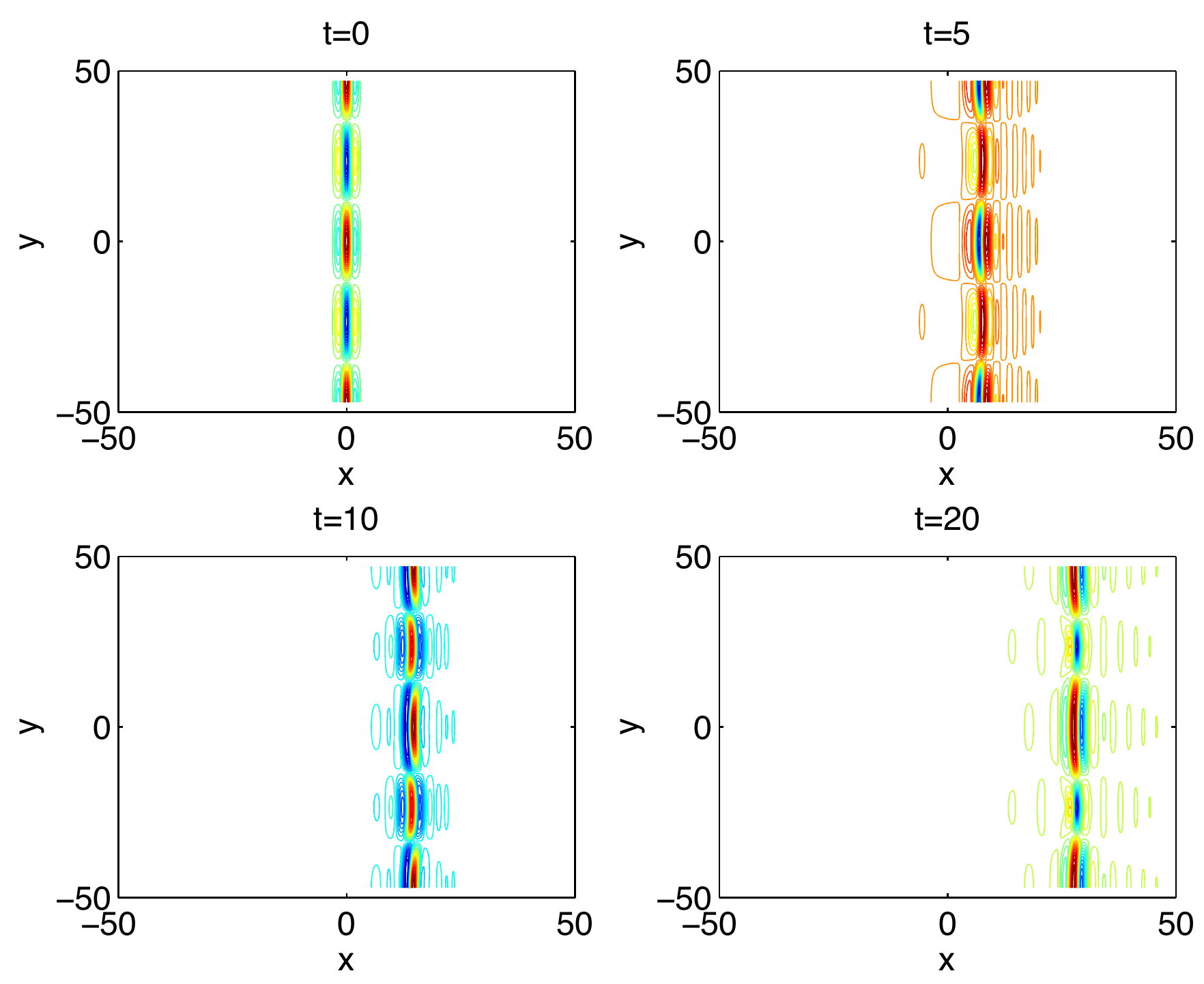}  
\caption{Difference between the solution to DS II for an initial 
condition of the form (\ref{perio}), with $\epsilon=0.2$, $b=2$ and the original 
soliton at several times}
\label{ds2periodiffts}
\end{figure}

\begin{figure}[htb!]
\centering
\includegraphics[width=0.45\textwidth]{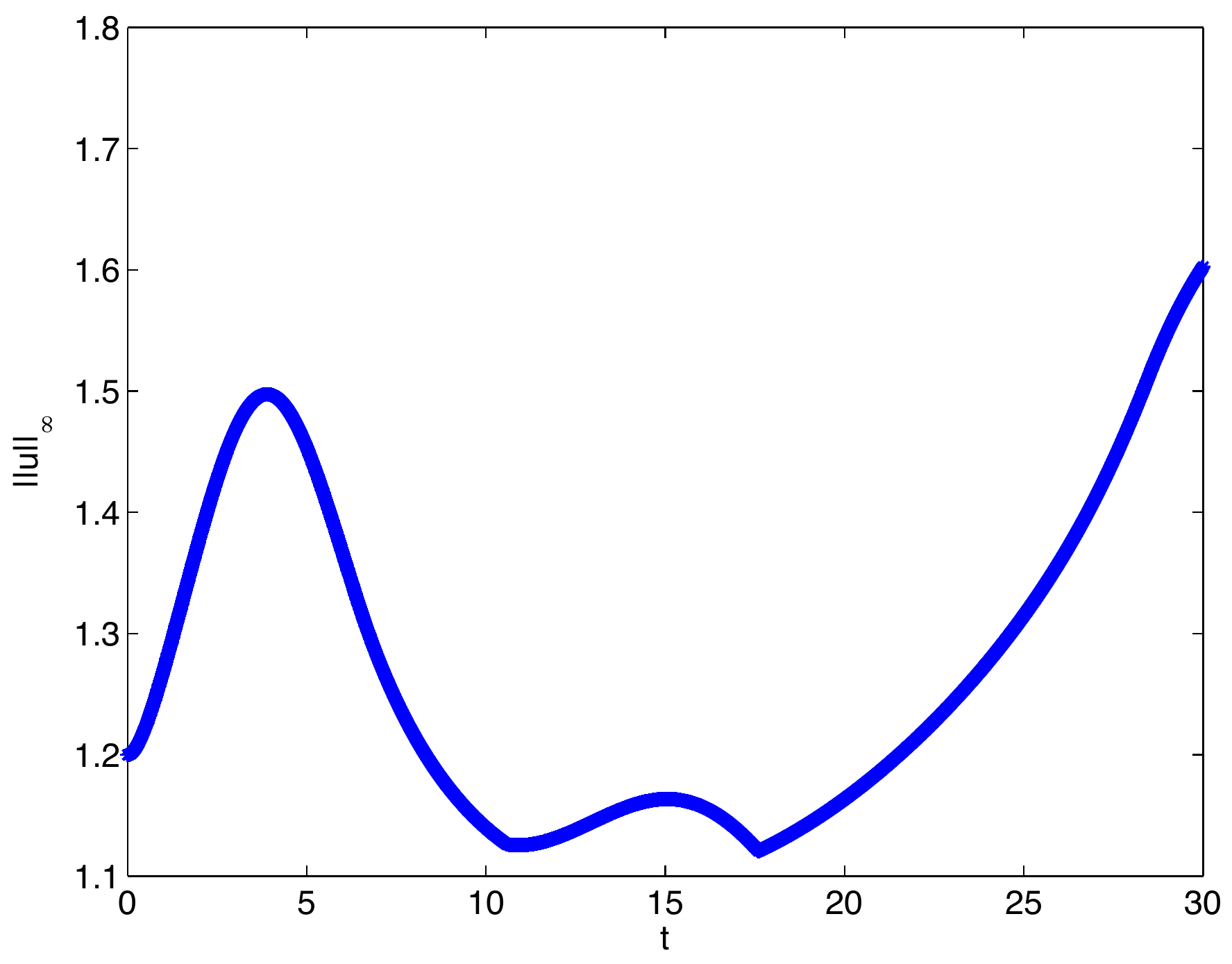} 
\includegraphics[width=0.45\textwidth]{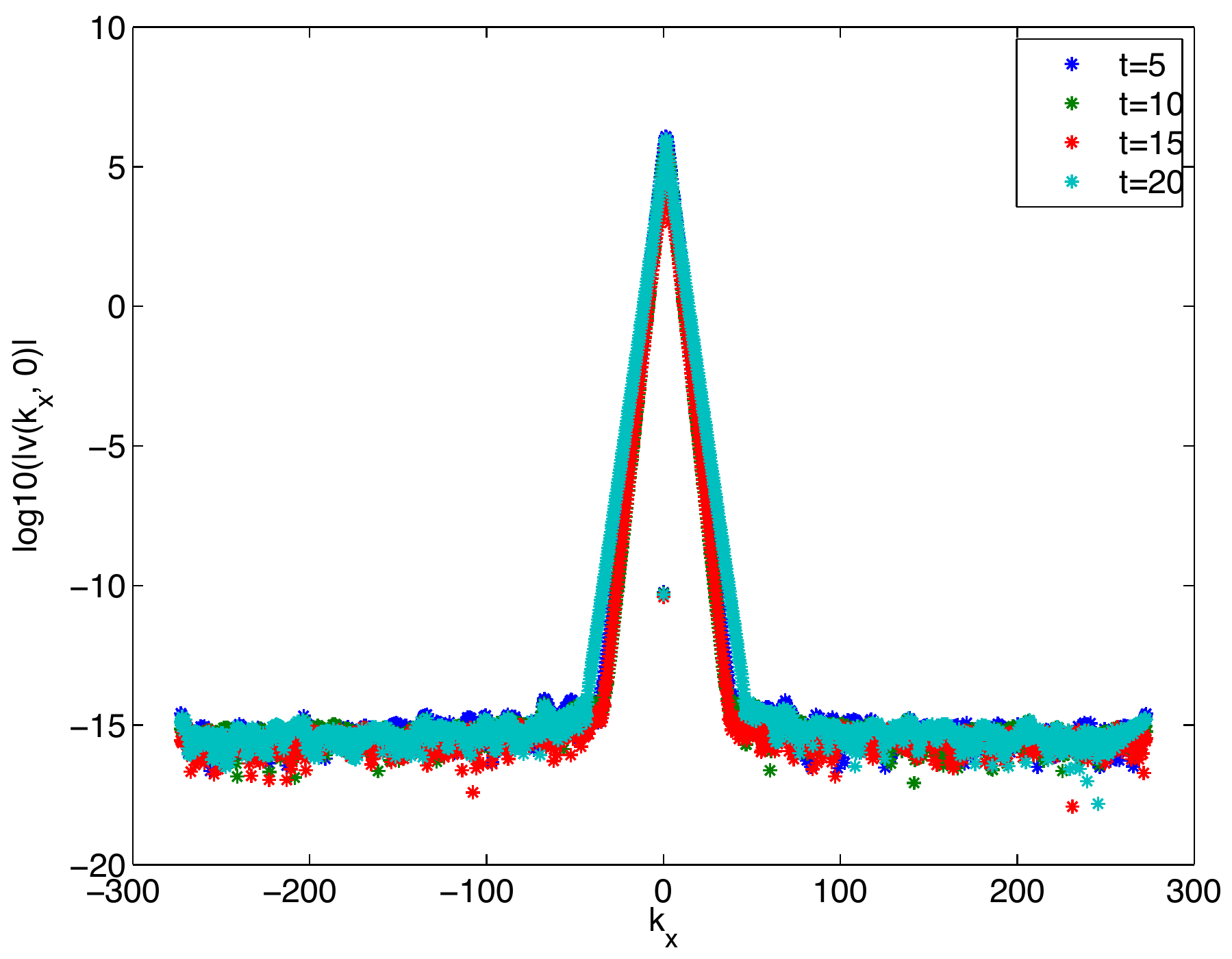} 
\caption{Time evolution of the $L_{\infty}$-norm of the solution shown in Fig. \ref{ds2periouts} and 
Fourier coefficients of the solution shown in Fig. \ref{nlshperiouts} at several times, plotted on the $k_x$-axis.}
\label{ds2perioampl}
\end{figure}

For the DS II equation we consider again an initial data of the form (\ref{perio})
with $b=2$ and $\epsilon=0.2$.
The solution is shown in Fig. \ref{ds2periouts} at several times.
The difference between the solution to DS II and the original 
soliton is shown in Fig. \ref{ds2periodiffts} at several times. We observe here that the perturbation is also dispersed around the soliton. 
The $L_{\infty}$-norm of the solution is shown in Fig. \ref{ds2perioampl}, together with the Fourier coefficients 
of the solution shown in Fig. \ref{ds2periouts}. The latter decrease to machine precision ($10^{-15}$) all along the computation, and $\Delta_E \sim 10^{-12} $ at the end of the computation.
Though it is difficult numerically to determine the long time behavior of the solution shown in Fig. \ref{ds2periouts},
the soliton appears clearly to be unstable under the time evolution of DS II for periodic deformations.

\section{Conclusion}

In this paper, we addressed numerically the question of the nonlinear transverse stability of the isolated soliton (\ref{trav}) of the 1d cubic 
NLS equation in higher dimensional models, admissible as transversally perturbed models for it, including elliptic and non-elliptic NLS equations.
Whereas for the 2d elliptic cubic NLS, a theory in \cite{RT09} has been successfully applied to this problem,
no such results are known for others models considered in this paper, in particular for hyperbolic NLS equations.
In addition, the features of the instability of the isolated soliton (\ref{trav}) under the flow of the 2d elliptic NLS remain unknown from the analysis point of view.

We investigated here these questions numerically, by using efficient numerical schemes for such dispersive PDEs which can develop singularity of blow up type in finite time. The spatial resolution as seen from the Fourier coefficients was always well beyond typical plotting accuracies of the order of $10^{-3}$. For the time integration we used a fourth order time stepping scheme already used in \cite{ckkdvnls, KR, DSdDS} for the study of NLS equations. As argued for instance in \cite{ckkdvnls, KR}, the numerically computed energy of the solution gives a valid indicator of the accuracy for sufficient spatial resolution. To ensure the latter, we always presented the Fourier coefficients of the solution.
The detection of the singularity formation has been performed thanks to a careful study of the asymptotics of the Fourier coefficients, 
as used in \cite{SSF, dkpsulart, DSdDS}. 

We found that
 the instability of the isolated soliton under the flow of the 2d elliptic NLS equation for localized perturbations occurs via a blow up in 
 the $L_{\infty}$-norm of the solution, in only one spatial point, and for periodic perturbations, this leads to blow up in multiple spatial points. 
 As expected, the same results were observed for the elliptic elliptic DS system, for which, however, neither the instability of 
 (\ref{trav}) have been proved, nor the existence of multiple blowing up solutions. 
 
 Another question that we addressed in this paper was the difference between 
 elliptic and hyperbolic variants of NLS in this context.
 We found that (\ref{trav}) is unstable for both localized and periodic perturbations under the flow of the 
 2d hyperbolic NLS equation, and that this instability occurs via the dispersion of the solution here.
 
 For the DS II equation, however, we found that the features of the soliton solutions, such as the velocity, shape, and amplitude, remain robust under localized perturbations considered here, and that the perturbation added to the soliton just appears to disperse away in this case. 
 In this sense, the soliton (\ref{trav}) appears to be somehow 'orbitally' stable under the time evolution of DS II.
 For periodic perturbations, however, it seems to be unstable.

\bibliographystyle{plain}
\bibliography{bibliof}{}

\def\cprime{$'$}
\begin{thebibliography}{10}

\bibitem{ASDS}
M.J. Ablowitz and H.~Segur.
\newblock {\em Solitons and the inverse scattering {T}ransform}, volume~4 of
  {\em SIAM Studies in Applied Mathematics}.
\newblock Society for Industrial and Applied Mathematics (SIAM), Philadelphia,
  Pa., 1981.

\bibitem{Agr}
G.~Agrawal.
\newblock Nonlinear fiber optics.
\newblock {\em Academic Press, San Diego}, 2006.

\bibitem{APPDS}
V.A. Arkadiev, A.K. Pogrebkov, and M.C. Polivanov.
\newblock Inverse scattering transform method and soliton solutions for the
  {D}avey-{S}tewartson {II} equation.
\newblock {\em Physica D}, 36:189--196, 1989.

\bibitem{arnold}
V.~I. Arnol{\cprime}d, V.~V. Kozlov, and A.~I. Ne{\u\i}shtadt.
\newblock {\em Dynamical {S}ystems. {III}}, volume~3 of {\em Encyclopaedia of
  Mathematical Sciences}.
\newblock Springer-Verlag, Berlin, 1988.
\newblock Translated from the Russian by A. Iacob.

\bibitem{FokS}
L.Y.~Sung A.S.~Fokas.
\newblock On the solvability of the {N}-wave, the {D}avey-{S}tewartson and the
  {K}adomtsev-{P}etviashmli equation.
\newblock {\em Inverse Problems}, 8:673--708, 1992.

\bibitem{BMS}
C.~Besse, N.J. Mauser, and H.P. Stimming.
\newblock Numerical {S}tudy of the {D}avey-{S}tewartson {S}ystem.
\newblock {\em M2AN}, 38(6):1035--1054, 2006.

\bibitem{Bourg99}
J.~Bourgain.
\newblock Global solutions of nonlinear {S}chr\"odinger equations.
\newblock {\em American Mathematical Society Colloquium Publications, American
  Mathematical Society, Providence, RI}, 46, 1999.

\bibitem{CR}
Russel~E. Caflisch.
\newblock Singularity formation for complex solutions of the {$3$}{D}
  incompressible {E}uler equations.
\newblock {\em Phys. D}, 67(1-3):1--18, 1993.

\bibitem{can}
C.~Canuto, M.~Y. Hussaini, A.~Quarteroni, and T.~A. Zang.
\newblock {\em Spectral methods}.
\newblock Scientific Computation. Springer-Verlag, Berlin, 2006.
\newblock Fundamentals in single domains.

\bibitem{asymbook}
G.E. Carrier and M.~Krook~C.E. Pearson.
\newblock {\em Functions of a {C}omplex {V}ariable, {T}heory and {T}echnique}.
\newblock Society for Industrial and Applied Mathematics (SIAM), Philadelphia,
  PA, 2005.

\bibitem{Cazenave12003}
T.~Cazenave.
\newblock Semilinear {S}chr\"odinger equations.
\newblock {\em Courant Lecture Notes in Mathematics, New York University
  Courant Institute of Mathematical Sciences, New York}, 10, 2003.

\bibitem{CazenaveLions}
T.~Cazenave and P.L. Lions.
\newblock Orbital stability of standing waves for some nonlinear
  {S}chr\"odinger equations.
\newblock {\em Comm. Math. Phys.}, 85:549--561, 1982.

\bibitem{CTDTVPJT}
C.~Conti, S.~Trillo, P.~Di Trapani, G.~Valiulis, A.~Piskarskas, O.~Jedrkiewicz,
  , and J.~Trull.
\newblock Nonlinear electromagnetic {X} waves.
\newblock {\em Physical Review Letters}, 90, 170406:1--4, 2003.

\bibitem{CH}
M.~Cross and P.~Hohenberg.
\newblock Pattern formation outside of equilibrium.
\newblock {\em Rev. Mod. Phys.}, 65, 1993.

\bibitem{Dris}
T.A. Driscoll.
\newblock A composite {R}unge-{K}utta {M}ethod for the spectral {S}olution of
  semilinear {PDE}s.
\newblock {\em Journal of Computational Physics}, 182:357--367, 2002.

\bibitem{Merlekblow}
Merle F.
\newblock Construction of solutions with exactly k blow-up points for the
  {S}chr\"odinger equation with critical nonlinearity.
\newblock {\em Communications in Mathematical Physics}, 129(2):223--240, 1990.

\bibitem{FL}
M.~Forest and J.~Lee.
\newblock Geometry and modulation theory for the periodic nonlinear
  {S}chr{\"o}dinger equation.
\newblock {\em in Oscillation Theory, Computation, and Methods of Compensated
  Compactness, Minneapolis, MN, 1985. The IMA Volumes in Mathematics and Its
  Applications, vol. 2, Springer, New York}, pages pp. 35--69, 1986.

\bibitem{FJ}
M.~Frigo and S.G. Johnson.
\newblock {\em {FFTW} for version 3.2.2}, July 2009.

\bibitem{FMB}
U.~Frisch, T.~Matsumoto, and J.~Bec.
\newblock Singularities of {E}uler flow? {N}ot out of the blue!
\newblock {\em J. Statist. Phys.}, 113(5-6):761--781, 2003.
\newblock Progress in statistical hydrodynamics (Santa Fe, NM, 2002).

\bibitem{RLSS}
Rocca G.D., Lombardo M.C., Sammartino M., and Sciacca V.
\newblock Singularity tracking for {C}amassa-{H}olm and {P}randtl's equations.
\newblock {\em Appl. Numer. Math.}, 56(8):1108--1122, August 2006.

\bibitem{GS}
J-M Ghidaglia and J-C. Saut.
\newblock On the initial value problem for the {D}avey-{S}tewartson systems.
\newblock {\em Nonlinearity}, 3, 1990.

\bibitem{GS96}
J.M. Ghidaglia and J.C. Saut.
\newblock Nonexistence of travelling wave solutions to nonelliptic nonlinear
  {S}chr\"odinger equations.
\newblock {\em Journal of Nonlinear Science}, 6(2):139--145, 1996.

\bibitem{GSS87a}
M.~Grillakis, J.~Shatah, and W.~Strauss.
\newblock Stability theory of solitary waves in the presence of symmetry, {I}.
\newblock {\em Journal of Functional Analysis}, 74(1):160 -- 197, 1987.

\bibitem{GSS87b}
M.~Grillakis, J.~Shatah, and W.~Strauss.
\newblock Stability theory of solitary waves in the presence of symmetry, {II}.
\newblock {\em Journal of Functional Analysis}, 94(2):308 -- 348, 1990.

\bibitem{GTL}
W.~Gropp, R.~Thakur, and E.~Lusk.
\newblock {\em Using {MPI}-2: Advanced Features of the Message Passing
  Interface}.
\newblock MIT Press Cambridge, MA, USA, second edition, 1999.

\bibitem{CazenaveBere}
Berestycki H. and Cazenave T.
\newblock Instabilité des états stationaires dans les équations de schrödinger
  et de klein-gordon non linéaires.
\newblock {\em C. R. Acad. Sci. Paris Sér. I Math.}, 293(9):489--492, 1981.

\bibitem{KassT}
A-K. Kassam and L.N. Trefethen.
\newblock Fourth-{O}rder {T}ime-{S}tepping for stiff {PDE}s.
\newblock {\em SIAM J. Sci. Comput}, 26(4):1214--1233, 2005.

\bibitem{KM77}
J.~P. Keener and D.~W. McLaughlin.
\newblock Solitons under perturbations.
\newblock {\em Phys. Rev. A}, 16:777--790, Aug 1977.

\bibitem{ckkdvnls}
C.~Klein.
\newblock Fourth order time-stepping for low dispersion {K}orteweg-de {V}ries
  and nonlinear {S}chr{\"o}dinger {E}quation.
\newblock {\em Electronic Transactions on Numerical Analysis.}, 39:116--135,
  2008.

\bibitem{KRM}
C.~Klein, B.~Muite, and K.~Roidot.
\newblock Numerical {S}tudy of {B}lowup in the {D}avey-{S}tewartson {S}ystem.
\newblock {\em DCDS-B}, 5:1361--1387, 2013.

\bibitem{KR}
C.~Klein and K.~Roidot.
\newblock Fourth order time-stepping for {K}adomtsev-{P}etviashvili and
  {D}avey-{S}tewartson equations.
\newblock {\em SIAM J. Sci. Comp.}, 2011.

\bibitem{dkpsulart}
C.~Klein and K.~Roidot.
\newblock Numerical study of shock formation in the dispersionless
  {K}adomtsev-{P}etviashvili equation and dispersive regularizations.
\newblock {\em Physica D: Nonlinear Phenomena}, 265:1--25, 2013.

\bibitem{DSdDS}
C.~Klein and K.~Roidot.
\newblock Numerical study of the semiclassical limit of the
  {D}avey-{S}tewartson {II} equations.
\newblock {\em preprint}, 2013.

\bibitem{LFSDHMC}
Y.~Lahini, E.~Frumker, Y.~Silberberg, S.~Droulias, K.~Hizanidis, R.~Morandotti,
  and D.N. Christodoulides.
\newblock Discrete {X}-wave formation in nonlinear waveguide arrays.
\newblock {\em Phys. Rev. Lett.}, 98, 023901:1--4, 2007.

\bibitem{PS98}
Pugh M. and Shelley M.
\newblock Singularity {F}ormation in thin {J} with {S}urface {T}ension.
\newblock {\em Comm. Pure Appl. Math.}, 51:733--795, 1998.

\bibitem{MBF}
T.~Matsumoto, J.~Bec, and U.~Frisch.
\newblock The analytic structure of 2{D} {E}uler flow at short times.
\newblock {\em Fluid Dynam. Res.}, 36(4-6):221--237, 2005.

\bibitem{MFP}
M.~McConnell, A.~S. Fokas, and B.~Pelloni.
\newblock Localised coherent solutions of the {DSI} and {DSII} equations---a
  numerical study.
\newblock {\em Math. Comput. Simulation}, 69(5-6):424--438, 2005.

\bibitem{MR}
F.~Merle and P.~Raphael.
\newblock On universality of blow-up profile for $l^2$ critical nonlinear
  {S}chr{\"o}dinger equation.
\newblock {\em Inventiones Mathematicae}, 156:565--672, 2004.

\bibitem{Oza}
T.~Ozawa.
\newblock Exact {B}low-up {S}olutions to the {C}auchy {P}roblem for the
  {D}avey-{S}tewartson {S}ystems.
\newblock {\em Proc. Roy. Soc. London Ser. A}, 436(1897):345--349, 1992.

\bibitem{Pelinovsky2001585}
D.~E. Pelinovsky.
\newblock A mysterious threshold for transverse instability of deep-water
  solitons.
\newblock {\em Mathematics and Computers in Simulation}, 55(4-6):585--594,
  2001.

\bibitem{RT08}
F.~Rousset and N.~Tzvetkov.
\newblock Transverse nonlinear instability of solitary waves for some
  hamiltonian pde's.
\newblock {\em J. Math. Pures Appl}, 90:550--590, 2008.

\bibitem{RT09}
F.~Rousset and N.~Tzvetkov.
\newblock Transverse nonlinear instability for two-dimensional dispersive
  models.
\newblock {\em Annales de l'Institut Henri Poincare (C) Non Linear Analysis},
  26:2:477--496, 2009.

\bibitem{DHMC}
Droulias S., Hizanidis K., Meier J., and Christodoulides D.
\newblock {X}-waves in nonlinear normally dispersive waveguide arrays.
\newblock {\em Optics Express}, 13(6):1827--1832, 2005.

\bibitem{SCE96}
D.~Senouf, R.~Caflisch, and N.~Ercolani.
\newblock Pole dynamics and oscillations for the complex {B}urgers equation in
  the small-dispersion limit.
\newblock {\em Nonlinearity}, 9:1671--1702, 1996.

\bibitem{SS}
C.~Sulem and P.L. Sulem.
\newblock {\em The nonlinear {S}chr{\"o}dinger equation}.
\newblock Springer, 1999.

\bibitem{SSF}
C.~Sulem, P.L. Sulem, and H.~Frisch.
\newblock Tracing complex singularities with spectral methods.
\newblock {\em J. Comp. Phys.}, 50:138--161, 1983.

\bibitem{SSP}
P.-L. Sulem, C.~Sulem, and A.~Patera.
\newblock Numerical simulation of singular solutions to the two-dimensional
  cubic {S}chr\"odinger equation.
\newblock {\em Comm. Pure Appl. Math.}, 37(6):755--778, 1984.

\bibitem{Sun}
L.Y. Sung.
\newblock Long-time decay of the solutions of the {D}avey-{S}tewartson {II}
  equations.
\newblock {\em J. Nonlinear Sci}, 5:433--452, 1995.

\bibitem{DTVPJTCT}
P.~Di Trapani, G.~Valiulis, A.~Piskarskas, O.~Jedrkiewicz, J.~Trull, C.~Conti,
  and S.~Trillo.
\newblock Spontaneously generated {X}-shaped light bullets.
\newblock {\em Physical Review Letters}, 91, 093904, 2003.

\bibitem{tref}
L.N. Trefethen.
\newblock {\em Spectral {M}ethods in {MATLAB}}, volume~10 of {\em Software,
  Environments, and Tools}.
\newblock Society for Industrial and Applied Mathematics (SIAM), Philadelphia,
  PA, 2000.

\bibitem{ZS72}
A.~Shabat V.~Zakharov.
\newblock Exact theory of two-dimensional self-focusing and one-dimensional
  self-modulation of waves in nonlinear media.
\newblock {\em Sov. Phys. JETP}, 34:62--69, 1972.

\bibitem{Weins83}
M.I. Weinstein.
\newblock Nonlinear {S}chr\"odinger equations and sharp interpolation
  estimates.
\newblock {\em Comm. Math. Phys.}, 87:567--576, 1983.

\bibitem{W86}
M.I. Weinstein.
\newblock Lyapunov stability of ground states of nonlinear dispersive evolution
  equations.
\newblock {\em Comm. on Pure and Appl. Math.}, 39:1:51--67, 1986.

\bibitem{WW}
P.W. White and J.A.C. Weideman.
\newblock Numerical simulation of solitons and dromions in the
  {D}avey-{S}tewartson system.
\newblock {\em Math. Comput. Simul.}, 37(4-5):469--479, December 1994.

\bibitem{Yajima01091974}
N.~Yajima.
\newblock Stability of envelope soliton.
\newblock {\em Progress of Theoretical Physics}, 52(3):1066--1067, 1974.

\bibitem{ZakRu74}
V.~E. Zakharov and A.~M. Rubenchik.
\newblock Instability of waveguides and solitons in nonlinear media.
\newblock {\em Sov. Phys. JETP}, 38:494?500, 1974.

\end{thebibliography}

\end{document}